\pdfoutput=1

\documentclass[11pt,twoside,a4paper,cmspaper,final,collab]{cms-tdr}
\def\svnVersion{311837}\def\cmsCernNo{2013-037}\def\cmsCernDate{\today}\def\cmsMessage{Published in the European Physical Journal C as \href{http://dx.doi.org/10.1140/epjc/s10052-015-3709-x}{\doi{10.1140/epjc/s10052-015-3709-x.}}}

\begin{document}\cmsNoteHeader{TOP-12-028}

\hyphenation{had-ron-i-za-tion}
\hyphenation{cal-or-i-me-ter}
\hyphenation{de-vices}
\RCS$HeadURL: svn+ssh://svn.cern.ch/reps/tdr2/papers/TOP-12-028/trunk/TOP-12-028.tex $
\RCS$Id: TOP-12-028.tex 304048 2015-09-20 15:23:47Z alverson $
\newlength\cmsFigWidth
\ifthenelse{\boolean{cms@external}}{\providecommand{\suppMaterial}{the supplemental material [URL will be inserted by publisher]}}{\providecommand{\suppMaterial}{Appendix~\ref{app:suppMat}}}
\ifthenelse{\boolean{cms@external}}{\providecommand{\breakhere}{\linebreak[4]}}{\providecommand{\breakhere}{\relax}}
\ifthenelse{\boolean{cms@external}}{\setlength\cmsFigWidth{0.85\columnwidth}}{\setlength\cmsFigWidth{0.4\textwidth}}
\ifthenelse{\boolean{cms@external}}{\providecommand{\cmsLeft}{top}}{\providecommand{\cmsLeft}{left}}
\ifthenelse{\boolean{cms@external}}{\providecommand{\cmsRight}{bottom}}{\providecommand{\cmsRight}{right}}
\ifthenelse{\boolean{cms@external}}{\providecommand{\suppRef}[2]{#2}}{\providecommand{\suppRef}[2]{#1}}
\ifthenelse{\boolean{cms@external}}{\providecommand{\supplemental}{Supplemental\xspace}}{\providecommand{\supplemental}{\relax}}
\newcommand{\mumu}{\ensuremath{\PGmp\PGmm}\xspace}
\newcommand{\ee}{\ensuremath{\Pep\Pem}\xspace}
\newcommand{\emu}{\ensuremath{\Pe^\pm\PGm^{\mp}}\xspace}
\newcommand{\ljets}{\ensuremath{\ell}+jets\xspace}
\newcommand{\Madspin}{\textsc{MadSpin}\xspace}
\newcommand{\Whizard}{\textsc{Whizard}\xspace}

\cmsNoteHeader{TOP-12-028}
\title{Measurement of the differential cross section for top quark pair production in pp collisions at \texorpdfstring{$\sqrt{s} = 8\TeV$}{sqrt(s) = 8 TeV}}

\date{\today}

\abstract{The normalized differential cross section for top quark pair (\ttbar) production is measured in pp collisions at a centre-of-mass energy of 8\TeV at the CERN LHC using the CMS detector in data corresponding to an integrated luminosity of 19.7\fbinv. The measurements are performed in the lepton+jets ($\Pe/\mu$+jets) and in the dilepton (\ee, \mumu, and \emu) decay channels. The \ttbar cross section is measured as a function of the kinematic properties of the charged leptons, the jets associated to b quarks, the top quarks, and the \ttbar system. The data are compared with several predictions from perturbative QCD up to approximate next-to-next-to-leading-order precision. No significant deviations are observed relative to the standard model predictions.
}

\hypersetup{%
pdfauthor={CMS Collaboration},%
pdftitle={Measurement of the differential cross section for top quark pair production in pp collisions at sqrt(s) = 8 TeV},%
pdfsubject={CMS},%
pdfkeywords={CMS, physics, top quarks, differential cross sections}}

\maketitle

\section{Introduction}

Understanding the production and properties of top quarks is fundamental for testing the quality of the standard model (SM) and for searching for new physical phenomena beyond its scope. The large top quark data samples produced in proton-proton (pp) collisions at the CERN LHC provide access to precision measurements that are crucial for checking the internal consistency of the SM at the LHC energy scale. In particular, measurements of the top quark pair (\ttbar) production cross section as a function of \ttbar kinematic observables are important for comparing with the state-of-the-art quantum chromodynamic (QCD) predictions within the SM, and thereby constrain QCD parameters. In addition, the top quark plays a relevant role in theories beyond the SM, and such differential measurements are therefore expected to be sensitive to new phenomena~\cite{bib:newphys}.

Differential \ttbar production cross sections have been measured previously at the Fermilab $\Pp\Pap$ Tevatron~\cite{bib:CDF,bib:D0}, and at the LHC at a centre-of-mass energy $\sqrt{s}=7$\TeV~\cite{bib:ATLAS,bib:TOP-11-013_paper,bib:ATLASnew}. We present here the first measurement of the normalized differential \ttbar production cross section with the CMS detector at $\sqrt{s}=8$\TeV. The analysis uses data recorded in 2012 corresponding to an integrated luminosity of $19.7 \pm 0.5\fbinv$, which is about a factor of four larger than the sample used in the measurement performed by the CMS Collaboration at 7\TeV~\cite{bib:TOP-11-013_paper}. The analysis largely follows the procedures of Ref.~\cite{bib:TOP-11-013_paper} and benefits from the increase in statistical precision together with improvements in kinematic reconstruction algorithms and extended systematic studies, leading to a significant reduction of the total uncertainties.

The measurements are performed in \ljets channels ($\ell = \Pe \text{ or }\mu$), which contain a single isolated charged lepton and at least four jets in the final state, and in dilepton channels, with two oppositely charged leptons (\ee, \mumu, \emu) and at least two jets. The \ttbar cross section is determined as a function of the kinematic properties of the top quarks and of the \ttbar system, as well as of the leptons and jets associated with bottom (b) quarks (b jets) from top quark decays.

The kinematic properties of top quarks are obtained through kinematic-fitting and reconstruction algorithms. The normalized differential \ttbar cross section is determined by counting the number of \ttbar signal events in each bin of a given observable, correcting for detector effects and acceptance, and dividing by the measured total inclusive \ttbar event rate. The latter is evaluated by integrating over all bins in each observable.

The results for directly measured quantities, such as kinematic properties of leptons and b jets, are presented in a fiducial phase space defined by the kinematic and geometric acceptance of all selected final-state objects. This avoids extrapolating the measured cross section into regions that are not experimentally accessible. In addition, the top quark and \ttbar distributions are determined in the full phase space, in order to facilitate the comparison with higher-order perturbative QCD calculations.
The results are compared to several predictions obtained with the leading-order (LO) \MADGRAPH~\cite{bib:madgraph} generator interfaced to \PYTHIA~\cite{bib:pythia} for parton evolution and hadronization, the next-to-leading-order (NLO) generators \POWHEG~\cite{bib:powheg1,bib:powheg2,bib:powheg3}, interfaced to both \PYTHIA and \HERWIG~\cite{bib:HERWIG}, and \MCATNLO \cite{bib:mcatnlo} interfaced to \HERWIG, and the latest NLO calculations with next-to-next-to-leading-logarithm (NNLL) corrections \cite{bib:ahrens_mttbar, bib:ahrens_ptttbar}, and approximate next-to-next-to-leading-order (NNLO) predictions~\cite{bib:kidonakis_8TeV}. The approximate NNLO predictions can be computed with the \textsc{DiffTop}~\cite{bib:difftop} program.

This document is structured as follows. A brief description of the CMS detector is provided in Section~\ref{sec:detector}. Details of the event simulation are given in Section~\ref{sec:simulation}, and event reconstruction and selection are discussed in Section~\ref{sec:selection}. The estimated systematic uncertainties on the measurements of the cross section are described in Section~\ref{sec:errors}. The results of the measurement are discussed in Section~\ref{sec:diffxsec}, followed by a summary in Section~\ref{sec:concl}.

\section{CMS detector}
\label{sec:detector}

The central feature of the CMS apparatus is a superconducting solenoid of 13\unit{m} length and 6\unit{m} inner diameter, which provides an axial magnetic field of 3.8\unit{T}. Within the field volume are a silicon-pixel and strip tracker, a lead tungstate crystal electromagnetic calorimeter (ECAL), and a brass and scintillator hadron calorimeter (HCAL), each composed of a barrel and two endcap sections. Charged particle trajectories are measured by the inner tracking system, covering a pseudorapidity range of $\abs{\eta}<2.5$. The ECAL and the HCAL surround the tracking volume, providing high-resolution energy and direction measurements of electrons, photons, and hadronic jets up to $\abs{\eta}<3$. Muons are measured in gas-ionization detectors embedded in the steel flux return yoke outside the solenoid covering the region $\abs{\eta}<2.4$. Extensive forward calorimetry complements the coverage provided by the barrel and endcap detectors up to $\abs{\eta}<5.2$. The detector is nearly hermetic, allowing for energy balance measurements in the plane transverse to the beam directions. A two-tier trigger system selects the pp collisions for use in the analysis. A more detailed description of the CMS detector, together with a definition of the coordinate system and the relevant kinematic variables, can be found in Ref.~\cite{bib:JINST}.

\section{Event simulation and theoretical calculations}
\label{sec:simulation}

Event generators, interfaced with a detailed detector simulation, are used to model experimental effects, such as consequences of event reconstruction and choice of selection criteria, as well as detector resolution.
The \ttbar sample is simulated using the LO \MADGRAPH event generator (v.~5.1.5.11), which implements the relevant matrix elements with up to three additional partons. The \Madspin~\cite{bib:madspin} package is used to incorporate spin correlation effects with matrix elements for up to three additional partons. The value of the top quark mass is fixed to $m_{\PQt}=172.5\GeV$ and the proton structure is described by the parton distribution functions (PDF) CTEQ6L1~\cite{bib:cteq}. The generated events are subsequently processed with \PYTHIA (v.~6.426, referred to as \PYTHIA{6} in the following) for parton showering and hadronization, and the MLM prescription~\cite{bib:MLM} is used for matching of matrix-element jets to parton showers. The CMS detector response is simulated using \GEANTfour (v.~9.4)~\cite{bib:geant}.

In addition to the \MADGRAPH prediction, calculations obtained with the NLO generators \MCATNLO (v.~3.41) and \POWHEG (v.~1.0 r1380) are compared to the results presented in Section~\ref{sec:diffxsec}.  While \POWHEG and \MCATNLO are formally equivalent up to the NLO accuracy, they differ in the techniques used to avoid double counting of radiative corrections that can arise from interfacing with the parton showering generators. Two \POWHEG samples are used: one is processed through \PYTHIA{6} and the other through \HERWIG (v.~6.520, referred to as \HERWIG{6} in the following) for the subsequent parton showering and hadronization. The parton showering in \PYTHIA{6} is based on a transverse-momentum-ordered evolution scale, whereas in \HERWIG{6} it is angular-ordered. The events generated with \MCATNLO are interfaced with \HERWIG{6}. The \HERWIG{6} AUET2 tune~\cite{bib:auet2tune} is used to model the underlying event in the \POWHEG{}+\HERWIG{6} sample, while the default tune is used in the \MCATNLO{}+\HERWIG{6} sample. The proton structure is described by the PDF sets CT10~\cite{bib:CT10} and CTEQ6M~\cite{bib:cteq} for \POWHEG and \MCATNLO, respectively. In addition, the latest available NLO+NNLL~\cite{bib:ahrens_mttbar, bib:ahrens_ptttbar} and approximate NNLO QCD predictions~\cite{bib:kidonakis_8TeV} are also used to compare with the data. The NNLO MSTW2008~\cite{bib:MSTW} PDF set is used for both the NLO+NNLL and the approximate NNLO calculations.

Standard model background samples are simulated with \MADGRAPH (without the \Madspin package), \POWHEG, or \PYTHIA{6}, depending on the process. The main background contributions originate from the production of W and Z/$\gamma^{*}$ bosons with additional jets (referred to as W+jets and Z+jets, respectively, in the following), single top quark ($s$-, $t$-, and tW channels), diboson (WW, WZ, and ZZ), \ttbar production in association with a Z, W, or $\gamma$ boson (referred to as \ttbar+Z/W/$\gamma$ in the following), and QCD multijet events. The W+jets, Z+jets, and \ttbar+Z/W/$\gamma$ samples are simulated with \MADGRAPH with up to two additional partons in the final state. The \POWHEG generator is used for simulating single top quark production, while \PYTHIA{6} is used to simulate diboson and QCD multijet events. Parton showering and hadronization are also simulated with \PYTHIA{6} in all the background samples. The \PYTHIA{6} Z2* tune~\cite{bib:Z2startune} is used to characterize the underlying event in both the \ttbar and the background samples.

{\tolerance=600
For comparison with the measured distributions, the event yields in the simulated samples are normalized to an integrated luminosity of 19.7\fbinv, according to their predicted cross sections. These are taken from NNLO (W+jets~\cite{bib:Z,bib:W} and Z+jets~\cite{bib:Z}), NLO+NNLL (single top quark $s$-, $t$-, and tW channels~\cite{bib:kidonakis_8TeV}), NLO (diboson~\cite{bib:mcfm:diboson}, \ttbar+W~\cite{bib:ttW}, and \ttbar+Z~\cite{bib:ttZ}), and LO (QCD multijet~\cite{bib:pythia}) calculations. The predicted cross section for the \ttbar+$\gamma$ sample is obtained by scaling the LO cross section obtained with the \Whizard event generator~\cite{bib:whizard} by an NLO/LO correction $K$-factor~\cite{bib:ttgamma}. Correction factors described in Sections~\ref{sec:selection} and~\ref{sec:errors}, and subsequently referred to as scale factors, are applied when needed to improve the description of the data by the simulation. The \ttbar simulation is normalized to the data to present the expected rates in the figures in Section~\ref{sec:selection}.
\par}

\section{Event reconstruction and selection}
\label{sec:selection}

The event selection is similar to that described in Ref.~\cite{bib:TOP-11-013_paper} for the measurement of normalized differential \ttbar cross sections at $\sqrt{s} = 7$\TeV, and is based on the final-state topology of \ttbar events. The top quark decays almost exclusively into a W boson and a b quark, and only the subsequent decays of one or two of the W bosons into a charged lepton (electron or muon) and a neutrino are considered. These signatures imply the presence of isolated leptons with high transverse momentum \pt, large \pt imbalance caused by the neutrinos that escape detection, and highly energetic jets. The identification of b jets through b-tagging techniques is used to increase the purity of the selected sample. The event selection in each channel is optimized to maximize the content of \ttbar signal events and background rejection.

\subsection{Lepton, jet, and missing transverse energy reconstruction}
\label{subsec:lepjetreco}

Events are reconstructed using a particle-flow technique~\cite{bib:pf2009, bib:pf2010}, which combines signals from all subdetectors to enhance the reconstruction and identification of individual particles observed in pp~collisions. Charged hadrons from pileup events, \ie those originating from additional pp interactions within the same bunch crossing, are subtracted on an event-by-event basis. Subsequently, the remaining neutral-hadron component from pileup is accounted for through jet energy corrections~\cite{bib:PUSubtraction}.

Electron candidates are reconstructed from a combination of the track momentum at the main interaction vertex, the corresponding energy deposition in the ECAL, and the energy sum of all bremsstrahlung photons attached to the track~\cite{bib:ele2013}. The candidates are required to have $\pt > 33\GeV$ within the pseudorapidity interval $\abs{\eta} < 2.1$ for the \ljets channels, while electron candidates in the dilepton channels are required to have $\pt > 20\GeV$ and $\abs{\eta} < 2.4$. As an additional quality criterion, a relative isolation $I_{\text{rel}}(0.3) < 0.10$ in the \ljets channels and $I_{\text{rel}}(0.3) < 0.15$ in the dilepton channels is required, where $I_{\text{rel}}(x)$ is defined as the sum of the \pt of all neutral and charged reconstructed particle candidates inside a cone of $\Delta R\equiv\sqrt{\smash[b]{(\Delta\eta)^2 + (\Delta\phi)^2}} < x$ around the electron (excluding the electron itself) in $\eta$-$\phi$~space, divided by the \pt of the electron.

Muon candidates are reconstructed using the track information from the silicon tracker and the muon system. They are required to have $\pt > 33\GeV$ and $\abs{\eta} < 2.1$ in the \ljets channels, while in the dilepton channels the corresponding selection requires $\pt>20\GeV$ and $\abs{\eta}<2.4$. Isolated muon candidates are selected if they fulfill $I_{\text{rel}}(0.4) < 0.12$ and $I_{\text{rel}}(0.3) < 0.15$ in the \ljets and dilepton channels, respectively. The same definition of relative isolation described above is also used for muon candidates.

Jets are reconstructed by clustering the particle-flow candidates~\cite{bib:JME-10-011:JES} using the anti-\kt clustering algorithm with a distance parameter of $R = 0.5$~\cite{bib:antikt}. Electrons and muons passing less stringent selections on lepton kinematic quantities and isolation, relative to the ones specified above, are identified but excluded from clustering. A jet is selected if it has $\pt > 30\GeV$ and $\abs{\eta} < 2.4$ for both the \ljets and dilepton channels. Jets originating from b quarks are identified through a ``combined secondary vertex'' algorithm~\cite{bib:btag004}, which provides a b-tagging discriminant by combining secondary vertices and track-based lifetime information. The chosen working point in the \ljets channels has an efficiency for tagging a b jet of $\approx$60\%, while the probability to misidentify light-flavour jets as b jets (mistag rate) is only ${\approx}$1.5\%. In the dilepton channels, the working point is selected to provide b-tagging efficiency and mistag rate of ${\approx}$80--85\% and ${\approx}$10\%, respectively~\cite{bib:btag004}. These requirements are chosen to reduce the background contribution in the corresponding channels while keeping a large fraction of the \ttbar signal.

The missing transverse energy \ETslash is defined as the magnitude of the imbalance in the transverse momentum $\vec{\PTslash}$ in the event, which is the negative of the vectorial sum of the momenta in the transverse plane of all the particles reconstructed with the particle-flow algorithm~\cite{bib:MET}. To mitigate the effect of contributions from pileup on the resolution in \ETslash, we use a multivariate correction where the input is separated into components that originate from the primary and other collision vertices~\cite{bib:mvamet}. This correction improves the \ETslash resolution by $\approx$5\%.

\subsection{Event selection}
\label{subsec:evsel}

Events in the \ljets channels that are triggered by the presence of a single electron (muon) with $\pt > 27\GeV$ ($\pt > 24\GeV$, $\abs{\eta} < 2.1$), are selected if they contain exactly one reconstructed lepton fulfilling the requirements described in Section~\ref{subsec:lepjetreco}. Events are rejected if there are additional electron candidates with $\pt > 20\GeV$, $\abs{\eta} < 2.5$, and $I_{\text{rel}}(0.3) < 0.15$, or additional muon candidates with $\pt > 10\GeV$, $\abs{\eta} < 2.5$, and $I_{\text{rel}}(0.4) < 0.2$. Additionally, an event must contain at least four reconstructed jets satisfying the criteria described in Section~\ref{subsec:lepjetreco}. To suppress background contribution mainly from W+jets events, at least two of these jets are required to be tagged as b jets, and at least two must not be tagged as b jets, as they are used to reconstruct $\PW\to \PQq\PAQq'$ decays. In the dilepton channels, events are triggered using combinations of two leptons with \pt thresholds of 8 and 17\GeV, and are selected if they contain at least two isolated leptons of opposite electric charge and at least two jets. At least one of the jets is required to be b-tagged. In events with more than two leptons, we choose the lepton pair with opposite charge and largest value in the sum of their scalar \pt. Events with an invariant mass of the lepton pair smaller than 20\GeV are removed to suppress events from decays of heavy-flavour resonances and low-mass Drell--Yan processes. Backgrounds from Z+jets processes in the \ee and \mumu channels are also suppressed by requiring the dilepton invariant mass to be outside a Z boson mass window of $91 \pm 15\GeV$, and to have $\ETslash > 40\GeV$.

\begin{figure*}[htbp]
  \centering
    \includegraphics[width=0.48\textwidth]{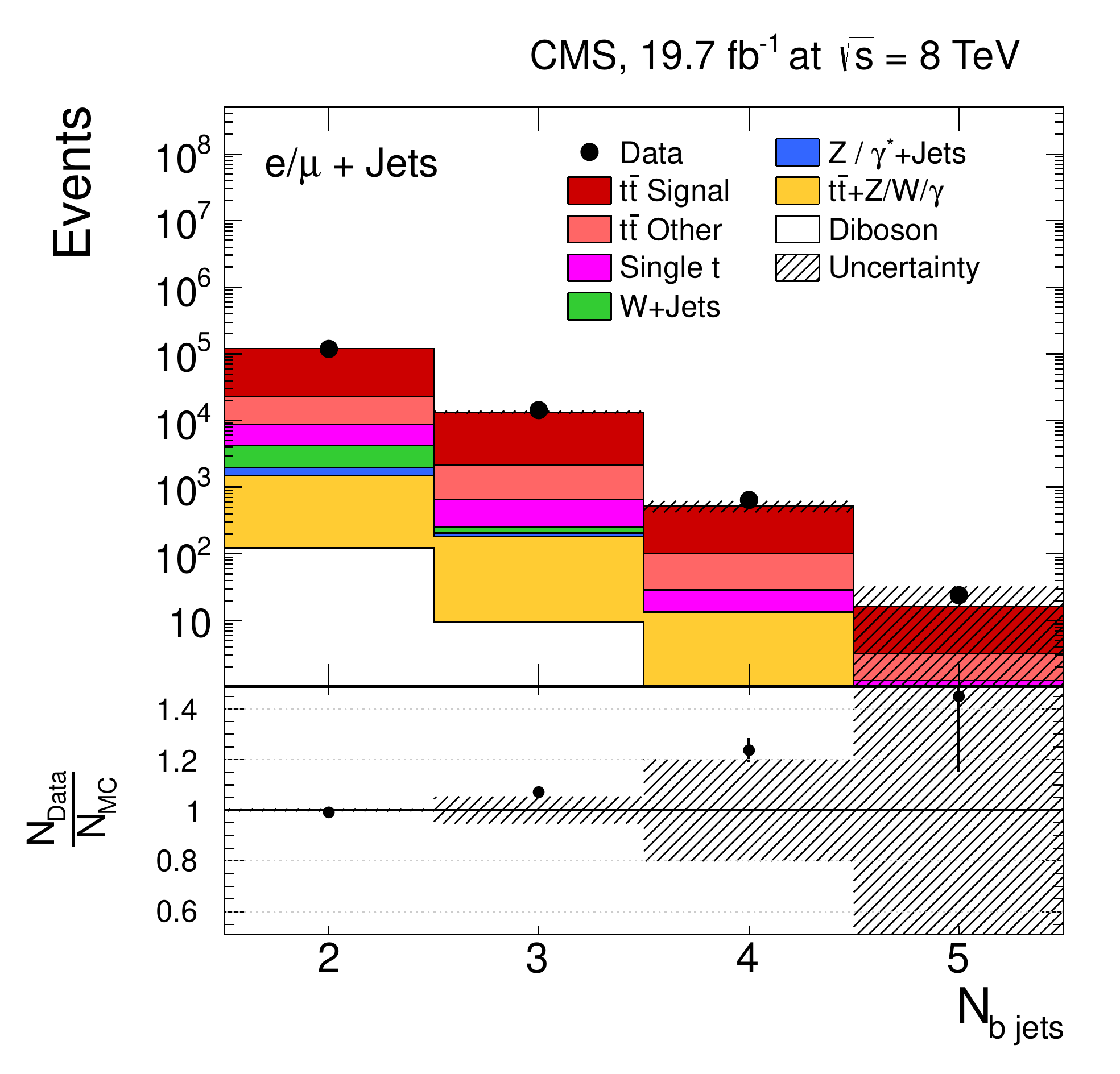}
    \includegraphics[width=0.48\textwidth]{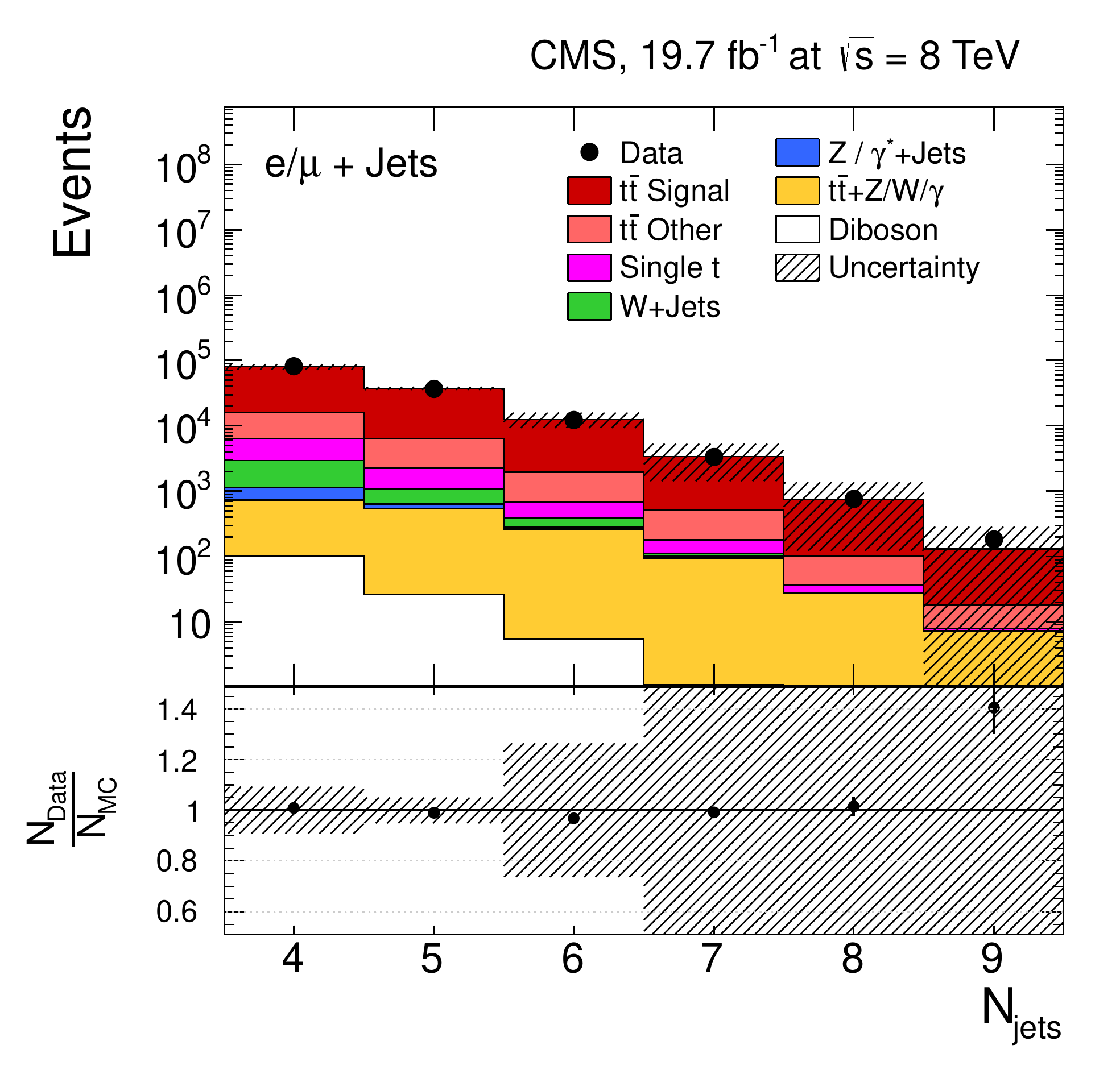}
    \includegraphics[width=0.48\textwidth]{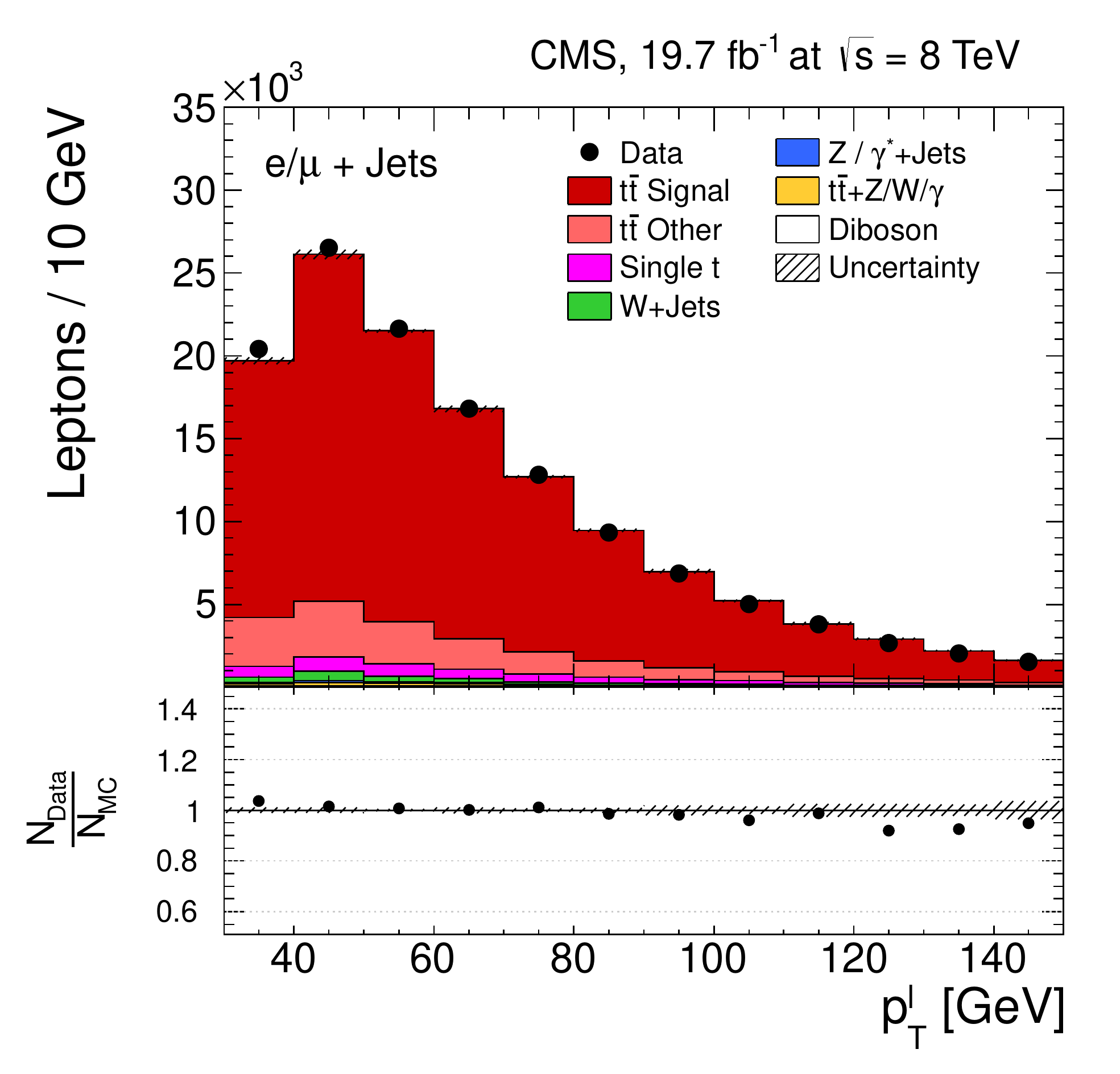}
    \includegraphics[width=0.48\textwidth]{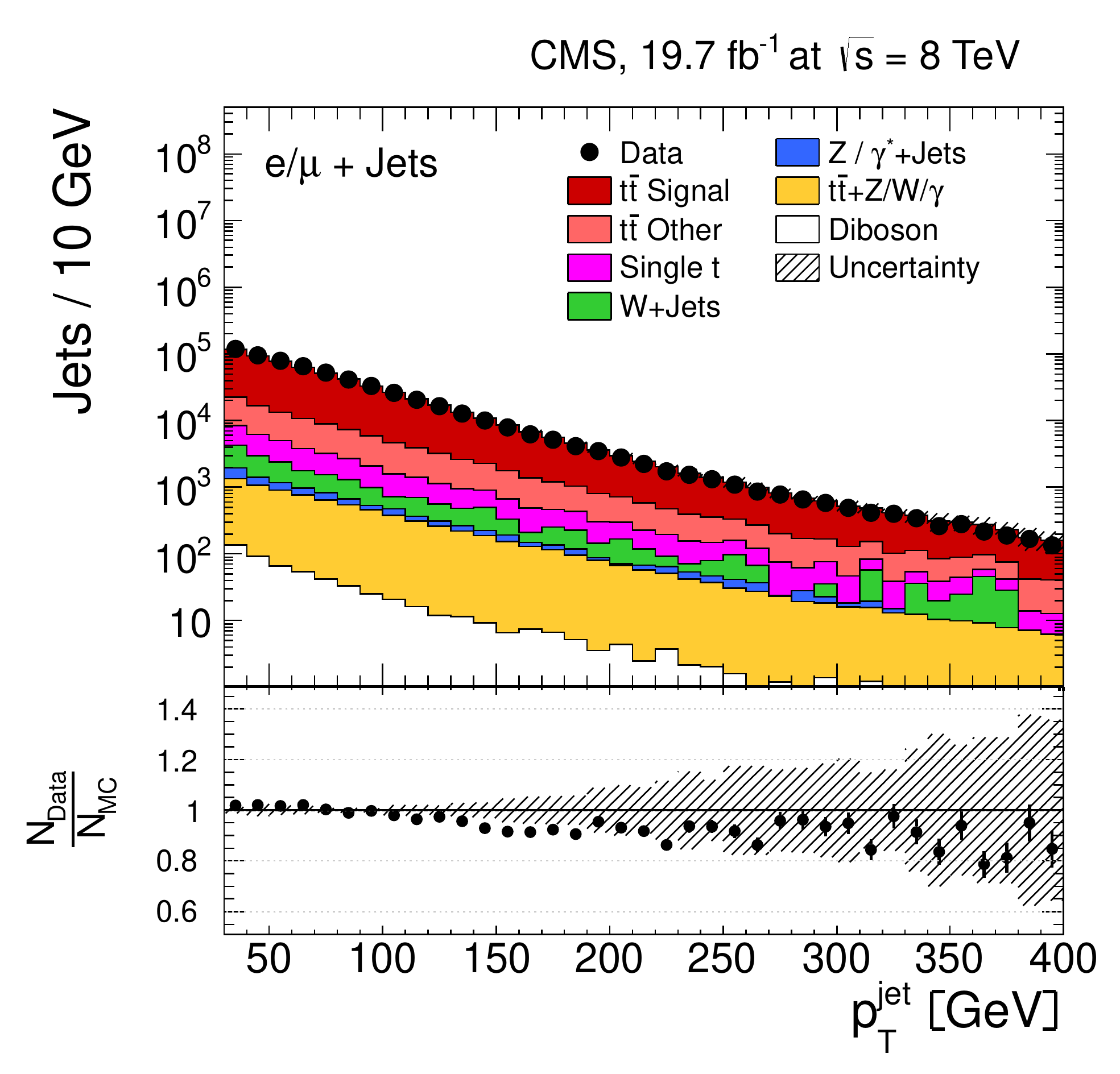}
	\setlength{\unitlength}{\textwidth}
    \caption{Kinematic distributions after event selection and before the kinematic reconstruction of the \ttbar system in the \ljets channels: the multiplicity in the reconstructed number of b-tagged jets (top left), the multiplicity in the reconstructed number of jets (top right), the \pt of the selected isolated leptons (bottom left), and the \pt of all reconstructed jets (bottom right). The QCD multijet background is negligible and not shown. The hatched regions correspond to the shape uncertainties for the signal and backgrounds (cf. Section~\ref{sec:errors}). The lower part of each plot shows the ratio of data to the predictions.}
    \label{fig:ctrl:ljets}

\end{figure*}

\begin{figure*}[htbp]
  \centering
	\includegraphics[width=0.48\textwidth]{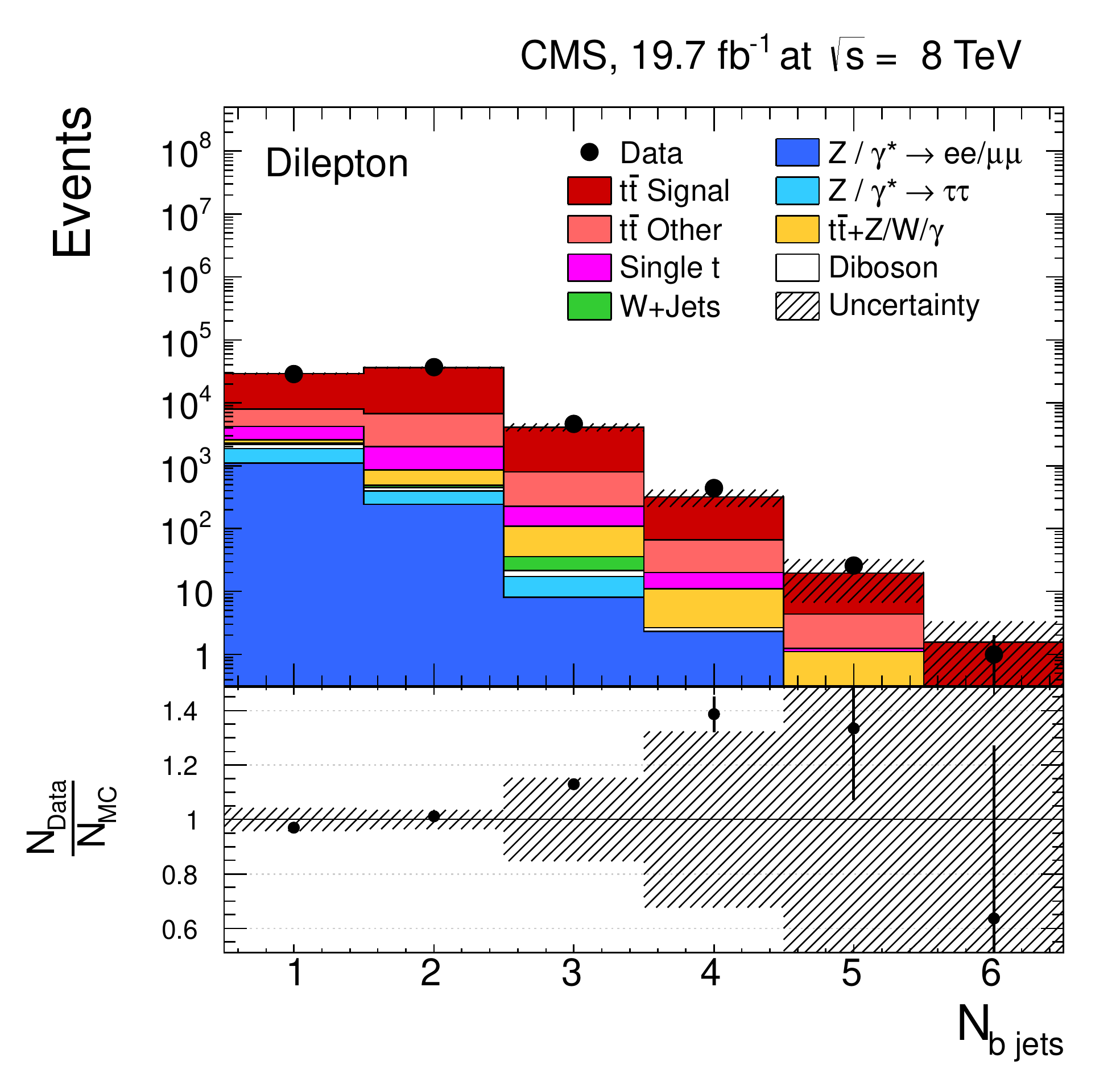}
	\includegraphics[width=0.48\textwidth]{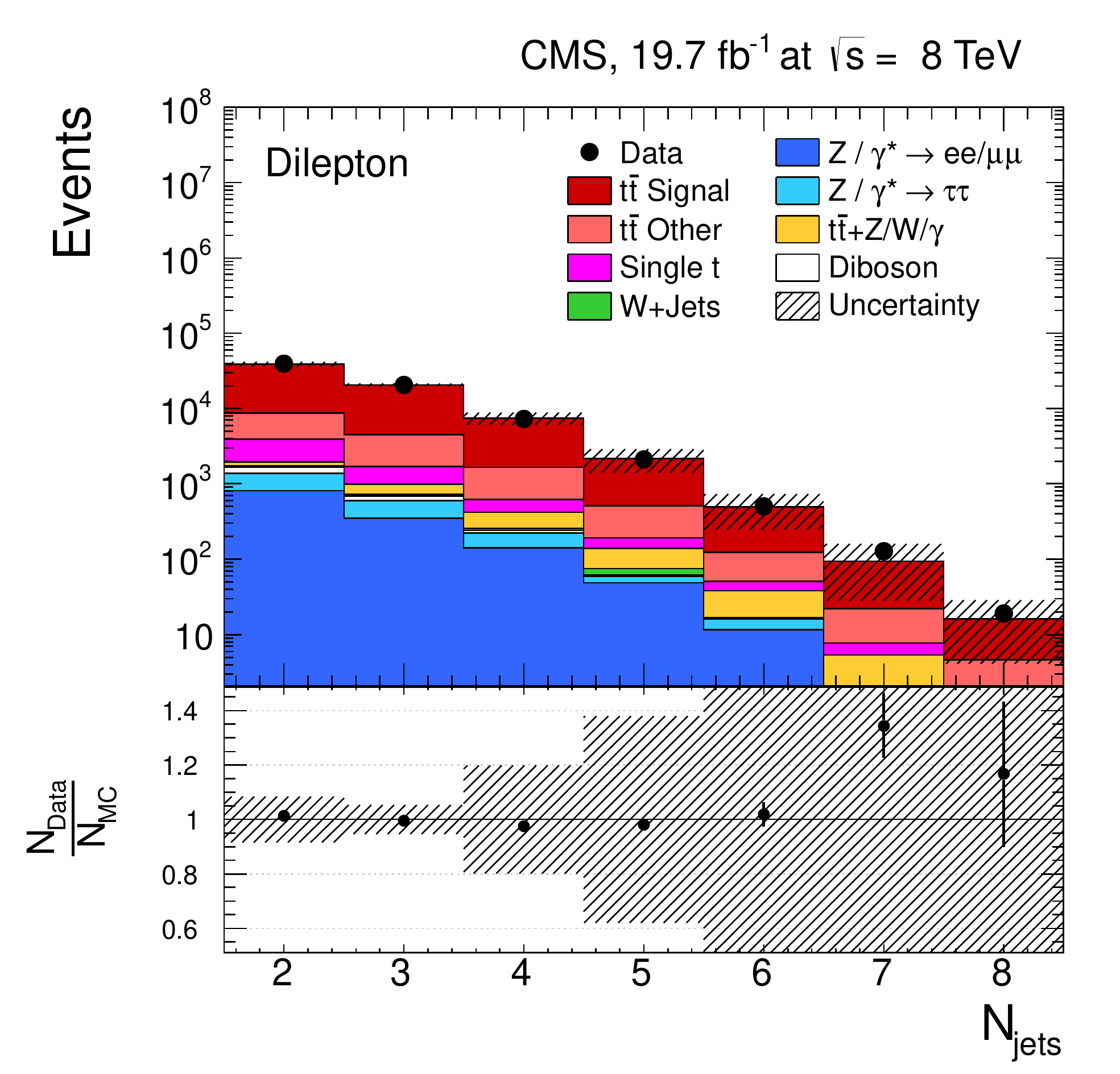}
	\includegraphics[width=0.48\textwidth]{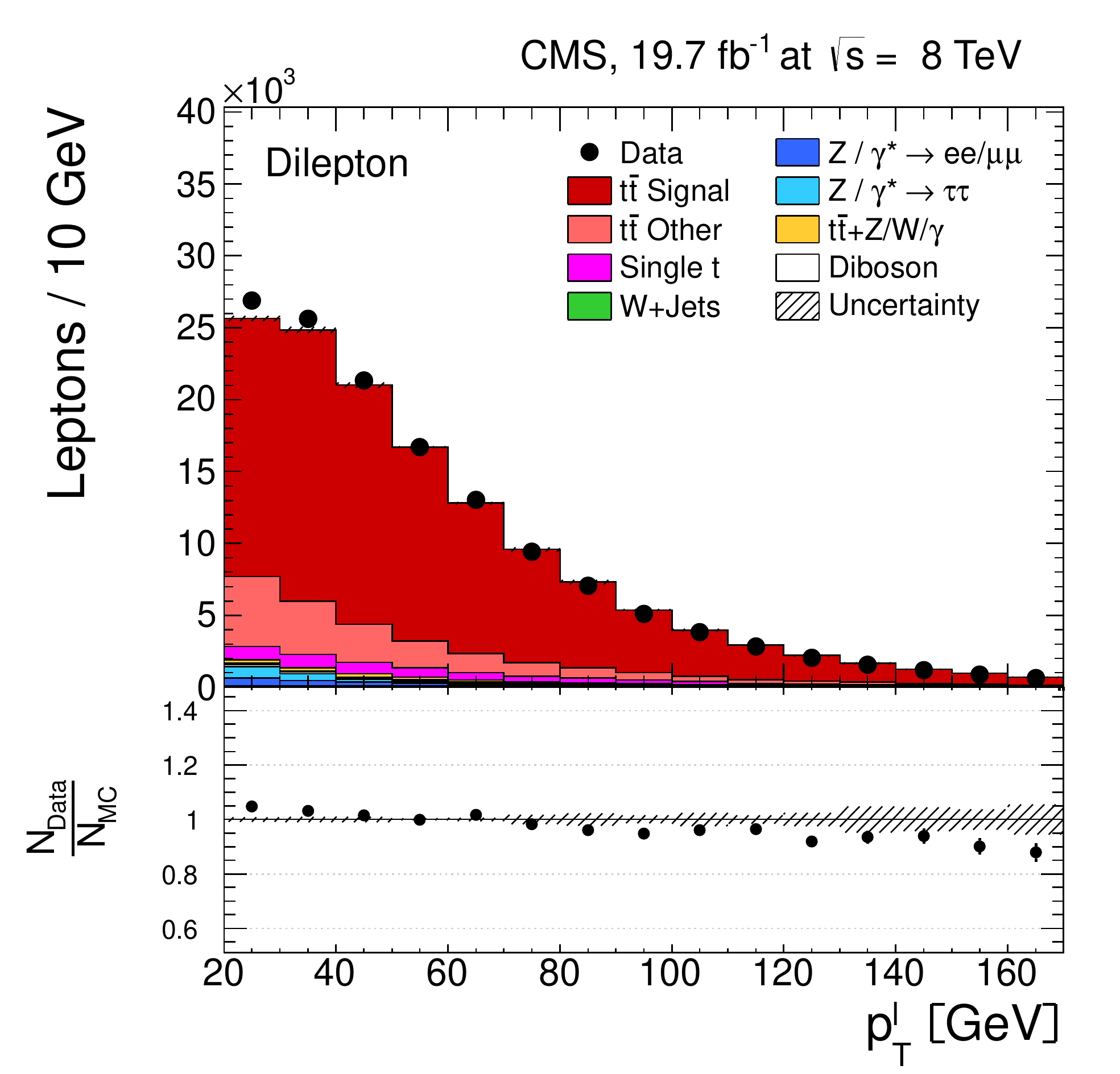}
	\includegraphics[width=0.48\textwidth]{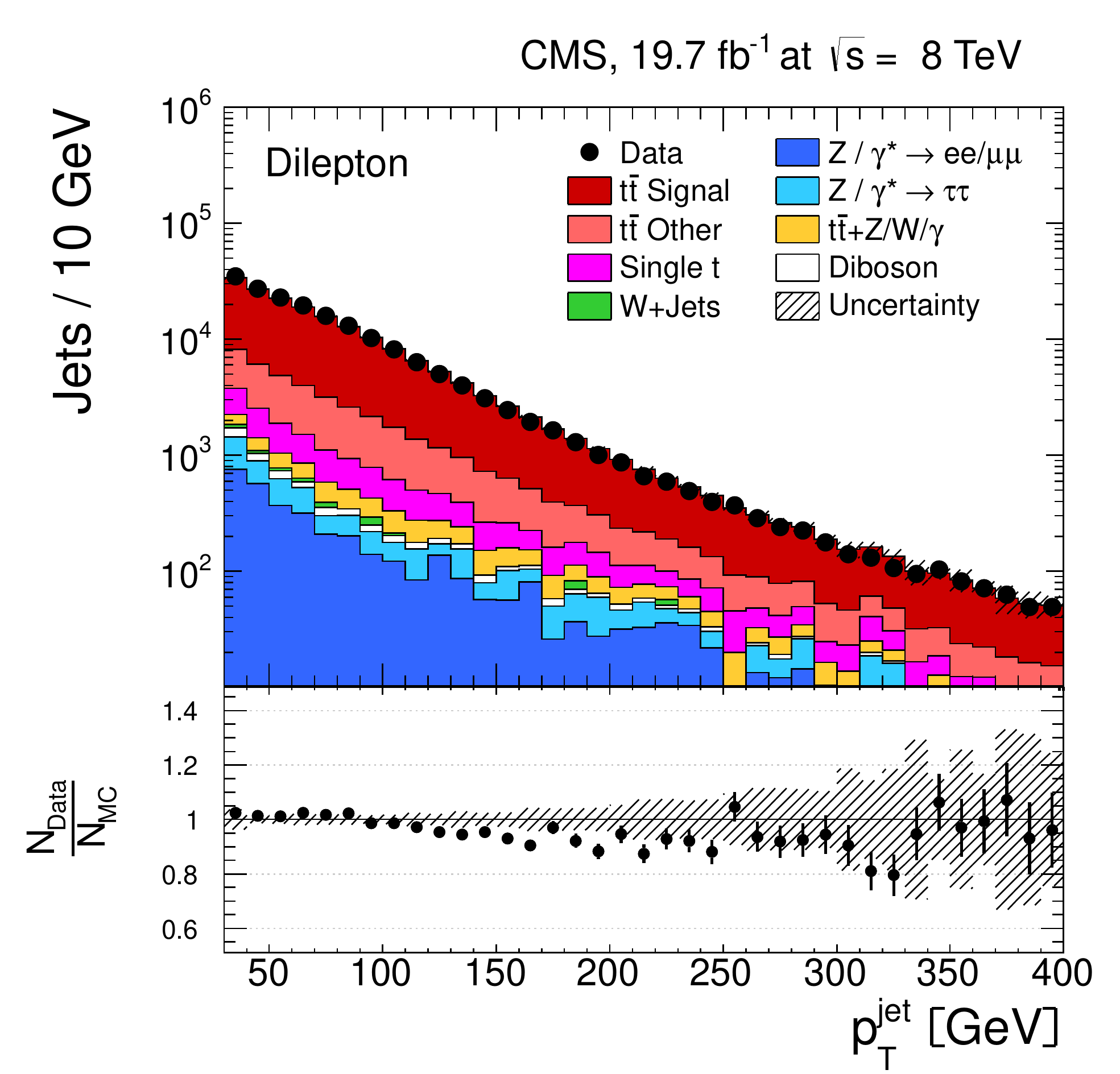}
    \setlength{\unitlength}{\textwidth}
    \caption{Kinematic distributions after event selection and before the kinematic reconstruction of the \ttbar system for the dilepton channels: the multiplicity in the reconstructed number of b-tagged jets (top left), the multiplicity in the number of reconstructed jets (top right), the \pt of the selected isolated leptons (bottom left), and the \pt of the reconstructed jets (bottom right). The QCD multijet background is negligible and not shown. The Z/$\gamma^{*}$+jets background is determined from data~\cite{bib:topPAS11_002,bib:TOP-11-013_paper}. The hatched regions correspond to the shape uncertainties for the signal and backgrounds (cf. Section~\ref{sec:errors}). The lower part of each plot shows the ratio of data to the predictions.}
  \label{fig:ctrl:dileptons}

\end{figure*}

{\tolerance=600
After these selection steps, several basic distributions in \ljets and dilepton events are shown in Figs.~\ref{fig:ctrl:ljets} and~\ref{fig:ctrl:dileptons}, respectively. The hatched regions correspond to the shape uncertainties for the signal and background (cf. Section~\ref{sec:errors}), and are dominated by the former. The data are reasonably well described by the simulation, as shown in the lower part of each plot, where the ratio of data to simulation is presented to better indicate the level of agreement between data and the default \ttbar signal (\MADGRAPH{}+\PYTHIA{6}) and background samples used in the analysis. For both channels, however, data tend to have lower \pt values than predicted by the simulation. It has been verified that the results presented in Section~\ref{sec:diffxsec} are not affected by these remaining differences between data and simulation. A better data-to-simulation agreement in the lepton and jet \pt distributions is obtained by scaling the top quark \pt spectrum in simulation to match the data. However, the impact on the measurement of the cross sections is negligible.
\par}

\subsection{Kinematic reconstruction of the \texorpdfstring{\ttbar}{ttbar} system}
\label{sec:kinfit}

The kinematic properties of the top quark pair are determined from the four-momenta of all final-state objects through kinematic reconstruction algorithms. These algorithms are improved versions of those described in Ref.~\cite{bib:TOP-11-013_paper}.

In the \ljets channels, a constrained kinematic fitting algorithm is applied~\cite{bib:CMSNOTE2006023,bib:TOP-11-013_paper} to the four-momenta of the selected lepton and up to five leading jets, and the $\vec{\PTslash}$ representing the transverse momentum of the neutrino, which are changed according to their resolutions. The fit is constrained to reconstruct two W bosons, each with a mass of $80.4\GeV$. In addition, the reconstructed top quark and antiquark masses are required to be equal. To reduce the number of permutations in the association of jets to quarks, only b-tagged jets are considered as b quarks, and only untagged jets are considered as light quarks. In events with several combinatorial solutions, only the one with the minimum $\chi^{2}$ in the fit is accepted. The main improvement relative to the method described in Ref.~\cite{bib:TOP-11-013_paper} is the increase in the number of correct assignments of b jets to b quarks.
 This is achieved by applying the kinematic fit twice, sequentially, in each event. In the first fit, the top quark mass is fixed to a value of 172.5\GeV. The jet combination that provides the minimum $\chi^{2}$ in the fit is then used as input to the second kinematic fit, in which the top quark mass is not fixed, and the solution to this fit is retained. A further improvement in the method is to require the $\chi^{2}$-probability of the second kinematic fit to be $>$2\%. This criterion is chosen to optimize the fraction of correctly reconstructed signal events, without increasing significantly the statistical uncertainty in the data. The efficiency of this requirement is about 87\% for signal events with the correct jet assignment. As a result, the number of correctly reconstructed events is increased by almost a factor of two relative to the method used in Ref.~\cite{bib:TOP-11-013_paper}, and effects from migration of events across bins, which are relevant for the measurements of the cross section, are reduced. It has been checked that any possible bias in the results that could be introduced by fixing the top quark mass to a specific value in the first kinematic fit is within the assigned systematic uncertainty on the dependence of the measurement on the top quark mass (cf.~Section~\ref{subsec:ModelUncertainties}).

The dilepton channels use an algebraic kinematic reconstruction method~\cite{bib:Abbott:1997fv,bib:TOP-11-013_paper}. The only unknowns are the three-momenta of the two neutrinos, which are reconstructed imposing the following kinematic constraints: \pt conservation in the event; the W bosons, and top quark and antiquark masses. In contrast to the method of Ref.~\cite{bib:TOP-11-013_paper}, the top quark mass is fixed to a value of 172.5\GeV. Each suitable pair of b jet candidates in the event, and both possible assignments of these two jets to the two selected leptons, are considered in the kinematic reconstruction. Combinations with two b-tagged jets are preferred to using single b-tagged jets. In the new method, events are reconstructed 100 times, each time randomly smearing the measured energies and directions of the reconstructed lepton and b jet candidates by their respective detector resolutions. This smearing recovers events that yielded no solution of the equations for the neutrino momenta, because of measurement fluctuations. The equations for the neutrino momenta can have up to four solutions. For a given smearing, the solution is identified by the one yielding the smallest invariant mass of the \ttbar system. For each solution, a weight is calculated based on the expected true lepton-b-jet invariant mass spectrum. The weights are summed over the 100 reconstruction attempts, and the kinematic quantities associated to the top quark and antiquark are calculated as a weighted average. Finally, the two jet and lepton-jet assignments that yield the maximum sum of weights are chosen for analysis. It has been checked that any bias introduced through the use of the lepton-b-jet and \ttbar invariant masses is negligible. This method yields on average a reconstruction efficiency of $\approx$94\%, which is 6\% higher than the one described in Ref.~\cite{bib:TOP-11-013_paper}, and reduces systematic migration effects.

\begin{figure*}[htbp]
  \centering
	\includegraphics[width=0.48\textwidth]{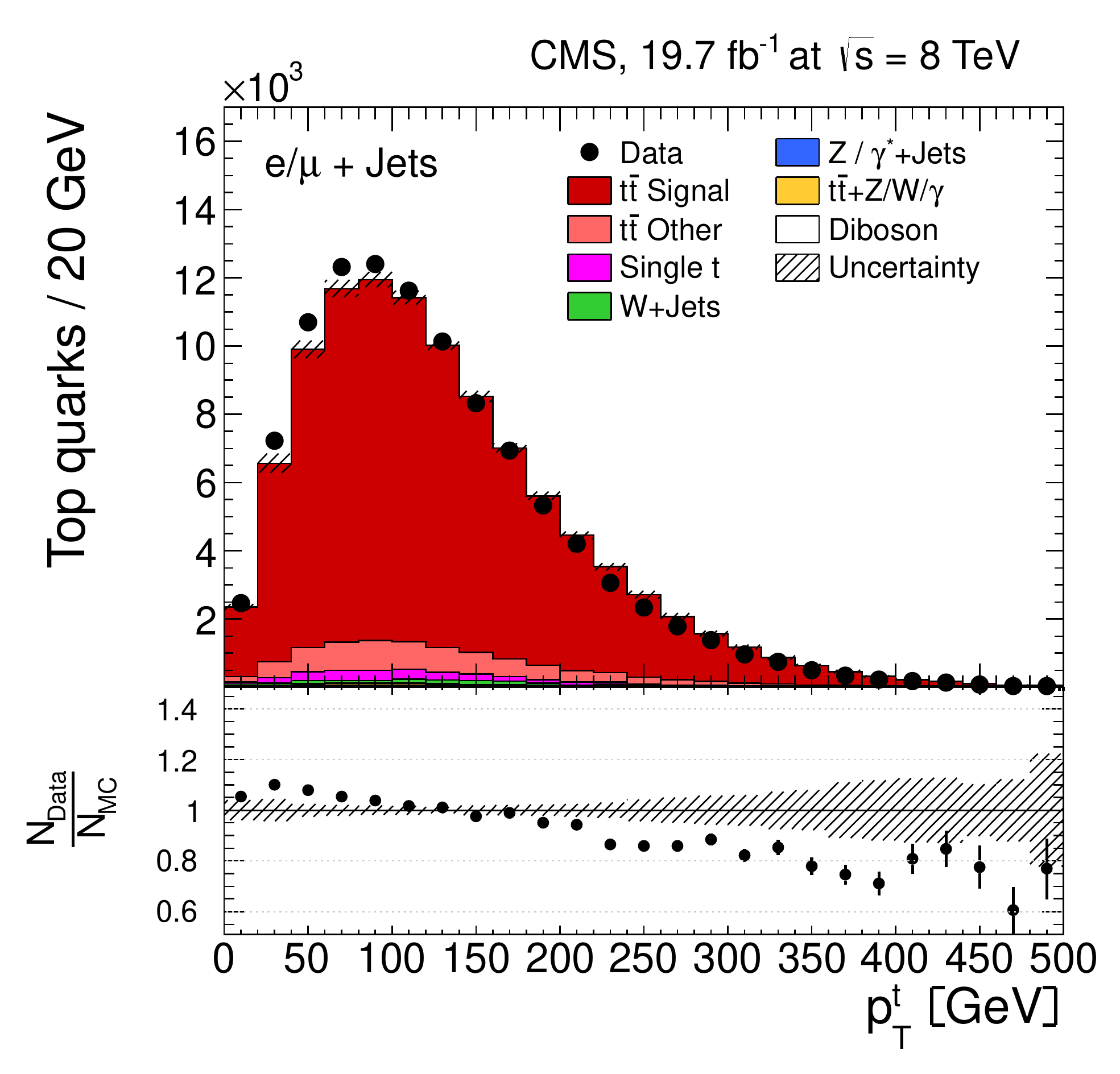}
	\includegraphics[width=0.48\textwidth]{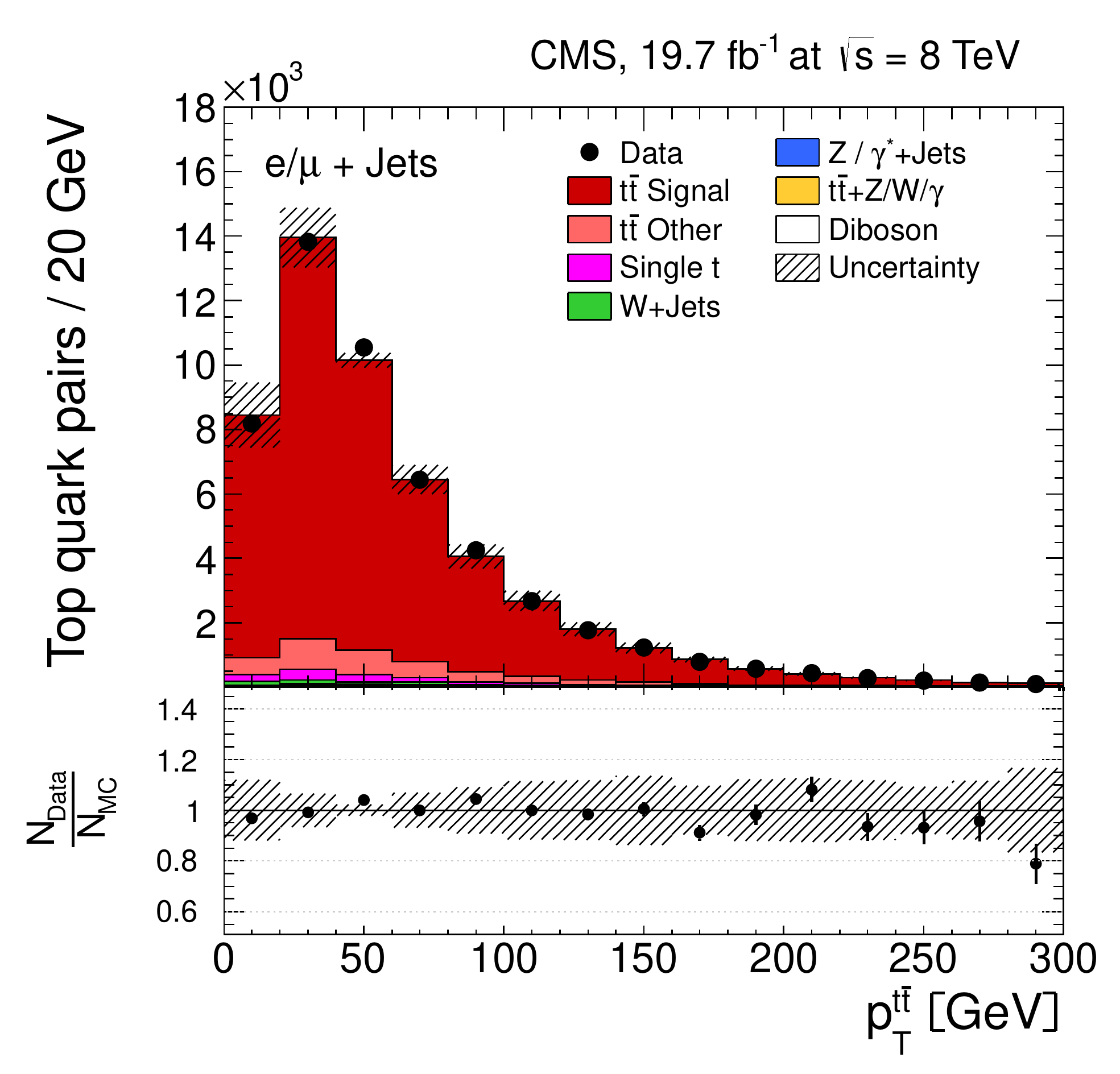}
	\includegraphics[width=0.48\textwidth]{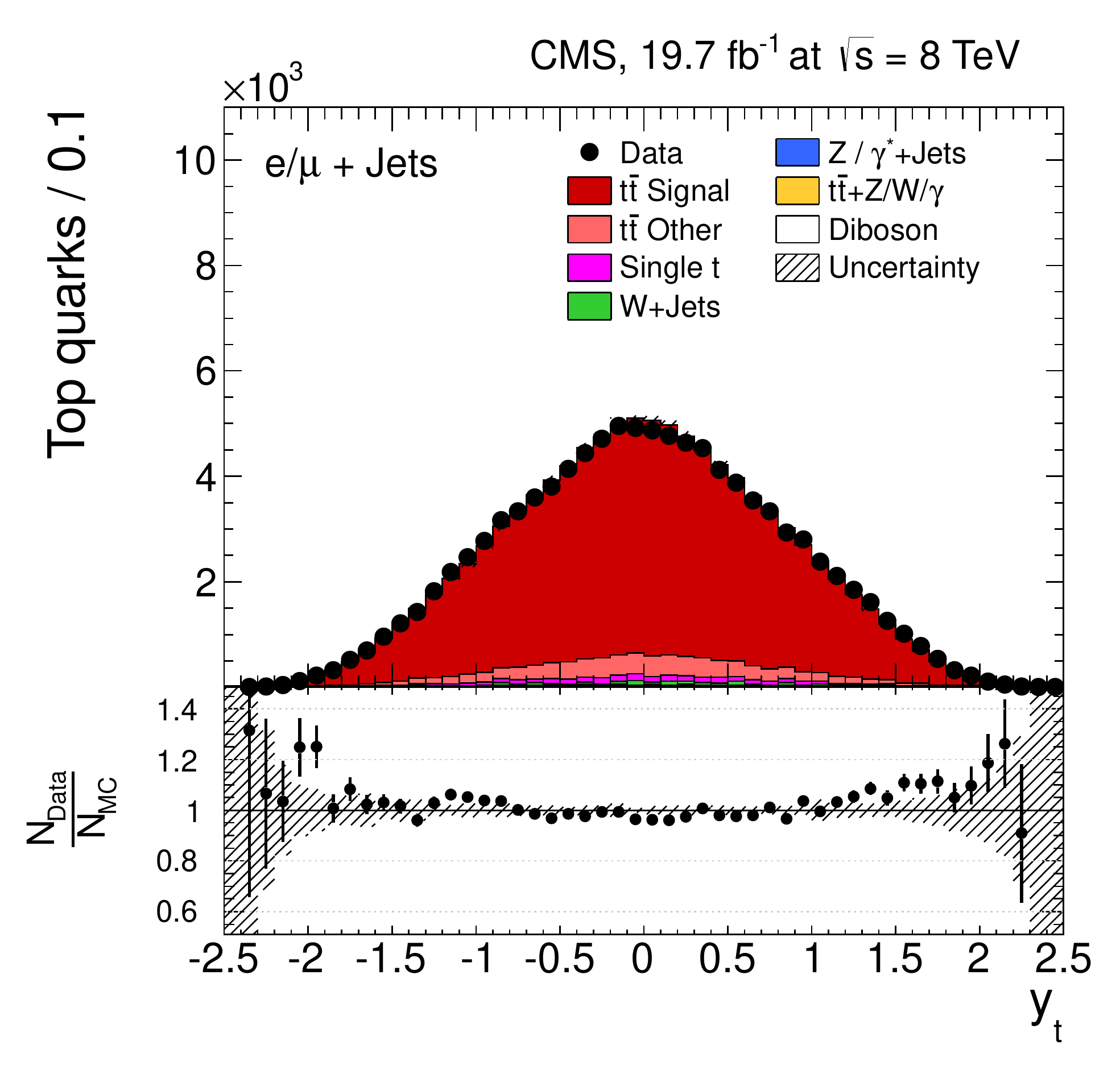}
	\includegraphics[width=0.48\textwidth]{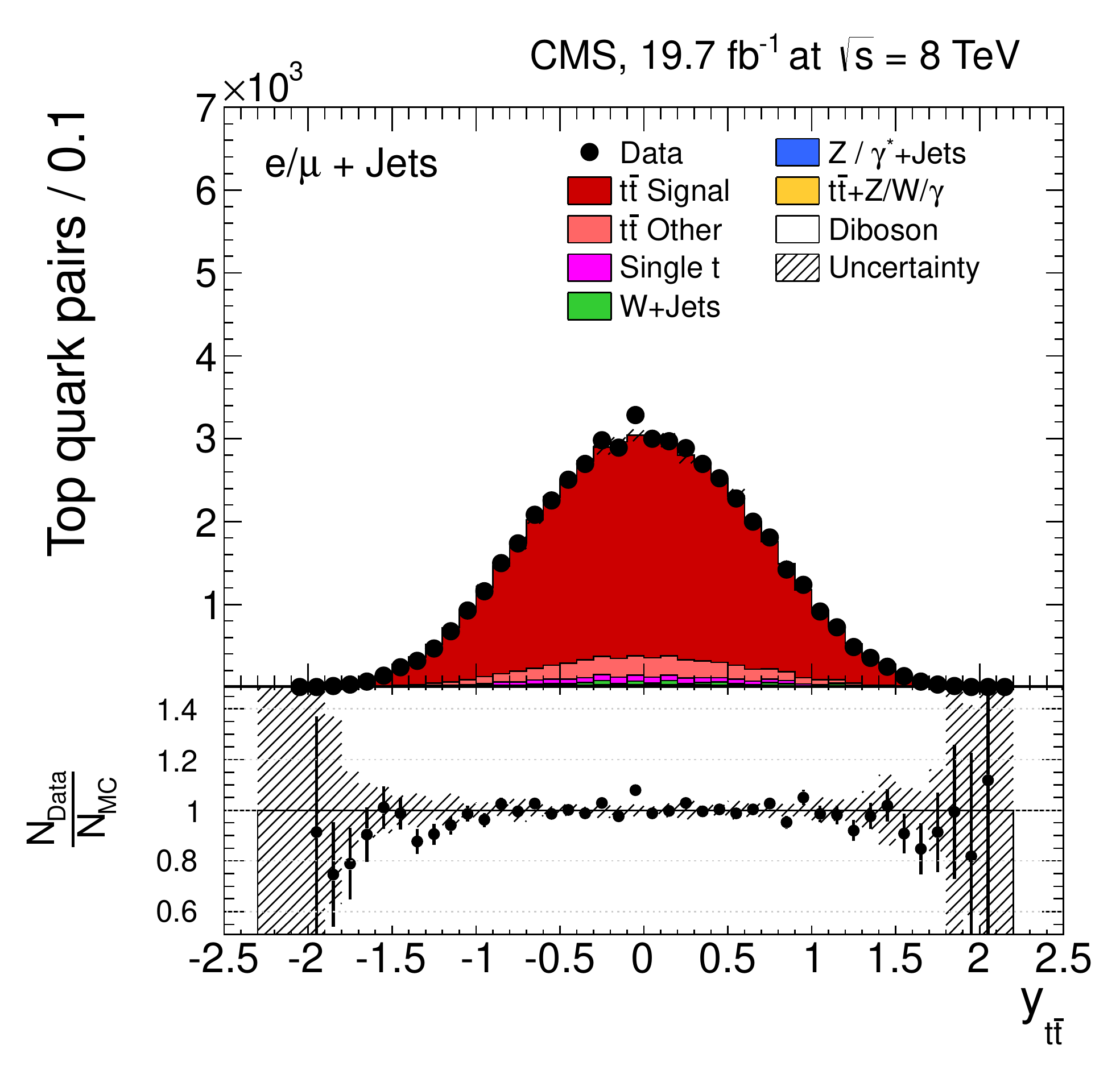}

    \caption{Distribution of top quark or antiquark (left) and \ttbar (right) quantities as obtained from the kinematic reconstruction in the \ljets channels. The top row shows the \pt, and the bottom row shows the rapidities. The QCD multijet background is negligible and not shown. The hatched regions correspond to the shape uncertainties for the signal and backgrounds (cf. Section~\ref{sec:errors}). The lower part of each plot shows the ratio of data to the predictions.}
    \label{fig:kinreco:ljets}

\end{figure*}

\begin{figure*}[htbp]
  \centering
	\includegraphics[width=0.48\textwidth]{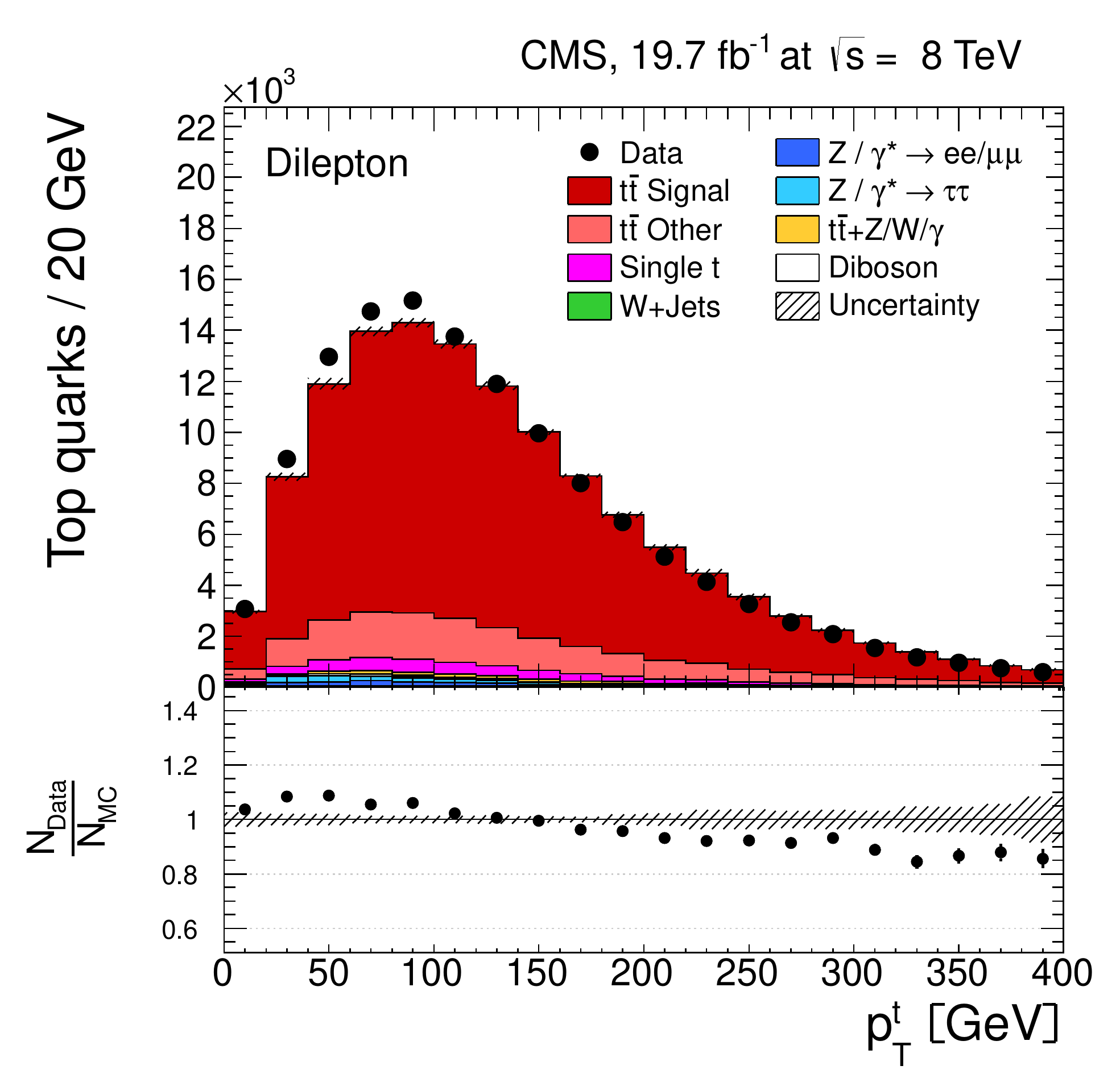}
    	\includegraphics[width=0.48\textwidth]{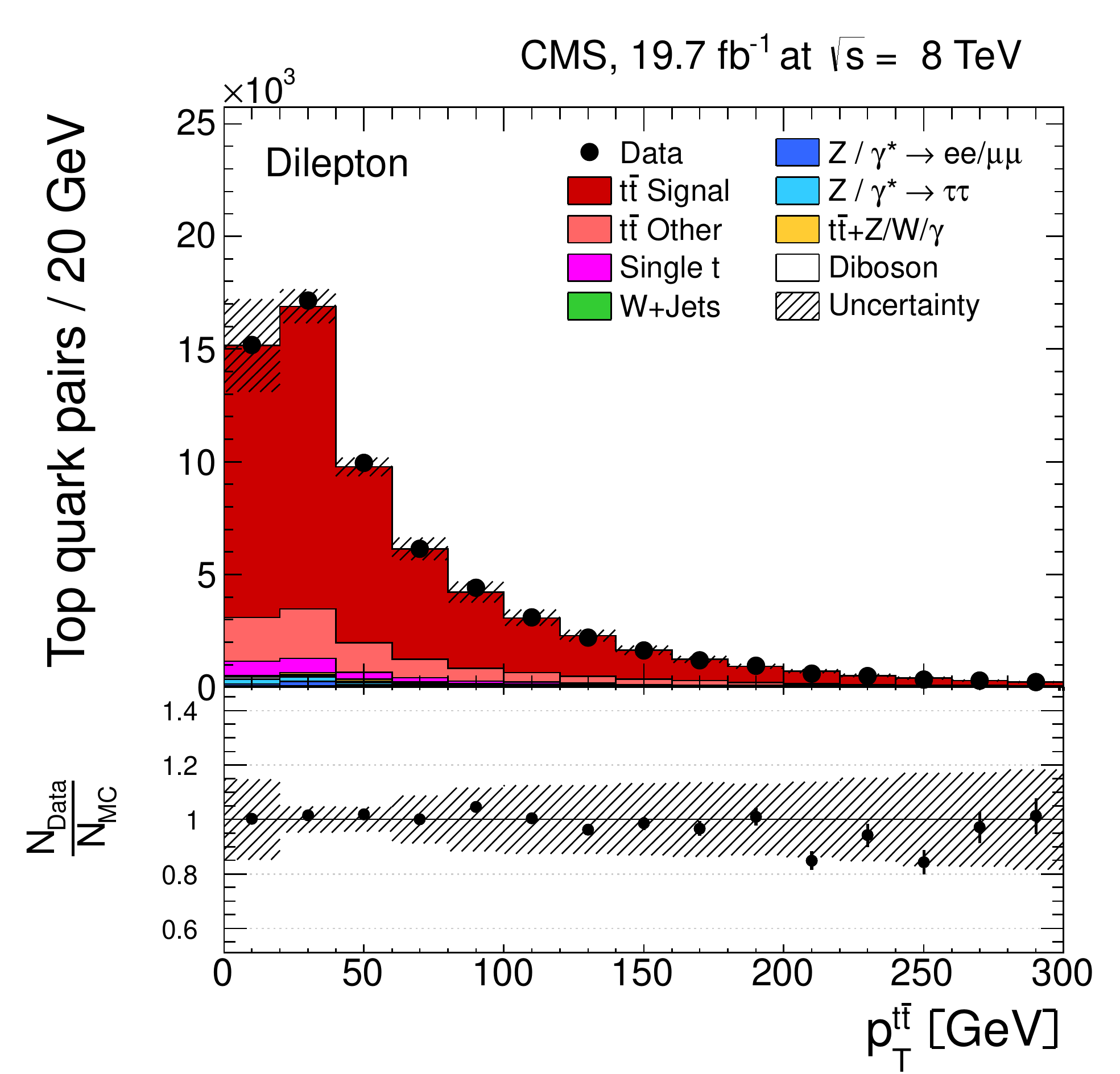}
	\includegraphics[width=0.48\textwidth]{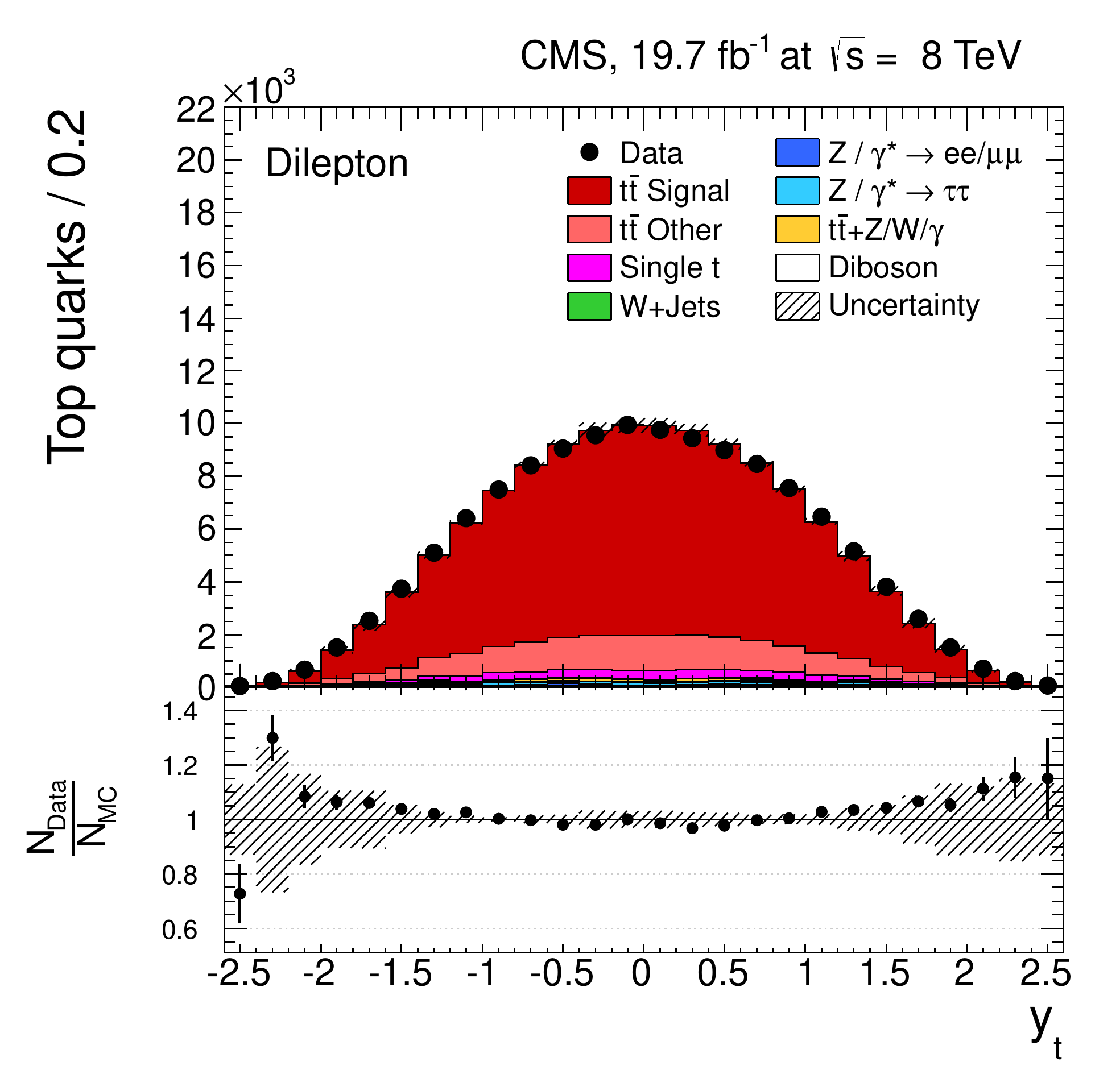}
	\includegraphics[width=0.48\textwidth]{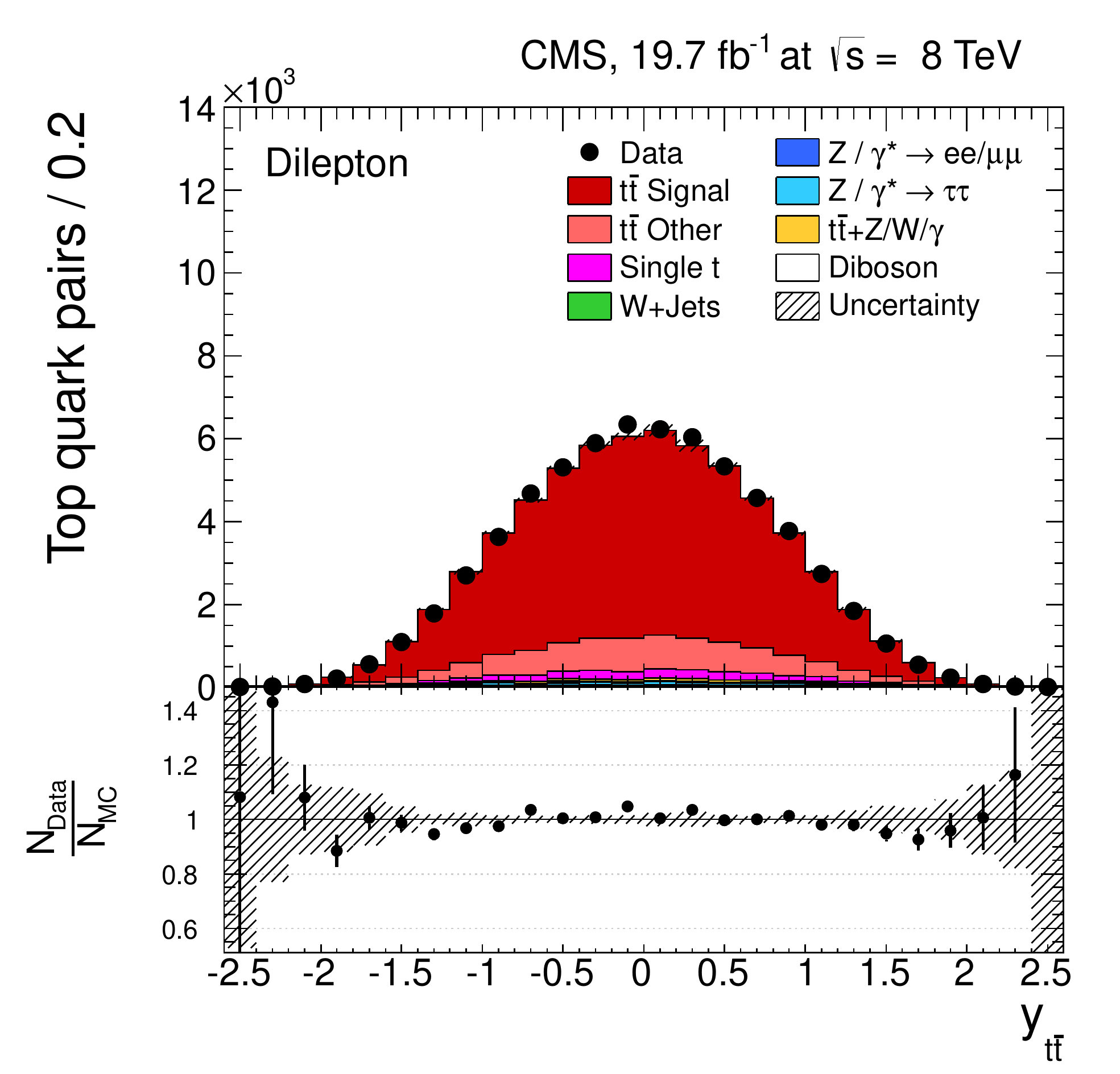}
    \caption{Distribution of top quark or antiquark (left) and \ttbar (right) quantities as obtained from the kinematic reconstruction in the dilepton channels. The top row shows the \pt, and the bottom row shows the rapidities. The QCD multijet background is negligible and not shown. The Z/$\gamma^{*}$+jets background is determined from data~\cite{bib:topPAS11_002,bib:TOP-11-013_paper}. The hatched regions correspond to the shape uncertainties for the signal and backgrounds (cf. Section~\ref{sec:errors}). The lower part of each plot shows the ratio of data to the predictions.}  	
    \label{fig:kinreco:dileptons}

\end{figure*}

Distributions of the top quark or antiquark and \ttbar kinematic observables (the transverse momenta $\pt^{\PQt}$, $\pt^{\ttbar}$, and the rapidities $y_{\PQt}$ and $y_{\ttbar}$) are presented in Figs.~\ref{fig:kinreco:ljets} and~\ref{fig:kinreco:dileptons} for the \ljets and dilepton channels, respectively. The hatched regions correspond to the shape uncertainties for the signal and background (cf. Section~\ref{sec:errors}), and are dominated by the former. The lower panel in each plot also shows the ratio of data relative to the simulated signal and background samples.

In general, the data are reasonably well described by the simulation within the uncertainties. For both channels, the measured \pt distributions, in particular $\pt^{\PQt}$, are somewhat softer than the simulated distributions: the data lie above the simulation for $\pt^{\PQt} < 60 (65)$\GeV in the \ljets (dilepton) channels, while they lie below for $\pt^{\PQt} > 200$\GeV.
This pattern was also observed at 7\TeV~\cite{bib:TOP-11-013_paper}. To ensure that the results presented in Section~\ref{sec:diffxsec} are not affected by such small remaining differences between data and simulation, the analysis has been repeated in different kinematic regions, with different selection requirements, and after scaling the top quark \pt spectrum in simulation to match the data. However, the impact on the measurement of the cross sections is negligible.

Following the event selection described in Section~\ref{subsec:evsel} and the kinematic reconstruction of the \ttbar system, the main contributions to the background in the \ljets channels arise from \ttbar decays into channel other than \ljets (including \ttbar decays into $\tau$ leptons originating from the primary interaction) and single top quark events. The contribution from W+jets and QCD multijet events are well suppressed after the b-tagging requirement, while other \ttbar events are somewhat reduced after the $\chi^{2}$-probability requirement. A total of 24\,927 events are found in the \Pe+jets channel and 26\,843 events in the \Pgm+jets channel. The contribution from \ttbar signal to the final event sample is 89.0\%. The remaining fraction of events contains 7.3\% \ttbar decays other than the \ljets channels, 2.4\% single top quark events, 0.9\% W+jets and \ttbar+Z/W/$\gamma$ events, and negligible fractions of Z+jets, diboson, and QCD multijet events. All background contributions are determined from simulation.

In the dilepton channels, 10\,678 events are found in the \ee channel, 14\,403 in the \mumu channel, and 39\,640 in the \emu channel. Only \ttbar events containing at least two leptons (electrons or muons) from W decays in the final state are considered as signal, and constitute 79.0\% of the final event sample. All other \ttbar candidate events, specifically those originating from decays via $\tau$ leptons, are considered as background and amount to 13.3\% of the final event sample. The fraction of Z+jets events is found to be 2.4\%. This background, which is dominant to the \ee and \mumu channels, is estimated from data using the number of events observed within the Z-peak region (which is removed from the candidate sample), and a correction needed for non-Z+jets backgrounds in this same control region is obtained from data in the \emu channel~\cite{bib:topPAS11_002,bib:TOP-11-013_paper}. Other sources of background, including single top quark production (3.4\%), \ttbar+Z/W/$\gamma$ production (1\%), the contribution arising from misidentified or genuine leptons within jets (0.6\%), or diboson events (0.3\%), are estimated from simulation.

\section{Systematic uncertainties}
\label{sec:errors}

The measurement is affected by systematic uncertainties that originate from detector effects and from theoretical assumptions. Each source of systematic uncertainty is assessed individually by changing the corresponding efficiency, resolution, or scale by its uncertainty, using a prescription similar to the one followed in Ref.~\cite{bib:TOP-11-013_paper}. For each change made, the measured normalized differential cross section is recalculated, and the difference of the changed result relative to its nominal value in each bin is taken as the systematic uncertainty. The overall uncertainty on the measurement is obtained by adding all the contributions in quadrature, and is of the order of 3--10\%, depending on the observable and the bin. A detailed description of this is given in Sections~\ref{subsec:ExpUncertainties} and~\ref{subsec:ModelUncertainties}. The typical representative values of the systematic uncertainties in the normalized differential cross sections are summarized in Table~\ref{tab:TypicalValSysUncertainties}.

\subsection{Experimental uncertainties}
\label{subsec:ExpUncertainties}

The efficiencies of the single-electron and single-muon triggers in the \ljets channels are determined using the ``tag-and-probe'' method of Ref.~\cite{bib:tp} using Z boson event samples. Scale factors close to unity within a few percent are extracted to account for the observed dependence on the $\eta$ and \pt of the lepton. The lepton identification and isolation efficiencies for the \ljets channels obtained with the tag-and-probe method agree well between data and simulation, so that the applied corrections are very close to unity. The systematic uncertainties are determined by shape-dependent changes in trigger and selection efficiencies by their uncertainties. Lepton trigger efficiencies in the dilepton channels are measured using triggers that are only weakly correlated to the dilepton triggers used in the analysis. A dependence on $\eta$ of a few percent is observed, and scale factors are extracted. The lepton identification and isolation uncertainties in the dilepton channels are also determined using the tag-and-probe method, and are again found to be described very well by the simulation for both electrons and muons. The overall difference between data and simulation in bins of $\eta$ and \pt is estimated to be $<$2\% for electrons, and scale factors for muons are found to be close to unity within 1.0\%.

The uncertainty due to the limited knowledge of the jet energy scale is determined by changes implemented in jet energy in bins of \pt and $\eta$~\cite{bib:JME-10-011:JES}. The uncertainty due to the limited accuracy of the jet energy resolution (JER) is determined by changing the simulated JER by ${\pm}1\sigma$ in different $\eta$ regions~\cite{bib:JME-10-011:JES}.

The uncertainty in b-tagging efficiency is determined by taking the maximum change in the shape of \pt and $\eta$ b jet distributions obtained by changing the scale factors. This is achieved by dividing the b jet distributions in \pt and $\eta$ into two bins at the median of the respective distributions. These correspond to $\pt=65\GeV$, and $\abs{\eta}=0.7$ and 0.75 for the \ljets and dilepton channels, respectively. The b-tagging scale factors for b jets in the first bin are scaled up by half of the uncertainties quoted in Ref.~\cite{bib:btag004}, while those in the second bin are scaled down, and vice versa, so that a maximum variation is assumed and the difference between the scale factors in the two bins reflects the full uncertainty. The changes are made separately in the \pt and $\eta$ distributions, and independently for heavy-flavour (b and c) and light (s, u, d, and gluon) jets, assuming that they are all uncorrelated.

The uncertainty in background normalization is determined by changing the background yields. In the \ljets channels, the background normalization for the diboson, QCD multijet, W+jets, and Z+jets samples is conservatively varied by $\pm50\%$~\cite{bib:TOP-11-013_paper}, since these backgrounds, being very small, are determined from simulation rather than from data. The normalization of the \ttbar+Z/W/$\gamma$ samples is changed by $\pm$30\%. For the single top quark sample, the uncertainty is covered by changing the normalization by $\pm$30\%, and the kinematic scales of the event process (renormalization and factorization scales) as described in~Section~\ref{subsec:ModelUncertainties}. In the \ee and \mumu channels, the dominant background from Z+jets determined from data~\cite{bib:topPAS11_002,bib:TOP-11-013_paper} is changed in normalization by $\pm$30\%. In addition, changes in the background contributions from single top quark, diboson, QCD multijet, \ttbar+Z/W/$\gamma$, and W+jets events of $\pm$30\% are used in dilepton channels~\cite{bib:TOP-11-013_paper}.

The kinematic reconstruction of top quarks is well described by the simulation, and the resulting uncertainties are small. In the case of the \ljets analysis, the uncertainty of the kinematic fit is included in the changes in jet energy scales and resolutions, and in the uncertainty on the dependence on the top quark mass (cf.~Section~\ref{subsec:ModelUncertainties}). In the dilepton analysis, the bin-to-bin uncertainty is determined from the small remaining difference in efficiency between simulation and data.

The pileup model estimates the mean number of additional pp interactions to be about 20 events per bunch crossing for the analyzed data. This estimate is based on the total inelastic proton-proton cross section, which is determined to be 69.4\unit{mb} following the measurement described in Ref.~\cite{bib:ppInelXSec}. The systematic uncertainty is determined by changing this cross section within its uncertainty of $\pm$5\%.

\subsection{Uncertainties in modelling}
\label{subsec:ModelUncertainties}

The impact of theoretical assumptions on the measurement is determined, as indicated previously, by repeating the analysis and replacing the standard \MADGRAPH \ttbar simulation by dedicated simulation samples with altered parameters.

The uncertainty in modelling of the hard-production process is assessed through changes in the renormalization and factorization scales in the \MADGRAPH sample by factors of two and 0.5 relative to their common nominal value, which is set to the $Q$ of the hard process. In \MADGRAPH, $Q$ is defined by $Q^2 = m^2_{\PQt} + \Sigma \pt^2$, where the sum is over all additional final state partons in the matrix element. The impact of the choice of the scale that separates the description of jet production through matrix elements (ME) or parton shower (PS) in \MADGRAPH is studied by changing its reference value of 20\GeV to 40\GeV and to 10\GeV. In the \ljets channels, changes in the renormalization and factorization scales are also applied to single top quark events to determine an uncertainty on the shape of this background contribution.
The dependence of the measurement on the top quark mass is also estimated from dedicated \MADGRAPH simulation samples in which the top quark mass is changed by $\pm$1\GeV relative to the value used in the default simulation.
The uncertainty from hadronization and parton showering is assessed by comparing the results obtained from samples simulated with \POWHEG and \MCATNLO interfaced with \PYTHIA{6} and \HERWIG{6}, respectively.
The uncertainty from the choice of PDF is determined by reweighting the sample of simulated \ttbar signal events according to the 52 CT10 PDF error sets~\cite{bib:CT10}, at a 90\% confidence level. The maximum variation is taken as uncertainty.
As mentioned in Sections~\ref{subsec:evsel} and~\ref{sec:kinfit}, the effect of scaling the top quark \pt spectrum in simulation to match the data has negligible impact on the measured cross sections, therefore no systematic uncertainty is taken into account for this effect.

\begin{table*}[phtb]
  \centering
    \topcaption{Breakdown of typical systematic uncertainties for the normalized differential cross sections. The uncertainty in the jet-parton matching threshold is indicated as ``ME-PS threshold''; ``PS'' refers to ``parton shower''. The medians of the distribution of uncertainties over all bins of the measurement are quoted. For the \ljets channels, the background from Z+jets is negligible and included in the ``Background (all other)'' category. }
    \label{tab:TypicalValSysUncertainties}
    \begin{tabular}{c|cc|cc}
    \hline
     \multicolumn{5}{c}{Relative systematic uncertainty (\%)} \\
     \hline
     Source  & \multicolumn{2}{c|}{Lepton and b jet observables} & \multicolumn{2}{c}{Top quark and \ttbar observables} \\

                   & \ljets & dileptons & \ljets & dileptons \\
      \hline
      Trigger eff. \& lepton selec.    & 0.1 & 0.1 & 0.1 & 0.1 \\
      Jet energy scale                 & 2.3 & 0.4 & 1.6 & 0.8 \\
      Jet energy resolution            & 0.4 & 0.2 & 0.5 & 0.3 \\
      Background (Z+jets)              & --- & 0.2 & --- & 0.1 \\
      Background (all other)           & 0.9 & 0.4 & 0.7 & 0.4 \\
      b tagging                        & 0.7 & 0.1 & 0.6 & 0.2 \\
      Kinematic reconstruction         & ---  & $<$0.1 & --- & $<$0.1 \\
      Pileup                           & 0.2 & 0.1 & 0.3 & 0.1 \\

      \hline
      Fact./renorm. scale              & 1.1 & 0.7 & 1.8 & 1.2 \\
      ME-PS threshold                  & 0.8 & 0.5 & 1.3 & 0.8 \\
      Hadronization \& PS               & 2.7 & 1.4 & 1.9 & 1.1 \\
      Top quark mass                   & 1.5 & 0.6 & 1.0 & 0.7 \\
      PDF choice                       & 0.1 & 0.2 & 0.1 & 0.5 \\
      \hline
    \end{tabular}

\end{table*}

\section{Normalized differential cross sections}
\label{sec:diffxsec}

The normalized \ttbar cross section in each bin $i$ of each observable $X$ is determined as a function of the kinematic properties of the leptons, the lepton pair, the b jets, the b jet system, the top quarks, and the \ttbar system through the relation~\cite{bib:TOP-11-013_paper}:
\begin{equation}
\label{eq:diffXsec}
\frac{1}{\sigma}\frac{\rd\sigma_{i}}{\rd X}=\frac{1}{\sum_{i}x_{i}}\frac{x_{i}}{\Delta^{\text{X}}_{i}}
\end{equation}

where $x_{i}$ represents the number of signal events measured in data in bin $i$ after background subtraction and corrected for detector efficiencies, acceptances, and migrations, and $\Delta_{i}^{X}$ is the bin width. The differential cross section is normalized by the sum of $x_i$ over all bins, as indicated in Eq.~(\ref{eq:diffXsec}). The integrated luminosity is omitted, as it cancels in the ratio. Because of the normalization, sources of systematic uncertainty that are correlated across all bins of the measurement, \eg the uncertainty in the integrated luminosity, also cancel. The contribution to the background from other \ttbar decays is taken into account, after subtracting all other background components, by correcting the number of signal events in data using the expected signal fraction. The expected signal fraction is defined as the ratio of the number of selected \ttbar signal events to the total number of selected \ttbar events (\ie signal and all other \ttbar events) in simulation. This procedure avoids the dependence on the total inclusive \ttbar cross section used in the normalization of the simulated signal sample.

Effects from trigger and detector efficiencies and resolutions leading to the migration of events across bin boundaries, and therefore to statistical correlations among neighbouring bins, are corrected by using a regularized unfolding method~\cite{bib:svd, bib:blobel,bib:TOP-11-013_paper}. For each measured distribution, a response matrix is defined that accounts for migrations and efficiencies using the simulated \MADGRAPH{}+\PYTHIA{6} \ttbar signal sample. The generalized inverse of the response matrix is used to obtain the unfolded distribution from the measured distribution by applying a $\chi^2$ minimization technique. A smoothing prescription (regularization) is applied to prevent large unphysical fluctuations that can be introduced when directly inverting the response matrix. The strength of the regularization is determined and optimized individually for each distribution using the averaged global correlation method~\cite{bib:james}.
To keep the bin-to-bin migrations small, the widths of bins in the measurement are chosen according to their purity (ratio of the number of events generated and reconstructed in a particular bin to the total number of events reconstructed in that bin; this quantity is sensitive to migrations into the bin) and stability (ratio of the number of events generated and reconstructed in a particular bin to the number of events generated in that bin; this is sensitive to migrations out of the bin). The purity and stability of the bins in this analysis are typically 60\% or larger, mainly due to the improvements in the kinematic reconstruction methods discussed in Section~\ref{sec:kinfit}.

The performance of the unfolding procedure is tested for possible biases from the choice of the input model (the \MADGRAPH{}+\PYTHIA{6} \ttbar signal simulation). It is verified that, either by reweighting the signal simulation or injecting a resonant \ttbar peak into the simulation of the signal, the unfolding procedure based on the nominal response matrices still recovers these altered shapes within statistical uncertainties. Moreover, \ttbar samples simulated with \POWHEG{}+\PYTHIA{6} and \MCATNLO{}+\HERWIG{6} are used to obtain the response matrices applied in the unfolding when determining the systematic uncertainties of the model (cf.~Section~\ref{subsec:ModelUncertainties}). Therefore, possible effects from the unfolding procedure are already taken into account in the systematic uncertainties. The unfolded results are found to be consistent with those obtained using other regularization techniques~\cite{bib:blobel}.

The measurement of the normalized differential cross sections proceeds as follows. For each kinematic distribution, the event yields in the separate channels are added together, the background is subtracted, and the unfolding is performed. It is verified that the measurements in separate channels yield results consistent within their uncertainties. The systematic uncertainties in each bin are determined from the changes in the combined cross sections. This requires the full analysis to be repeated for every systematic change, and the difference relative to the nominal combined value is taken as the systematic uncertainty for each bin of each observable. This method therefore takes into account the correlation among systematic uncertainties in different channels and bins.

{\tolerance=400
The normalized differential cross sections of leptons and b jets are unfolded to the particle level and determined in a fiducial phase space defined by the kinematic and geometric region in which the final-state leptons and jets are produced within the detector acceptance (cf.~Section~\ref{subsec:visiblePS}).
This minimizes model uncertainties from the extrapolation of the measurement outside of the experimentally well-described regions of phase space. In addition, the top quark and \ttbar-system quantities are unfolded to the parton level and presented in the full phase space (cf.~Section~\ref{subsec:fullPS}) to provide easier comparisons with recent QCD calculations. The measurements are compared to predictions from \MADGRAPH{}+\PYTHIA{6}, \POWHEG{}+\PYTHIA{6}, \POWHEG{}+\HERWIG{6}, and \breakhere \MCATNLO{}+\HERWIG{6}. The top quark and \ttbar results are also compared to the latest calculations at NLO+NNLL~\cite{bib:ahrens_mttbar,bib:ahrens_ptttbar} and approximate NNLO~\cite{bib:kidonakis_8TeV} precision, when available.
\par}

In addition to the measurements discussed in Ref.~\cite{bib:TOP-11-013_paper}, results for the \pt and invariant mass of the b jet pair, the \pt of the top quarks or antiquarks in the \ttbar rest frame, the \pt of the highest (\textit{leading}) and second-highest (\textit{trailing}) \pt of the top quark or antiquark, and the difference in the azimuthal angle between the top quark and antiquark are also presented.

{\tolerance=400
All values of normalized differential cross sections, including bin boundaries, are provided in tables in \suppMaterial.
\par}

\subsection{Lepton and b jet differential cross sections}
\label{subsec:visiblePS}

The normalized differential \ttbar cross section as a function of the lepton and b jet kinematic properties is measured at the particle level, where the objects are defined as follows. Leptons from W boson decays are defined after final-state radiation. A jet is defined at the particle level, following a procedure similar to that described in Section~\ref{subsec:lepjetreco} for reconstructed jets, by applying the anti-\kt clustering algorithm with a distance parameter of 0.5 to all stable particles (excluding the decay products from W boson decays into e$\nu$, $\mu\nu$, and final states with leptonic $\tau$ decays). A jet is defined as a b jet if it contains any of the decay products of a B hadron. Only the two b jets of highest \pt originating from different B hadrons are considered as arising from the top quark decays.

The measurements are presented in a fiducial phase space defined by geometric and kinematic requirements on these particle-level objects as follows. The charged leptons from the W boson decays must have $\abs{\eta}<2.1$ and $\pt > 33\GeV$ in the \ljets channels, and $\abs{\eta} < 2.4$ and $\pt > 20\GeV$ in the dilepton channels. Exactly one and two leptons are required, respectively, in the \ljets and the dilepton channels. At least four jets with $\abs{\eta} < 2.4$ and $\pt > 30\GeV$, two of which are b jets, are required in the \ljets channels. In the dilepton channels, both b jets from the top quark decays must satisfy $\abs{\eta} < 2.4$ and $\pt > 30\GeV$. The fiducial particle-level corrections are determined using simulated \ttbar events that fulfill these requirements; all other \ttbar events are classified as background and are removed.

\begin{figure*}[htb]
  \centering
       \includegraphics[width=0.48\textwidth]{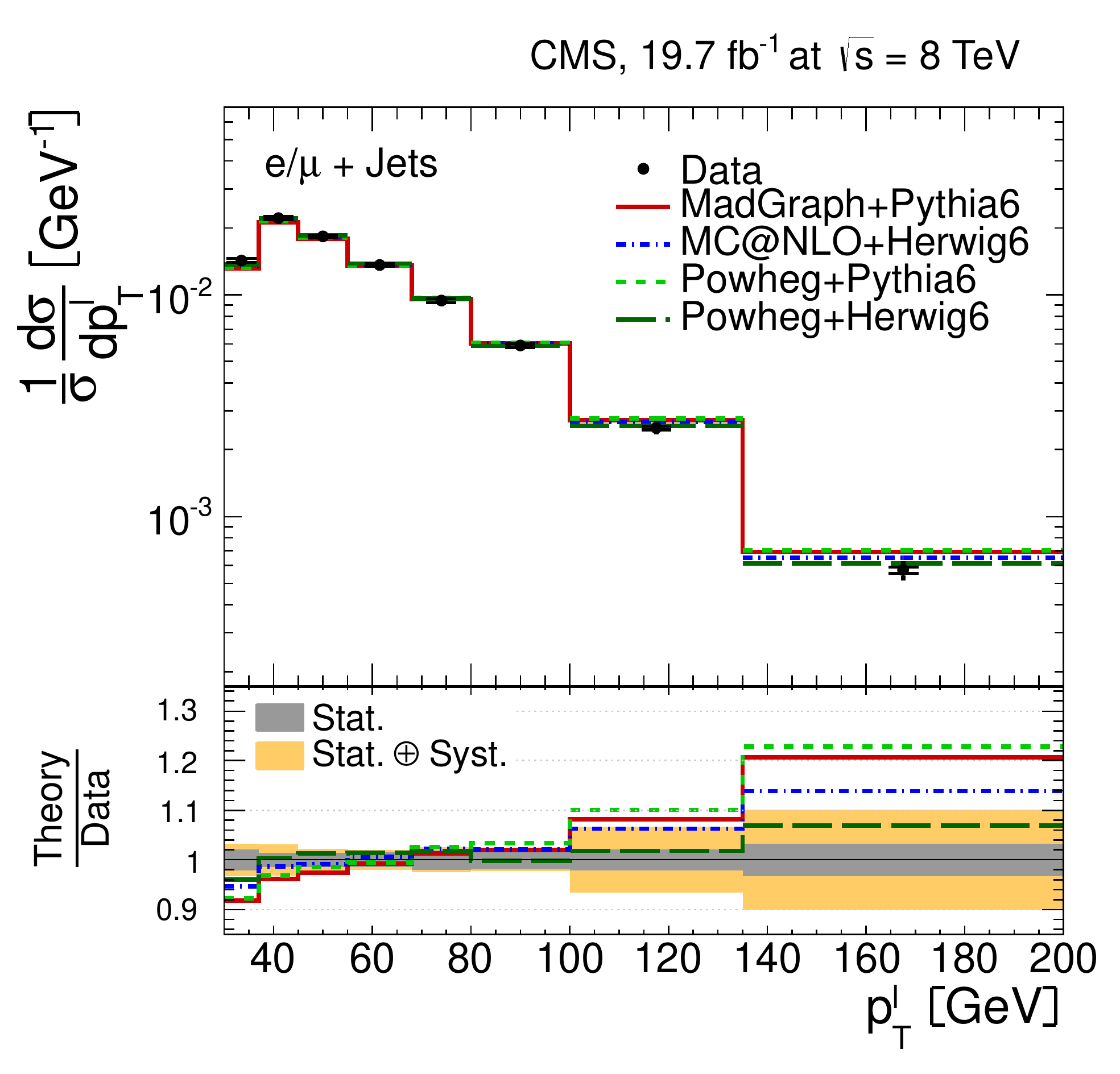}
	\includegraphics[width=0.48\textwidth]{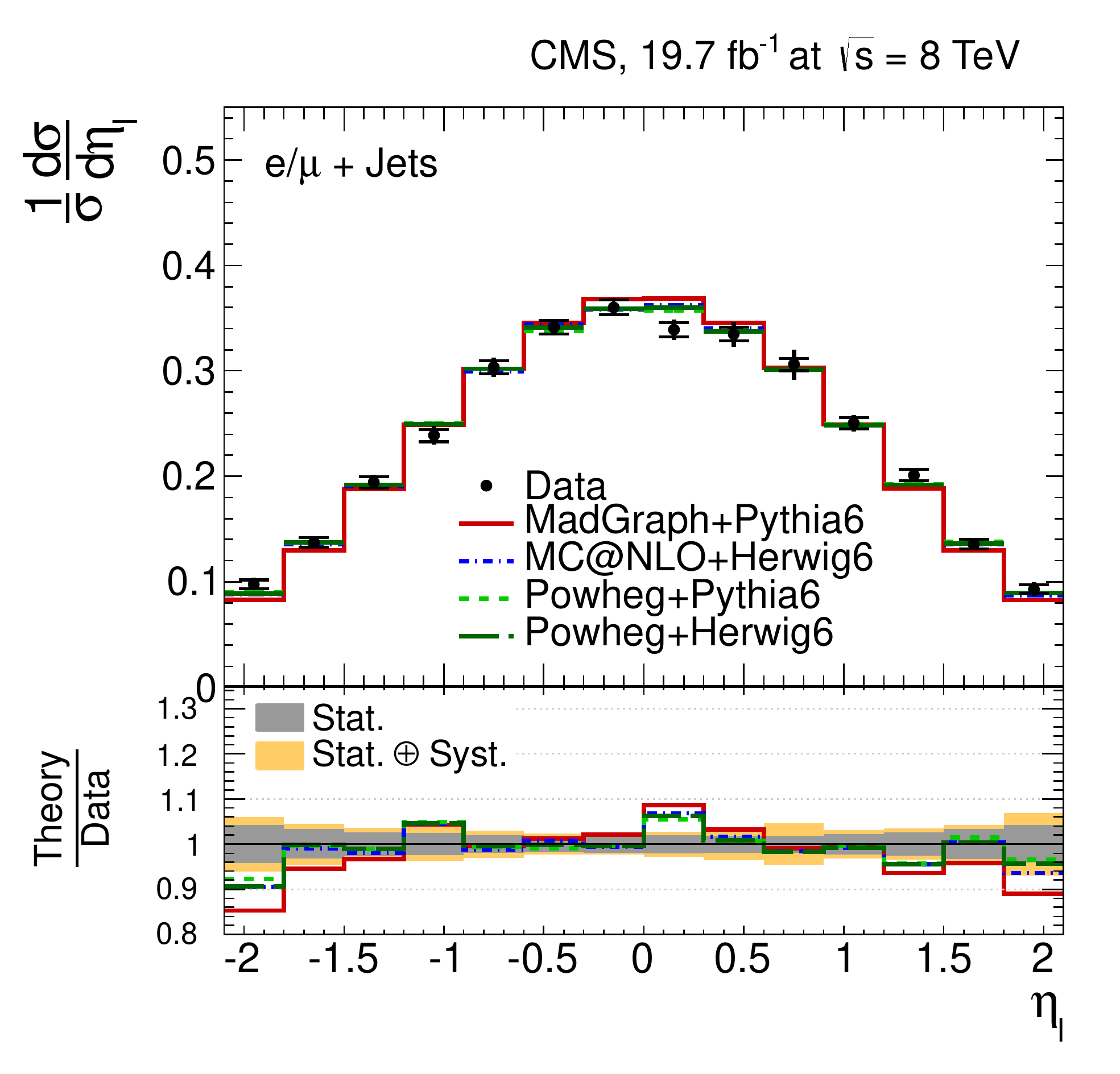}
    \caption{Normalized differential \ttbar production cross section in the \ljets channels as a function of the $\pt^{\ell}$ (left) and $\eta_{\ell}$ (right) of the charged lepton. The superscript `$\ell$' refers to both $\ell^{+}$ and $\ell^{-}$. The data points are placed at the midpoint of the bins. The inner (outer) error bars indicate the statistical (combined statistical and systematic) uncertainties. The measurements are compared to predictions from \MADGRAPH{}+\PYTHIA{6}, \POWHEG{}+\PYTHIA{6}, \POWHEG{}+\HERWIG{6}, and \MCATNLO{}+\HERWIG{6}. The lower part of each plot shows the ratio of the predictions to data.}
    \label{fig:diffXSec:l:ljets}

\end{figure*}

\begin{figure*}[htbp]
  \centering
        \includegraphics[width=0.48\textwidth]{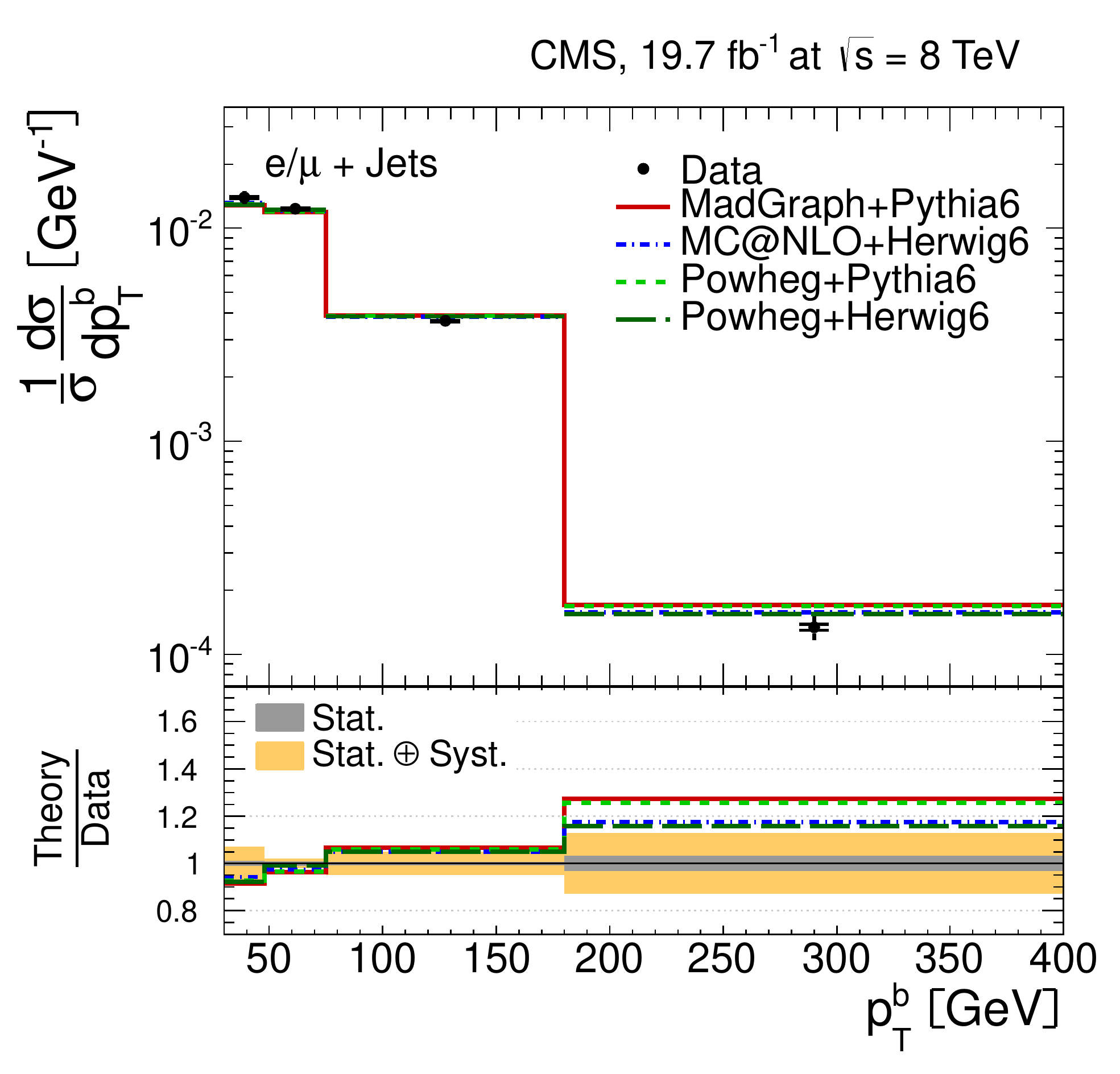}
	\includegraphics[width=0.48\textwidth]{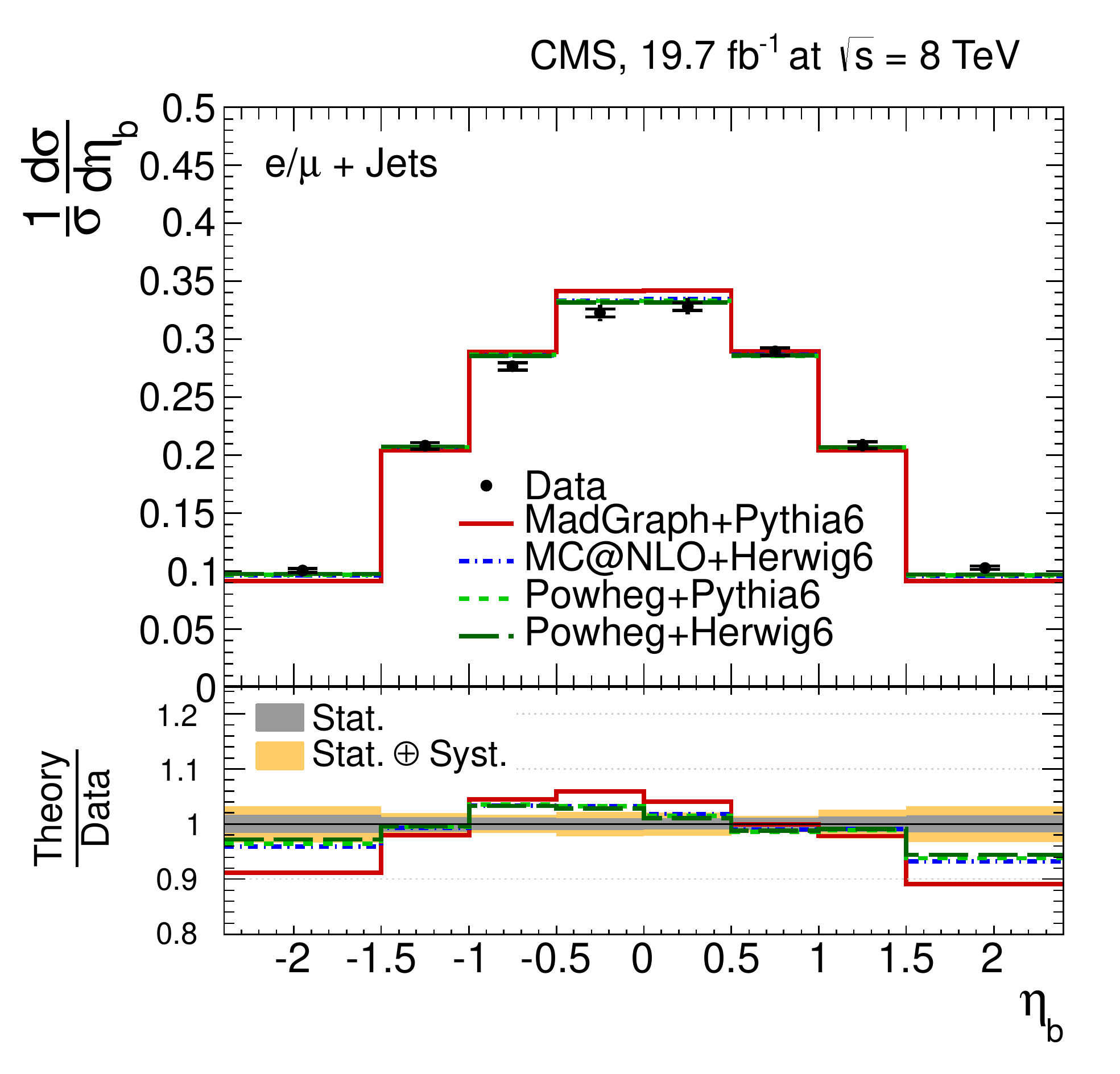}
        \includegraphics[width=0.48\textwidth]{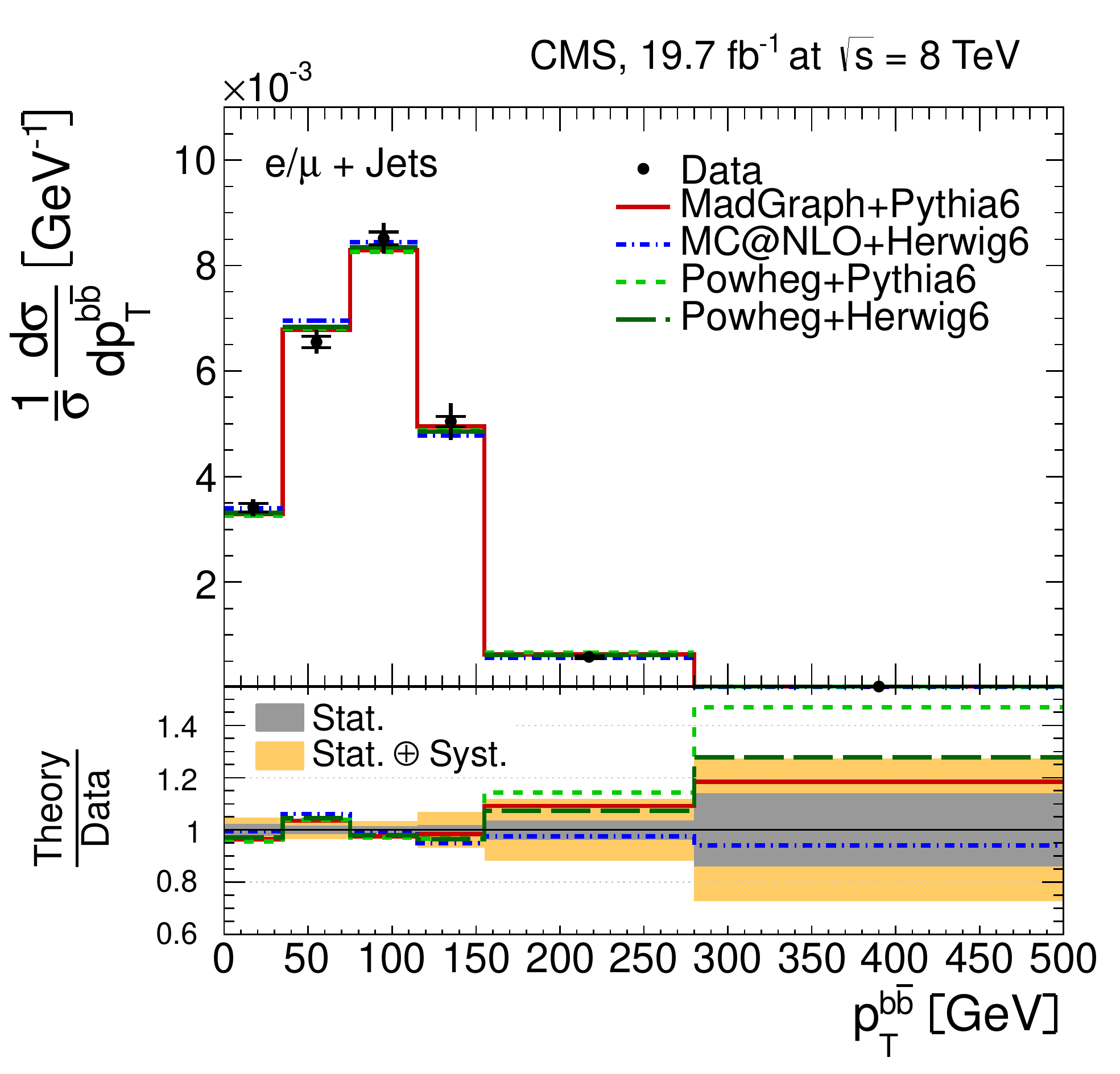}
	\includegraphics[width=0.48\textwidth]{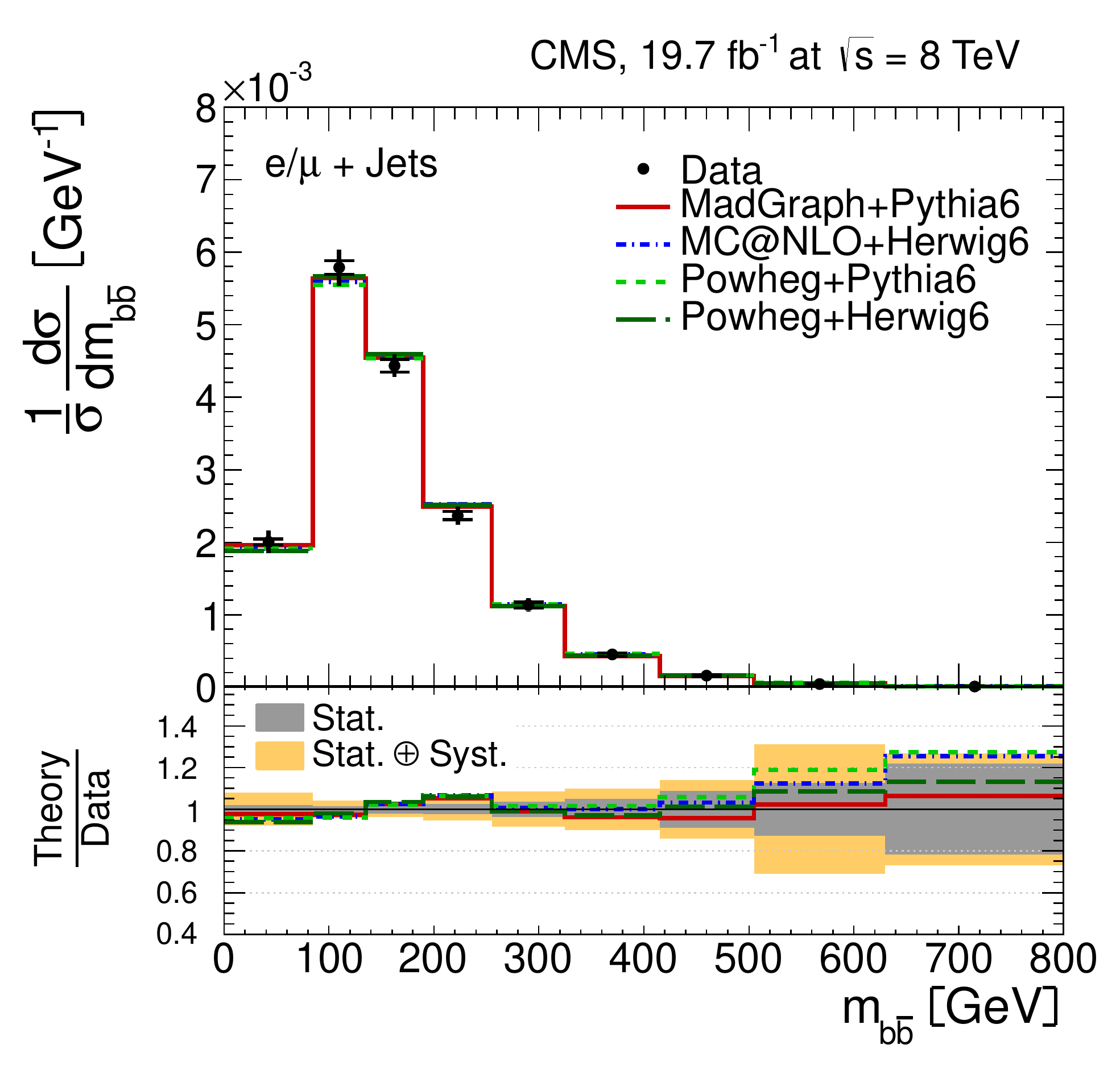}
    \caption{Normalized differential \ttbar production cross section in the \ljets channels as a function of the $\pt^{\PQb}$ (top left) and $\eta_{\PQb}$ (top right) of the b jets, and the $\pt^{\bbbar}$ (bottom left) and $m_{\bbbar}$ (bottom right) of the b jet pair. The superscript `b' refers to both b and $\PAQb$ jets. The data points are placed at the midpoint of the bins. The inner (outer) error bars indicate the statistical (combined statistical and systematic) uncertainties. The measurements are compared to predictions from \MADGRAPH{}+\PYTHIA{6}, \POWHEG{}+\PYTHIA{6}, \POWHEG{}+\HERWIG{6}, and \MCATNLO{}+\HERWIG{6}. The lower part of each plot shows the ratio of the predictions to data.}
    \label{fig:diffXSec:bjets:ljets}

\end{figure*}

{\tolerance=800
Figure~\ref{fig:diffXSec:l:ljets} presents the normalized differential cross section in the \ljets channels as a function of the lepton transverse momentum $\pt^{\ell}$ and pseudorapidity $\eta_{\ell}$. The distributions of the transverse momentum of the b jets $\pt^{\PQb}$ and their pseudorapidity $\eta_{\PQb}$ are given in Fig.~\ref{fig:diffXSec:bjets:ljets}, together with the transverse momentum $\pt^{\bbbar}$ and invariant mass $m_{\bbbar}$ of the b jet pair. Also shown are predictions from \MADGRAPH{}+\PYTHIA{6}, \POWHEG{}+\PYTHIA{6}, \POWHEG{}+\HERWIG{6}, and \MCATNLO{}+\HERWIG{6}. The lower panel in each plot shows the ratio of each of these predictions to data, in order to quantify their level of agreement relative to data.
\par}

\begin{figure*}[htbp]
  \centering
        \includegraphics[width=0.48\textwidth]{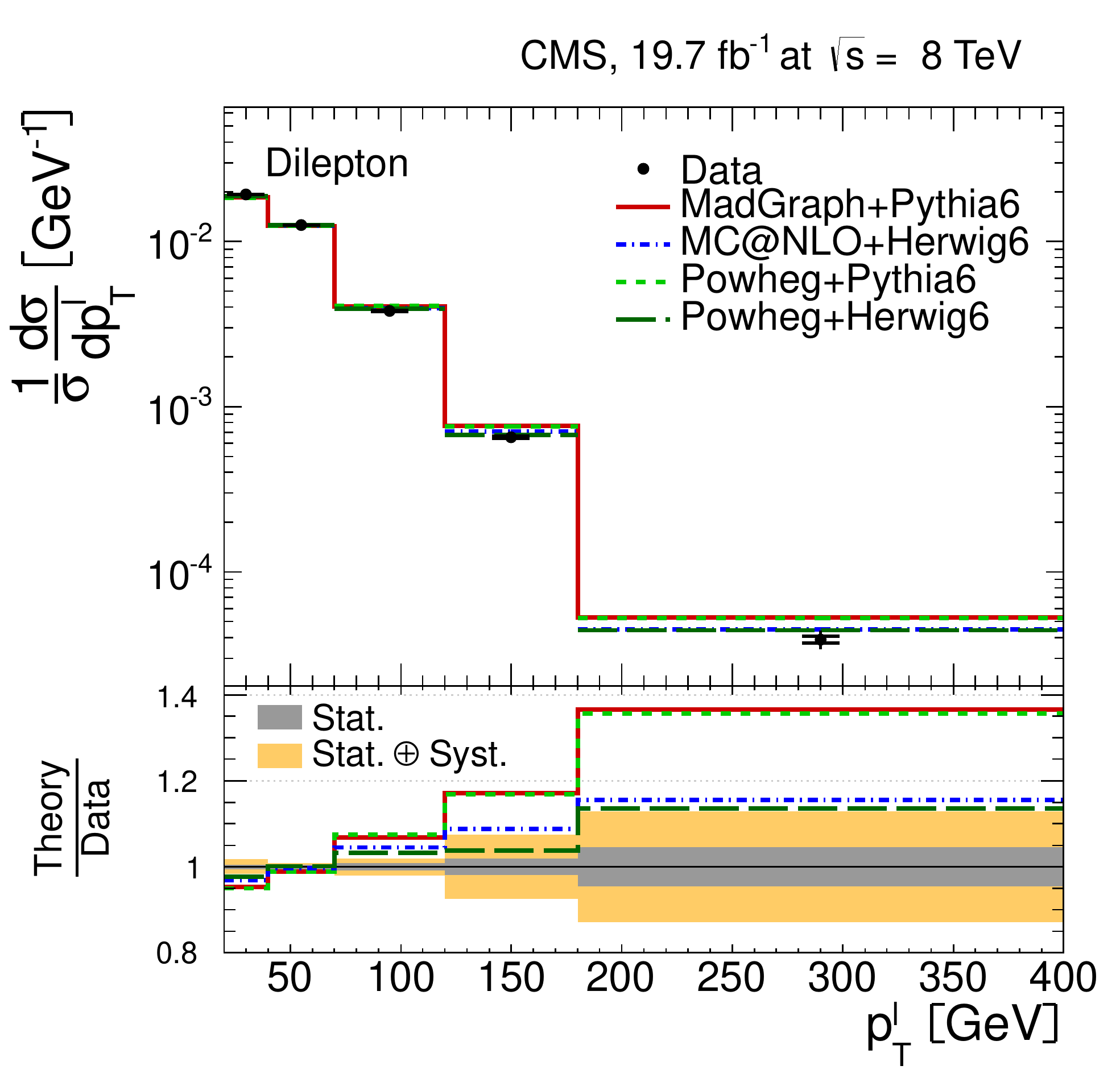}
	\includegraphics[width=0.48\textwidth]{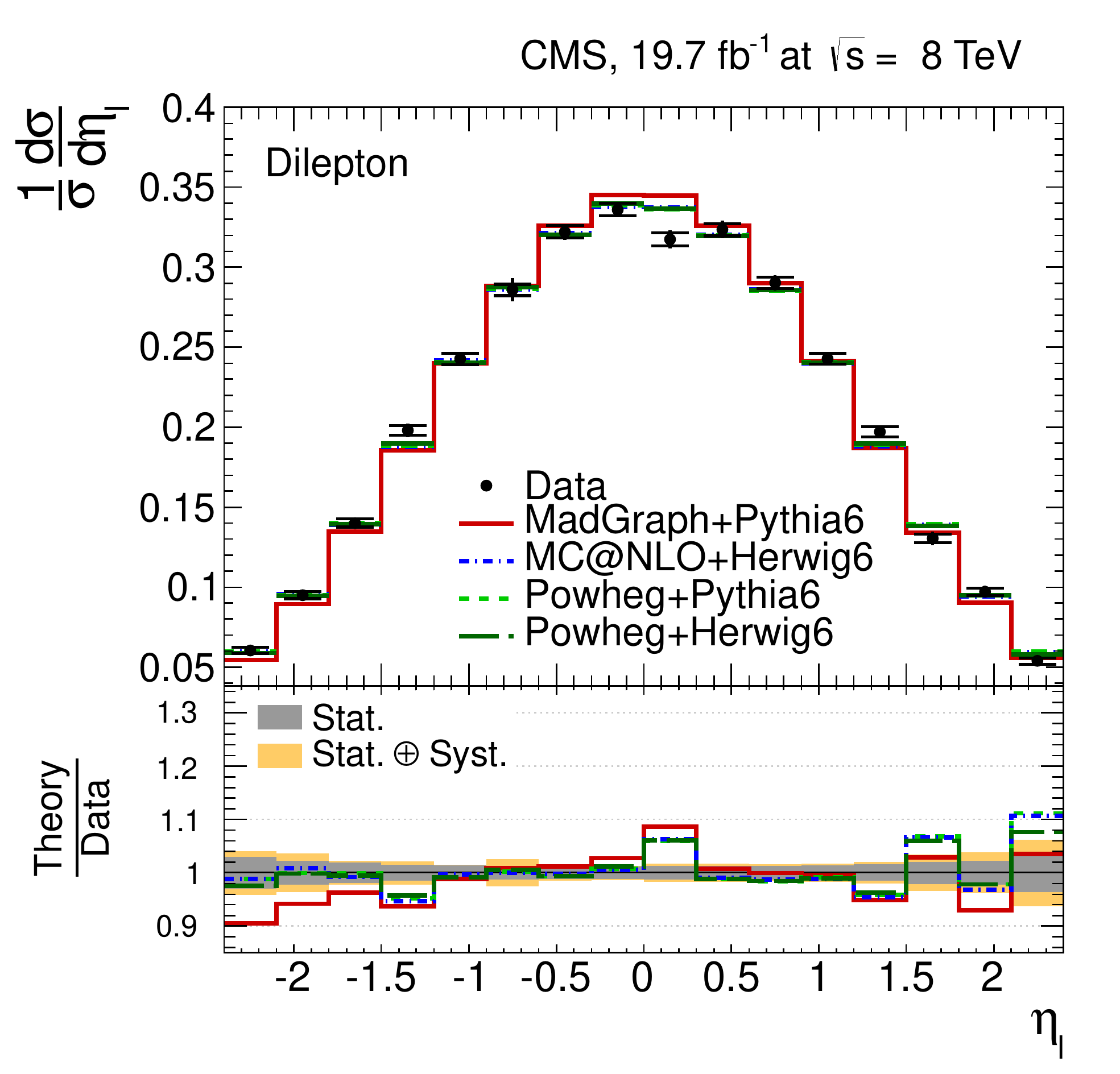}
	\includegraphics[width=0.48\textwidth]{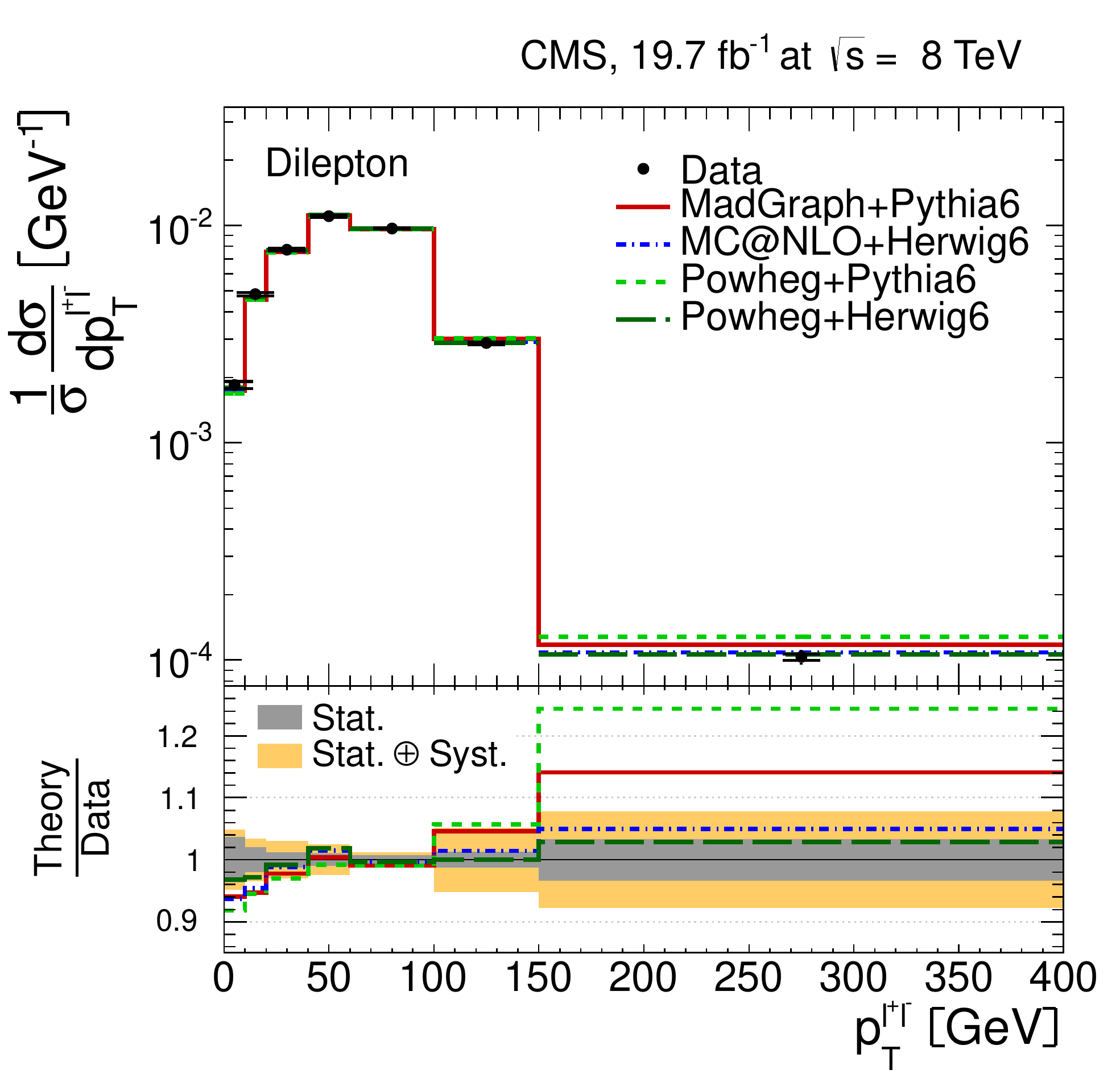}
	\includegraphics[width=0.48\textwidth]{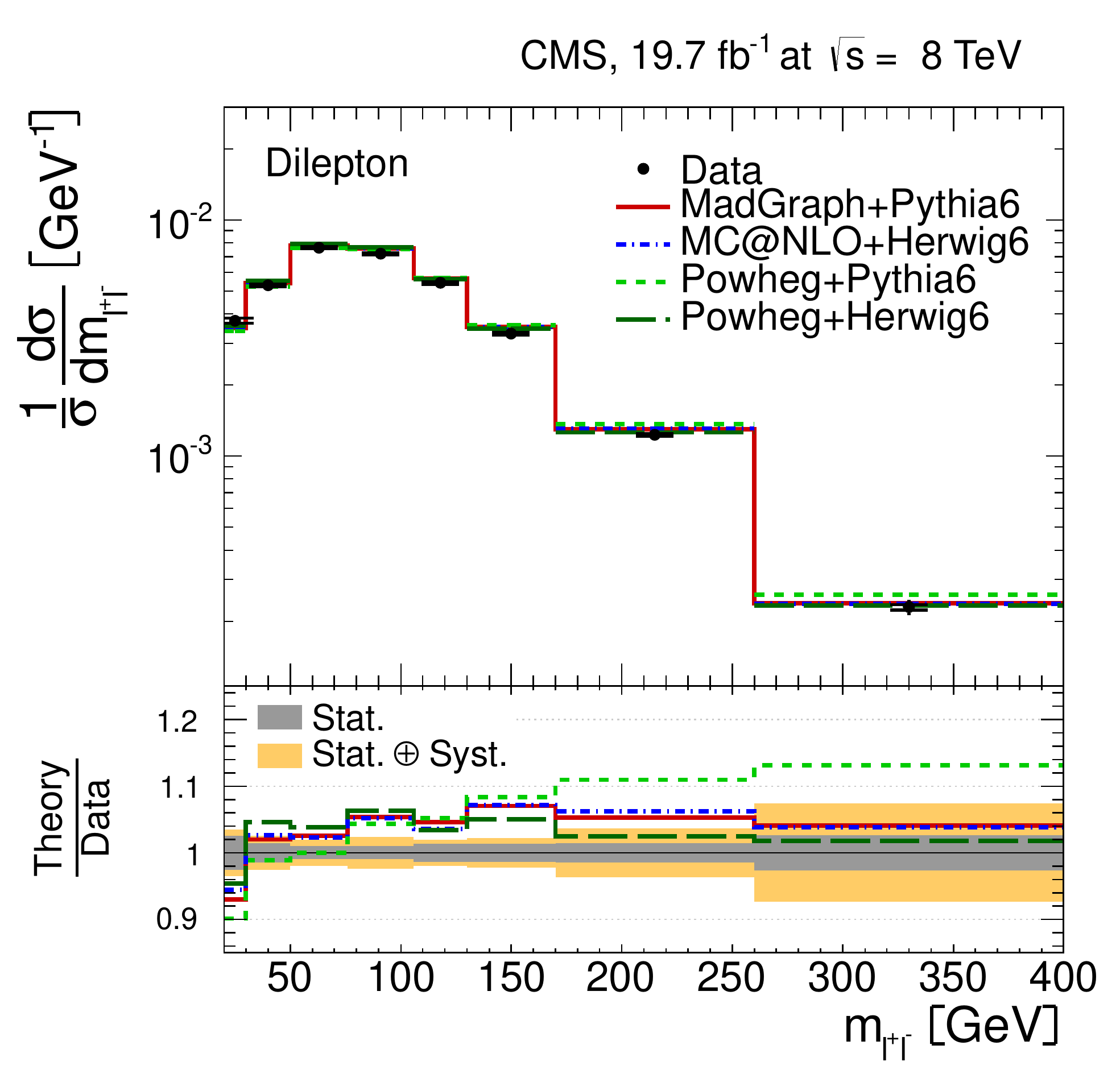}
    \caption{Normalized differential \ttbar production cross section in the dilepton channels as a function of the $\pt^{\ell}$ (top left) and $\eta_{\ell}$ (top right) of the charged leptons, and the $\pt^{\ell^{+}\ell^{-}}$ (bottom left) and $m_{\ell^{+}\ell^{-}}$ (bottom right) of the lepton pair. The superscript `$\ell$' refers to both $\ell^{+}$ and $\ell^{-}$. The data points are placed at the midpoint of the bins. The inner (outer) error bars indicate the statistical (combined statistical and systematic) uncertainties. The measurements are compared to predictions from \MADGRAPH{}+\PYTHIA{6}, \POWHEG{}+\PYTHIA{6}, \POWHEG{}+\HERWIG{6}, and \MCATNLO{}+\HERWIG{6}. The lower part of each plot shows the ratio of the predictions to data.}
    \label{fig:diffXSec:ll:dilepton}

\end{figure*}

\begin{figure*}[htbp]
  \centering
        \includegraphics[width=0.48\textwidth]{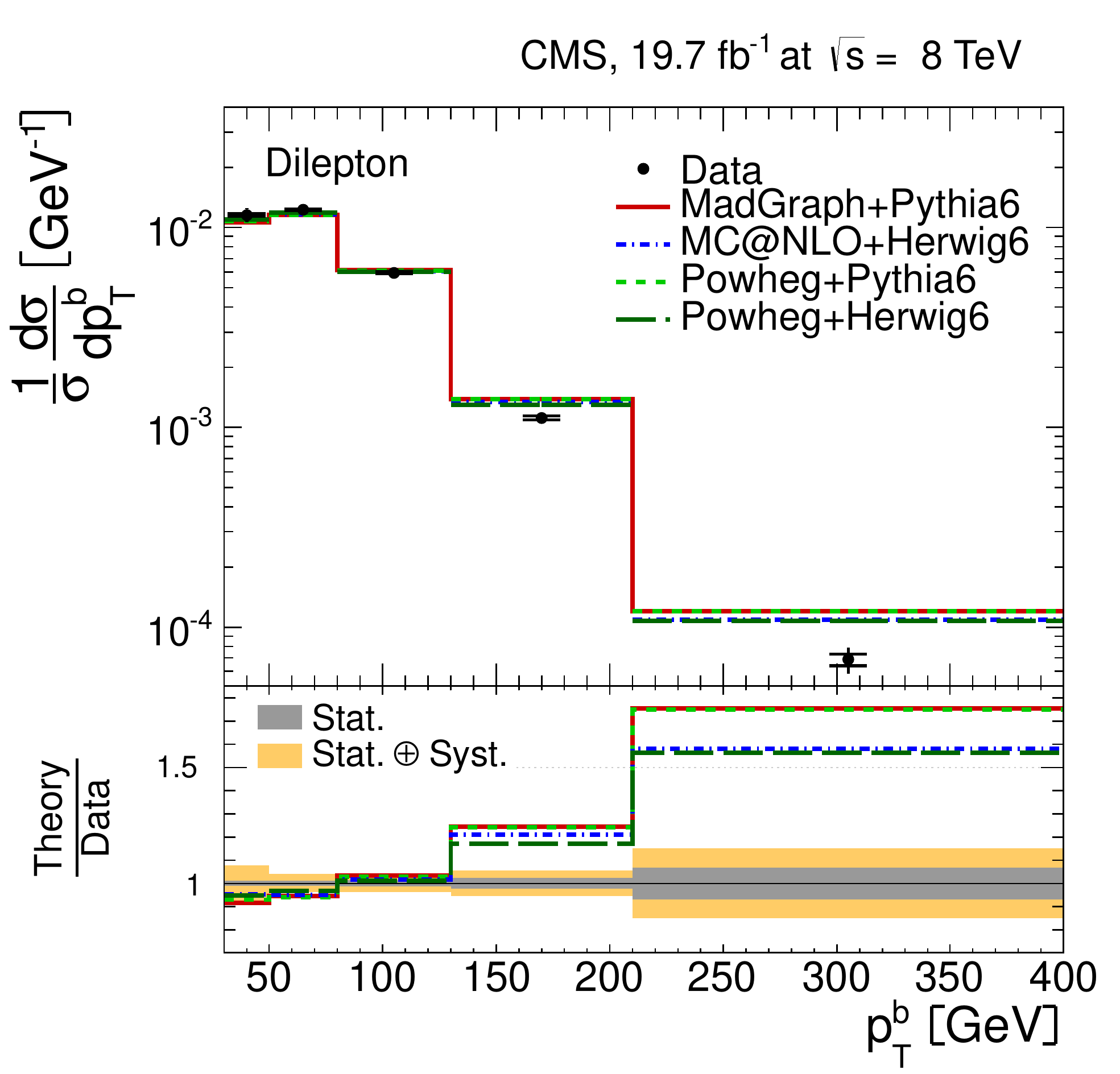}
	\includegraphics[width=0.48\textwidth]{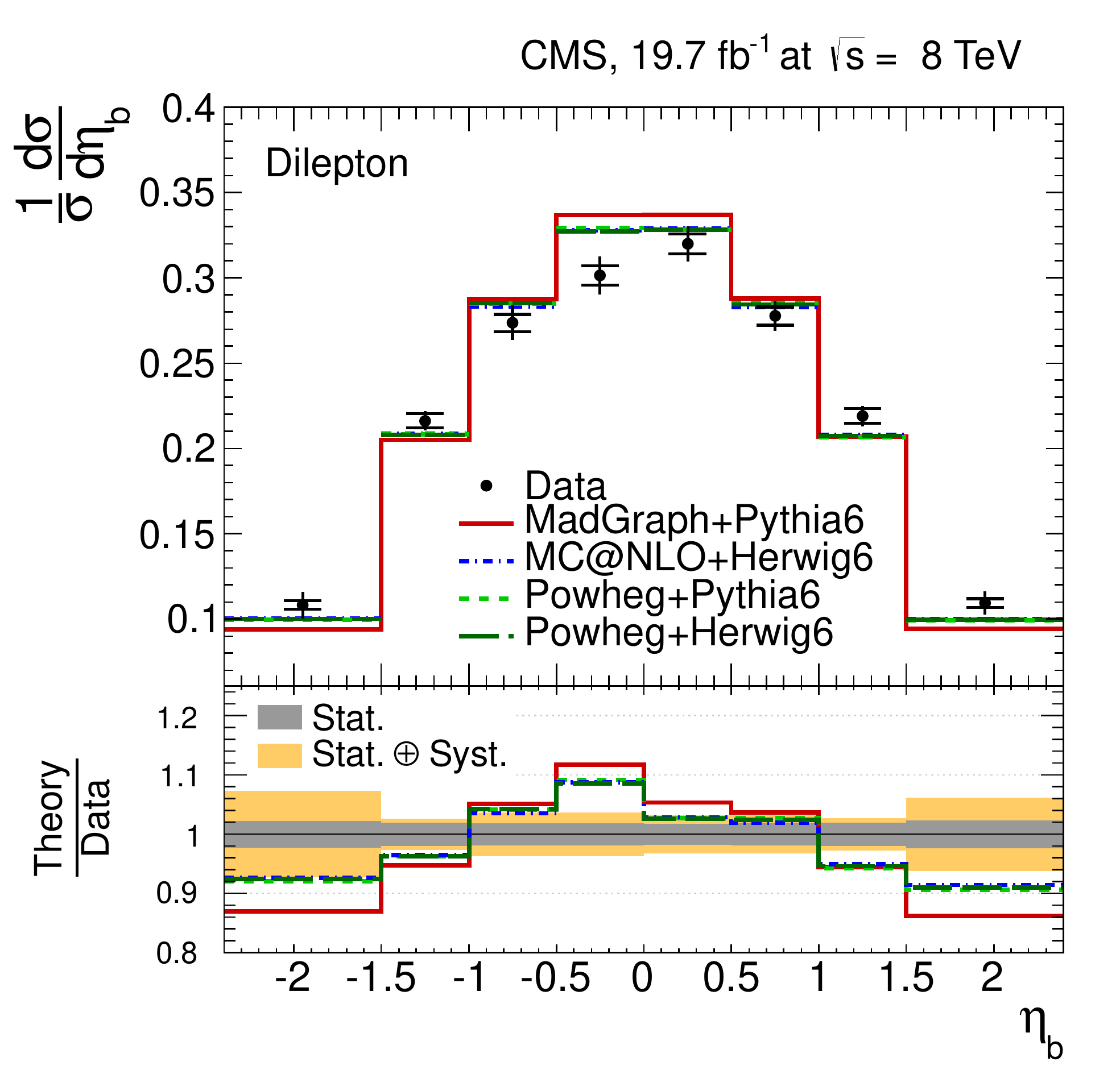}
	\includegraphics[width=0.48\textwidth]{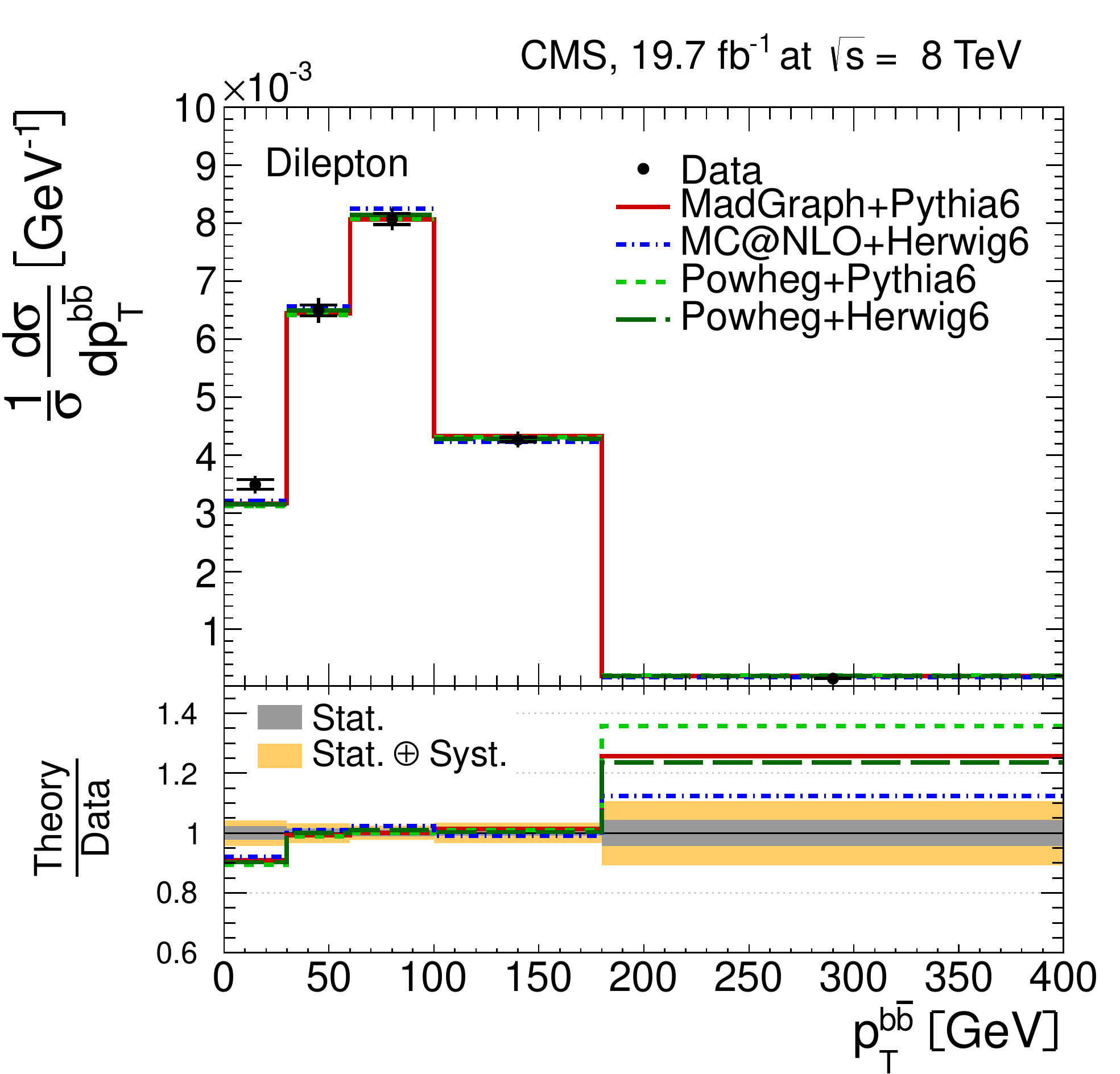}
	\includegraphics[width=0.48\textwidth]{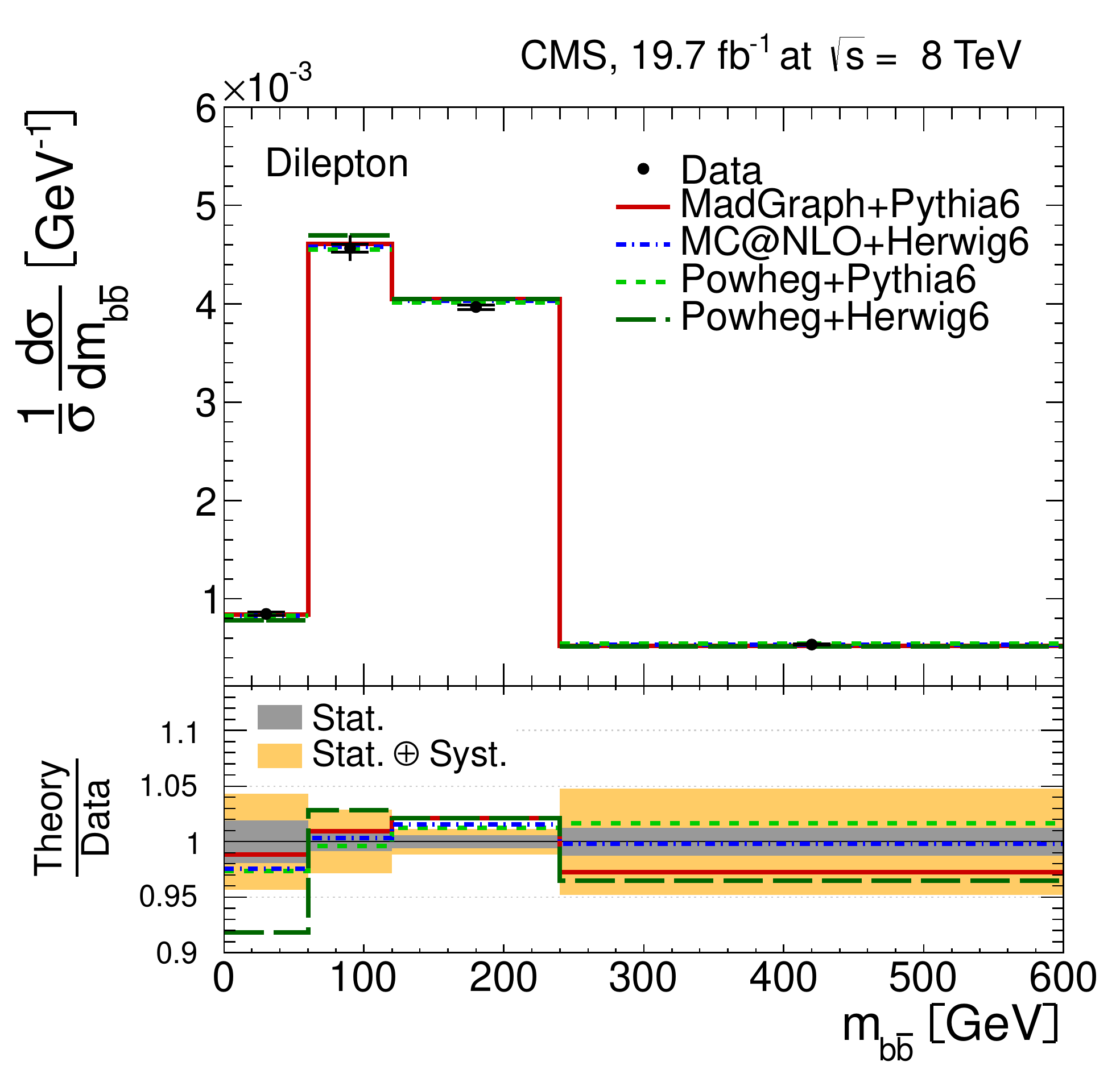}
        \caption{Normalized differential \ttbar production cross section in the dilepton channels as a function of the $\pt^{\PQb}$ (top left) and $\eta_{\PQb}$ (top right) of the b jets, and the $\pt^{\bbbar}$ (bottom left) and $m_{\bbbar}$ (bottom right) of the b jet pair. The superscript `b' refers to both b and $\PAQb$ jets. The data points are placed at the midpoint of the bins. The inner (outer) error bars indicate the statistical (combined statistical and systematic) uncertainties. The measurements are compared to predictions from \MADGRAPH{}+\PYTHIA{6}, \POWHEG{}+\PYTHIA{6}, \POWHEG{}+\HERWIG{6}, and \MCATNLO{}+\HERWIG{6}. The lower part of each plot shows the ratio of the predictions to data.}
    \label{fig:diffXSec:bjets:dilepton}

\end{figure*}

{\tolerance=800
Figure~\ref{fig:diffXSec:ll:dilepton} presents the normalized differential cross sections for the dilepton channels: the transverse momentum $\pt^{\ell}$ and the pseudorapidity $\eta_{\ell}$ of the leptons, and the transverse momentum $\pt^{\ell^{+}\ell^{-}}$ and the invariant mass $m_{\ell^{+}\ell^{-}}$ of the lepton pair. The distributions in the transverse momentum of the b jets $\pt^{\PQb}$ and their pseudorapidity $\eta_{\PQb}$ are shown in Fig.\,~\ref{fig:diffXSec:bjets:dilepton}, together with the transverse momentum $\pt^{\bbbar}$ and invariant mass $m_{\bbbar}$ of the b jet pair. Predictions from \MADGRAPH{}+\PYTHIA{6}, \POWHEG{}+\PYTHIA{6}, \POWHEG{}+\HERWIG{6}, and \MCATNLO{}+\HERWIG{6} are also presented for comparison.

In general, none of the examined predictions provides an accurate description of data for all measured lepton and b jet distributions. A steeper \pt spectrum is observed in data for the lepton and the b jet distributions compared to the predictions in both decay channels, which is best described by \POWHEG{}+\HERWIG{6}. The lepton \pt in data is above the predictions for $\pt^{\ell} < 40\GeV$, while it is below for $\pt^{\ell} > 100\GeV$. A similar behaviour is observed for $\pt^{\ell^{+}\ell^{-}}$, $\pt^{\PQb}$, and $\pt^{\bbbar}$. The $m_{\ell^{+}\ell^{-}}$ distribution in data is below all predictions for $m_{\ell^{+}\ell^{-}} > 30\GeV$. Worse agreement is found for \POWHEG{}+\PYTHIA{6}. The $\eta$ distributions in data are described by the predictions within the experimental uncertainties. The $\eta_{\PQb}$ distributions are slightly less central in data than in the predictions, and are worse described by \MADGRAPH{}+\PYTHIA{6}. The remaining distributions are described by the predictions within experimental uncertainties.
\par}

\subsection{Top quark and \texorpdfstring{\ttbar}{Top Quark Pair} differential cross sections}
\label{subsec:fullPS}

The normalized differential \ttbar cross section as a function of the kinematic properties of the top quarks and the \ttbar system is defined with respect to the top quarks or antiquarks before the decay (parton level) and after QCD radiation, and extrapolated to the full phase space using the \MADGRAPH{}+\PYTHIA{6} prediction for the \ljets and dilepton channels.

\begin{figure*}[htbp]
  \centering
       \includegraphics[width=0.48\textwidth]{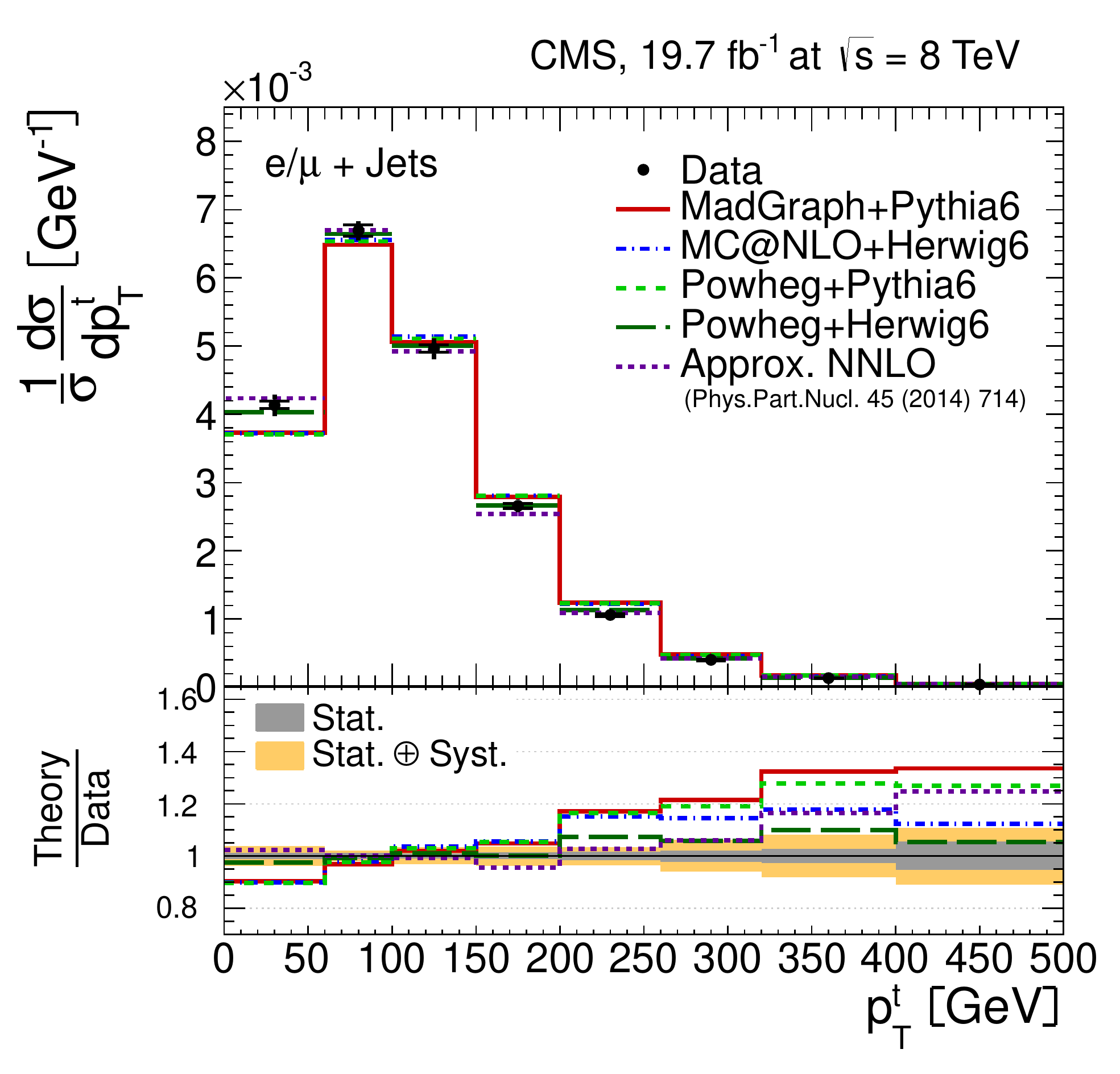}
	\includegraphics[width=0.48\textwidth]{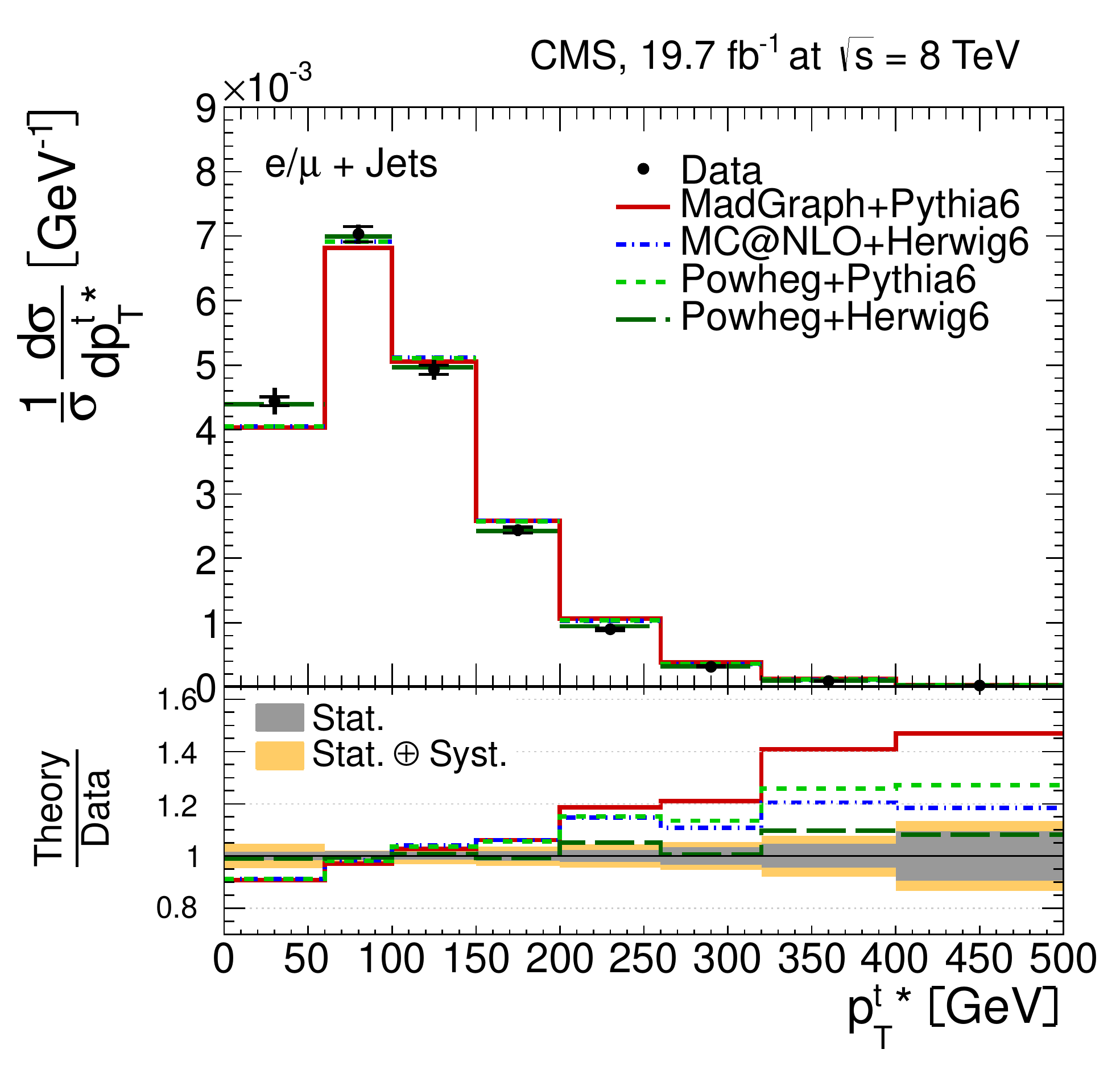}\\
	\includegraphics[width=0.48\textwidth]{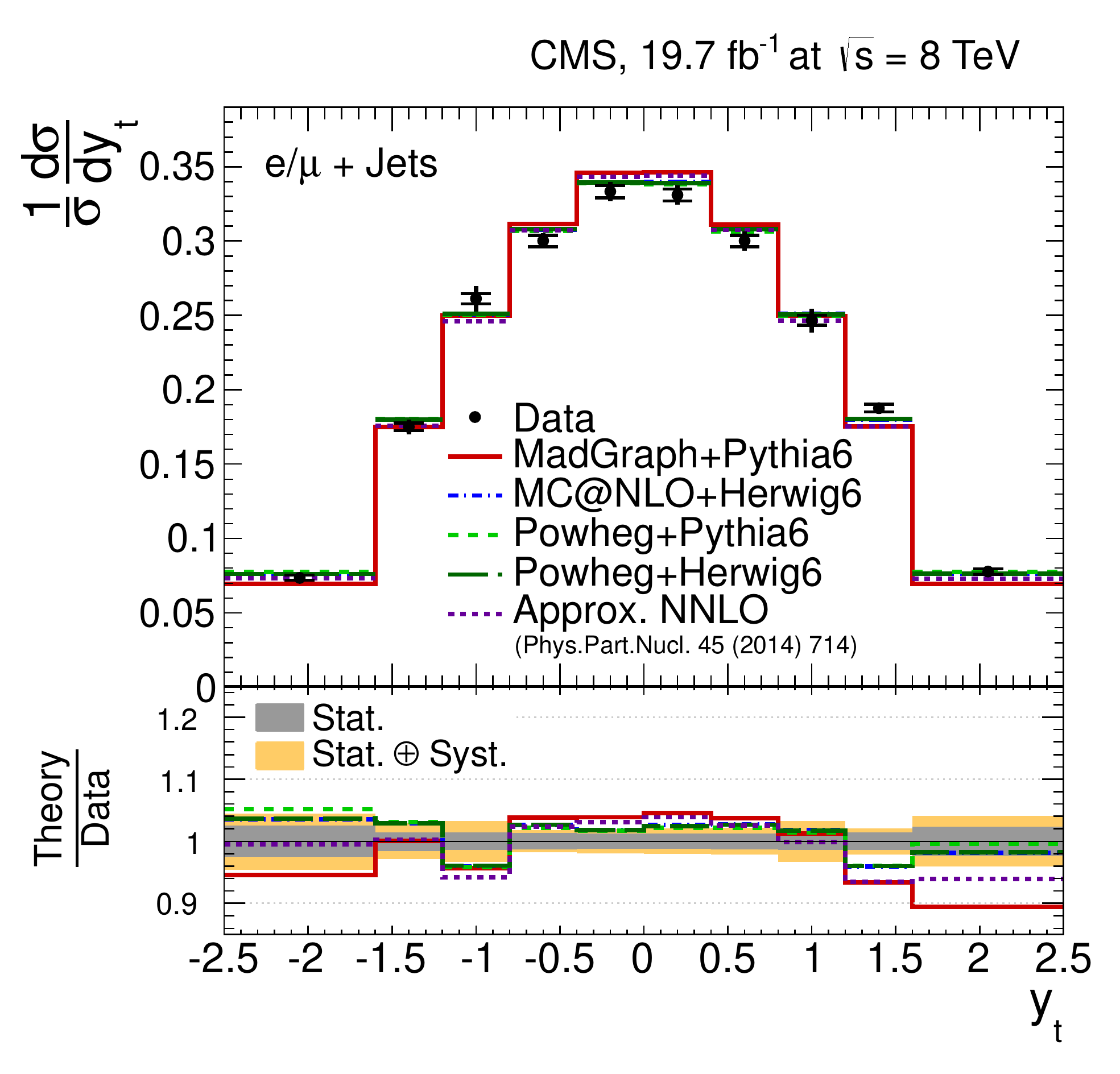}
	\includegraphics[width=0.48\textwidth]{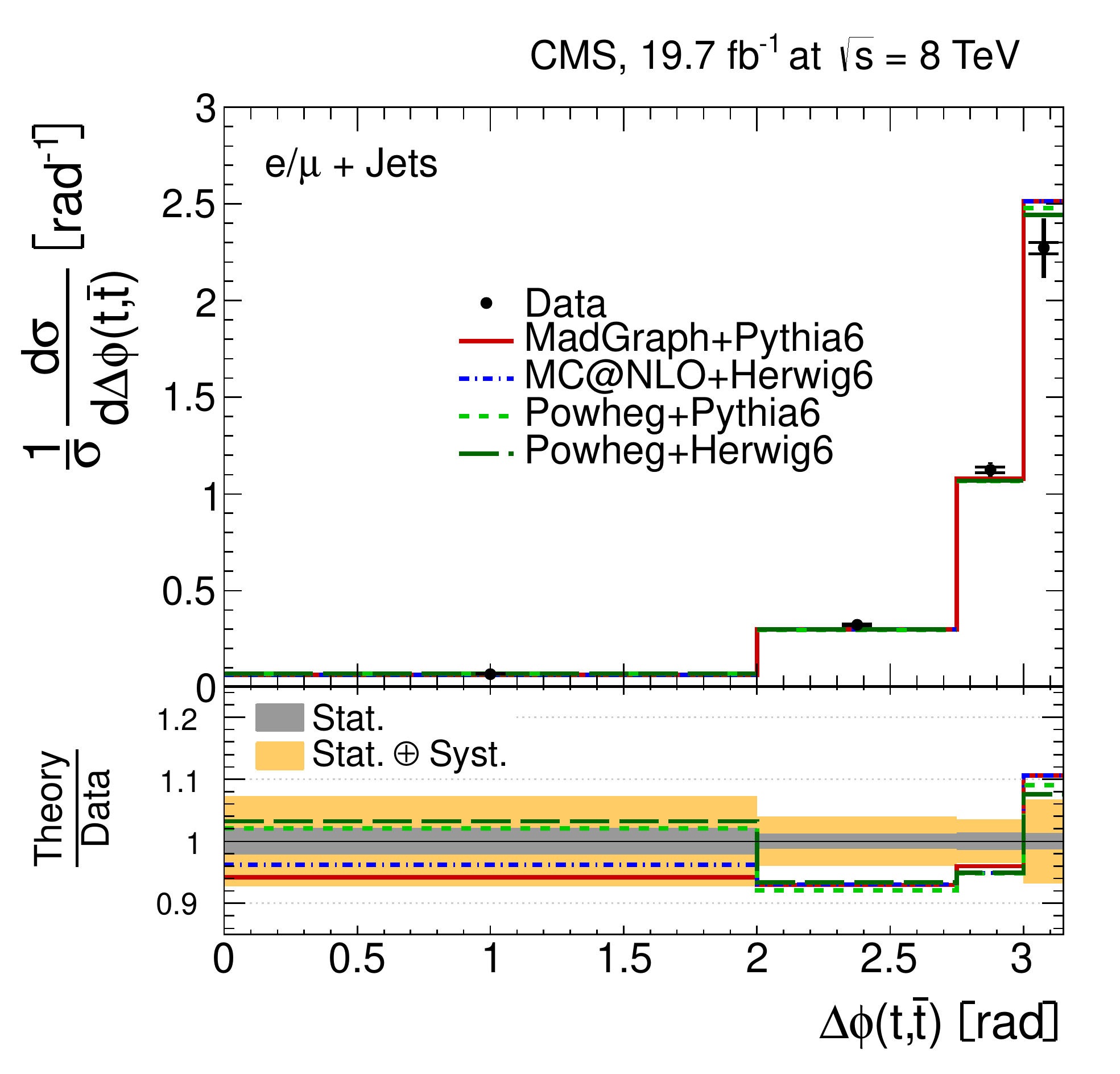}
    \caption{Normalized differential \ttbar production cross section in the \ljets channels as a function of the $\pt^{\PQt}$ (top left), the \ttbar rest frame $\pt^{\PQt\ast}$ (top right), and the rapidity $y_{\PQt}$ (bottom left) of the top quarks or antiquarks, and the difference in the azimuthal angle between the top quark and the antiquark $\Delta \phi(\text{t,}\bar{\PQt})$ (bottom right). The data points are placed at the midpoint of the bins. The inner (outer) error bars indicate the statistical (combined statistical and systematic) uncertainties. The measurements are compared to predictions from \MADGRAPH{}+\PYTHIA{6}, \POWHEG{}+\PYTHIA{6}, \POWHEG{}+\HERWIG{6}, \MCATNLO{}+\HERWIG{6}, and to approximate NNLO~\cite{bib:kidonakis_8TeV} calculations, when available. The lower part of each plot shows the ratio of the predictions to data.}
\label{fig:diffXSec:top:ljets}

\end{figure*}

\begin{figure*}[htb]
  \centering
       \includegraphics[width=0.48\textwidth]{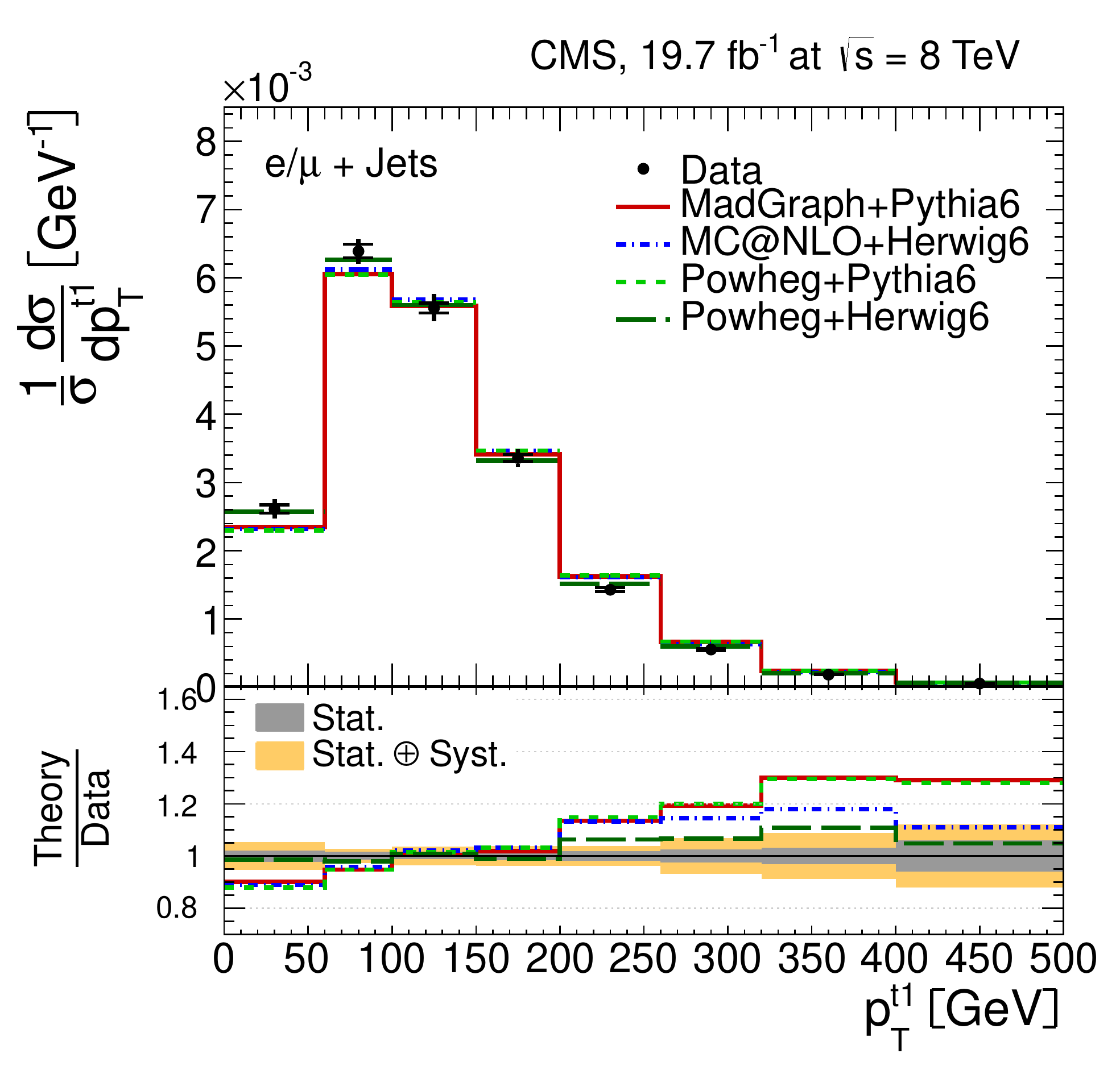}
       \includegraphics[width=0.48\textwidth]{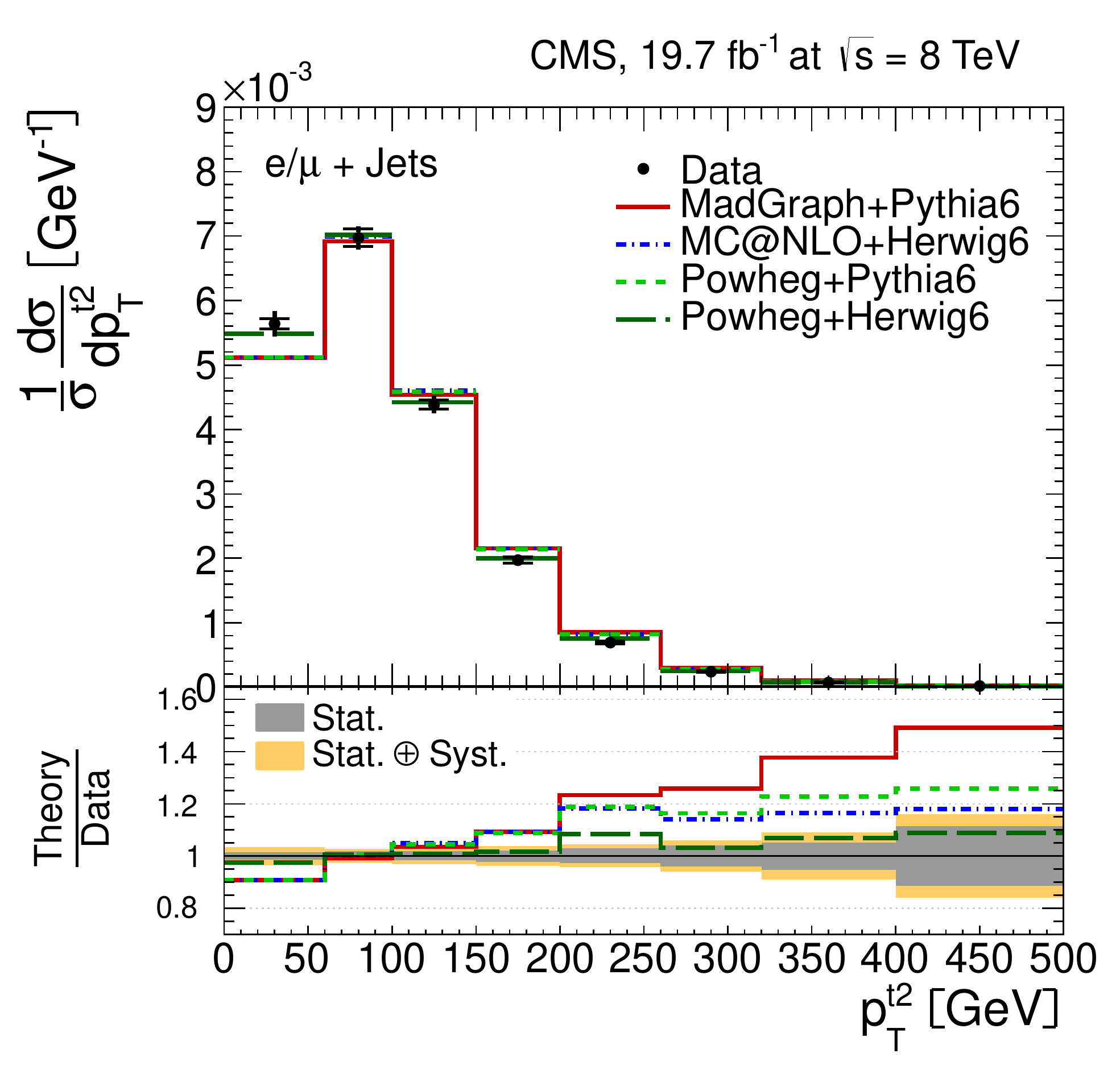}
    \caption{Normalized differential \ttbar production cross section in the \ljets channels as a function of the \pt of the leading (left) and trailing (right) top quarks or antiquarks. The data points are placed at the midpoint of the bins. The inner (outer) error bars indicate the statistical (combined statistical and systematic) uncertainties. The measurements are compared to predictions from \MADGRAPH{}+\PYTHIA{6}, \POWHEG{}+\PYTHIA{6}, \POWHEG{}+\HERWIG{6}, and \MCATNLO{}+\HERWIG{6}. The lower part of each plot shows the ratio of the predictions to data.}
\label{fig:diffXSec:top2:ljets}

\end{figure*}

\begin{figure*}[htbp]
  \centering
       \includegraphics[width=0.48\textwidth]{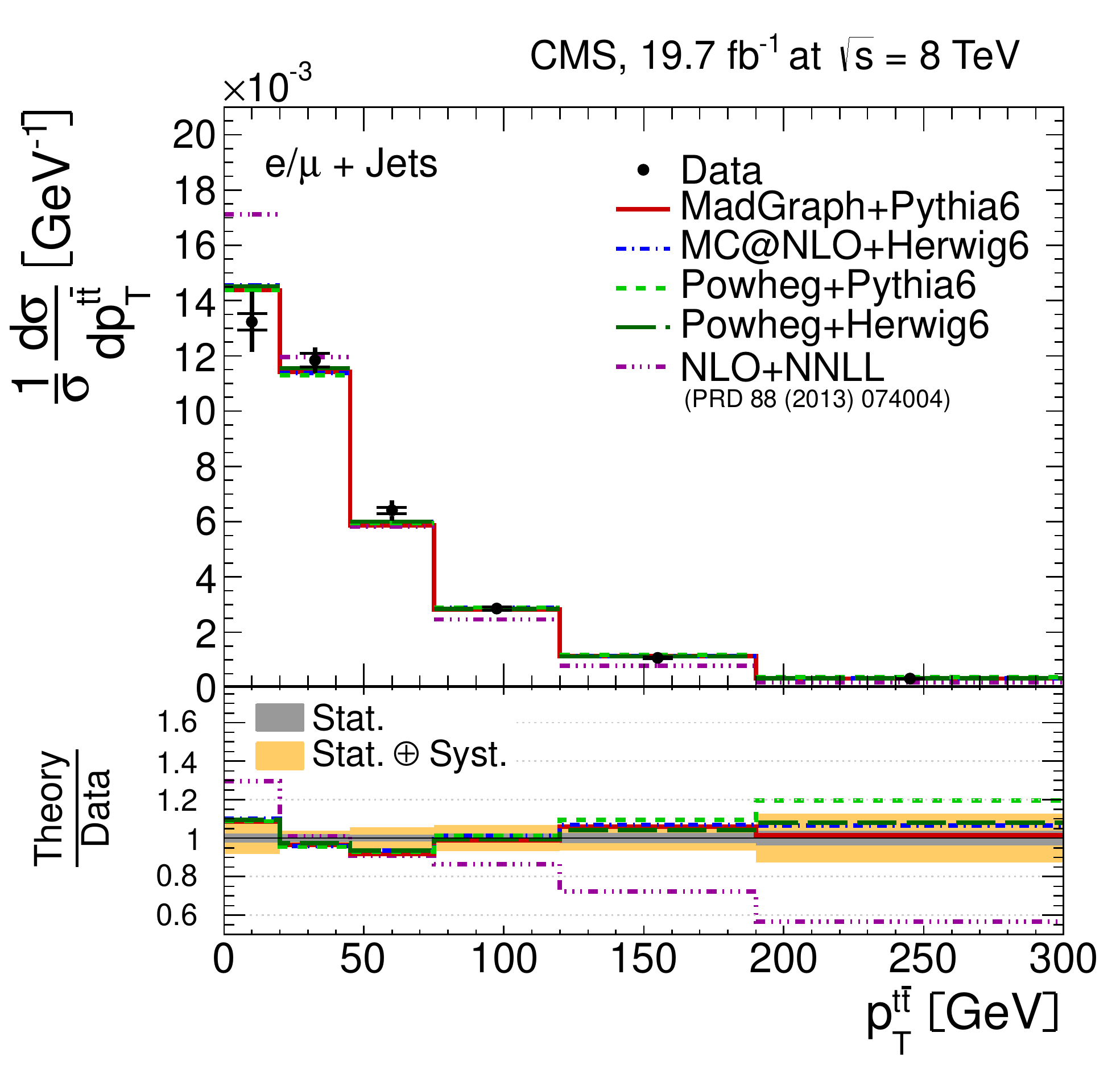}
	\includegraphics[width=0.48\textwidth]{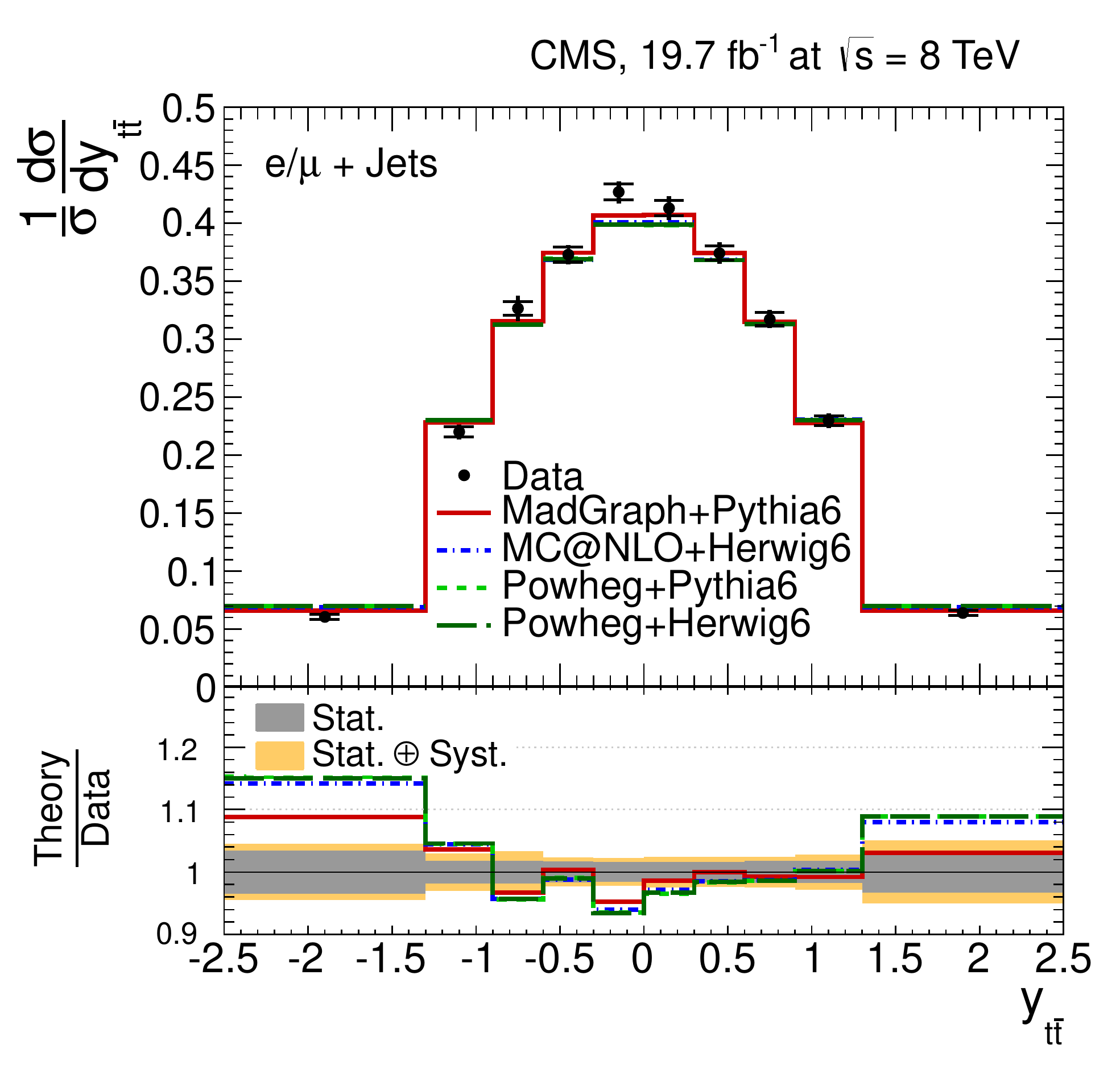}\\
	\includegraphics[width=0.48\textwidth]{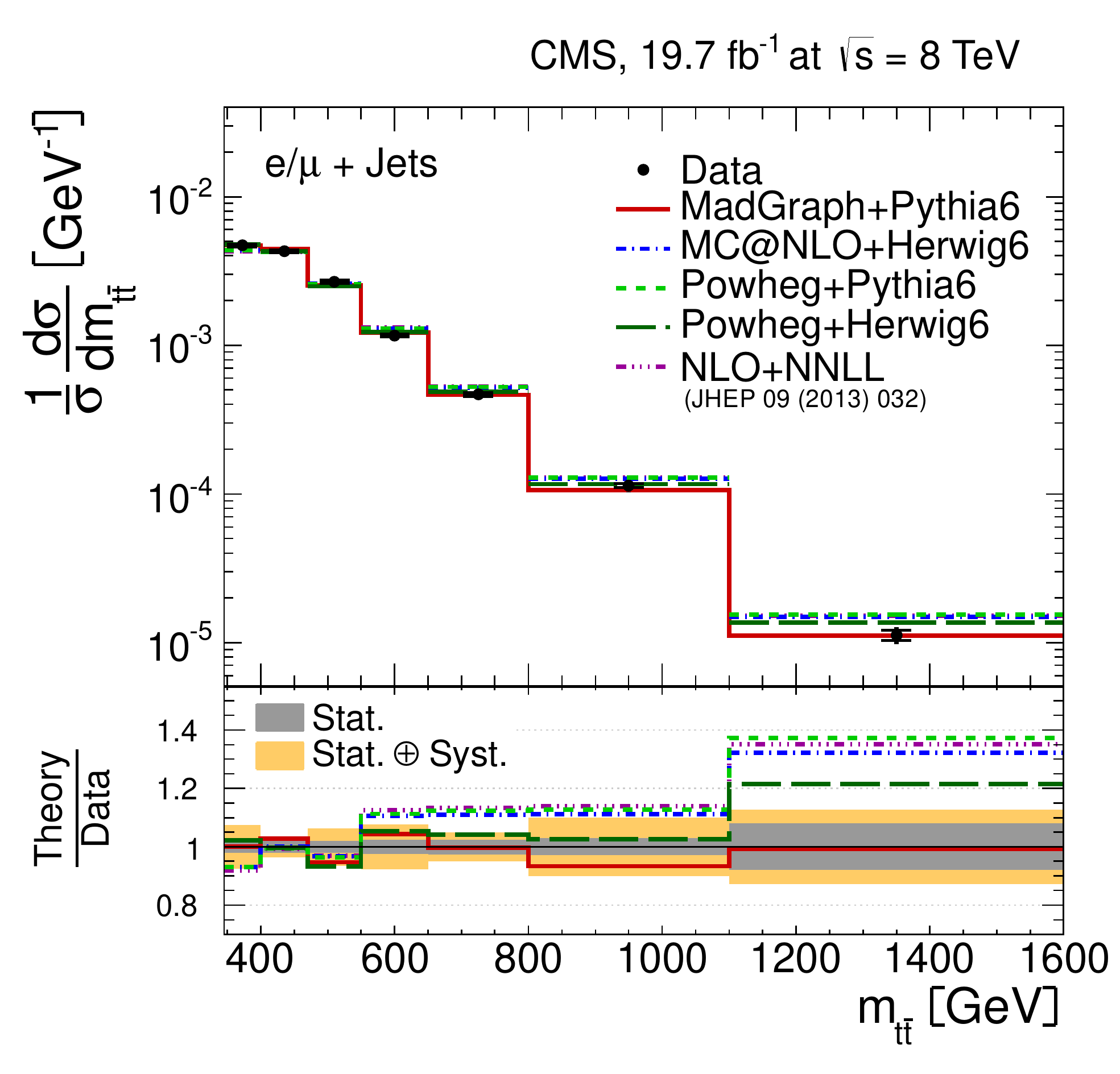}
    \caption{Normalized differential \ttbar production cross section in the \ljets channels as a function of the $\pt^{\ttbar}$ (top left), $y_{\ttbar}$ (top right), and $m_{\ttbar}$ (bottom) of the \ttbar system. The data points are placed at the midpoint of the bins. The inner (outer) error bars indicate the statistical (combined statistical and systematic) uncertainties. The measurements are compared to predictions from \MADGRAPH{}+\PYTHIA{6}, \POWHEG{}+\PYTHIA{6}, \POWHEG{}+\HERWIG{6}, \MCATNLO{}+\HERWIG{6}, and to NLO+NNLL~\cite{bib:ahrens_mttbar,bib:ahrens_ptttbar} calculations, when available. The lower part of each plot shows the ratio of the predictions to data.}
\label{fig:diffXSec:tt:ljets}

\end{figure*}

\begin{figure*}[htbp]
  \centering
        \includegraphics[width=0.48\textwidth]{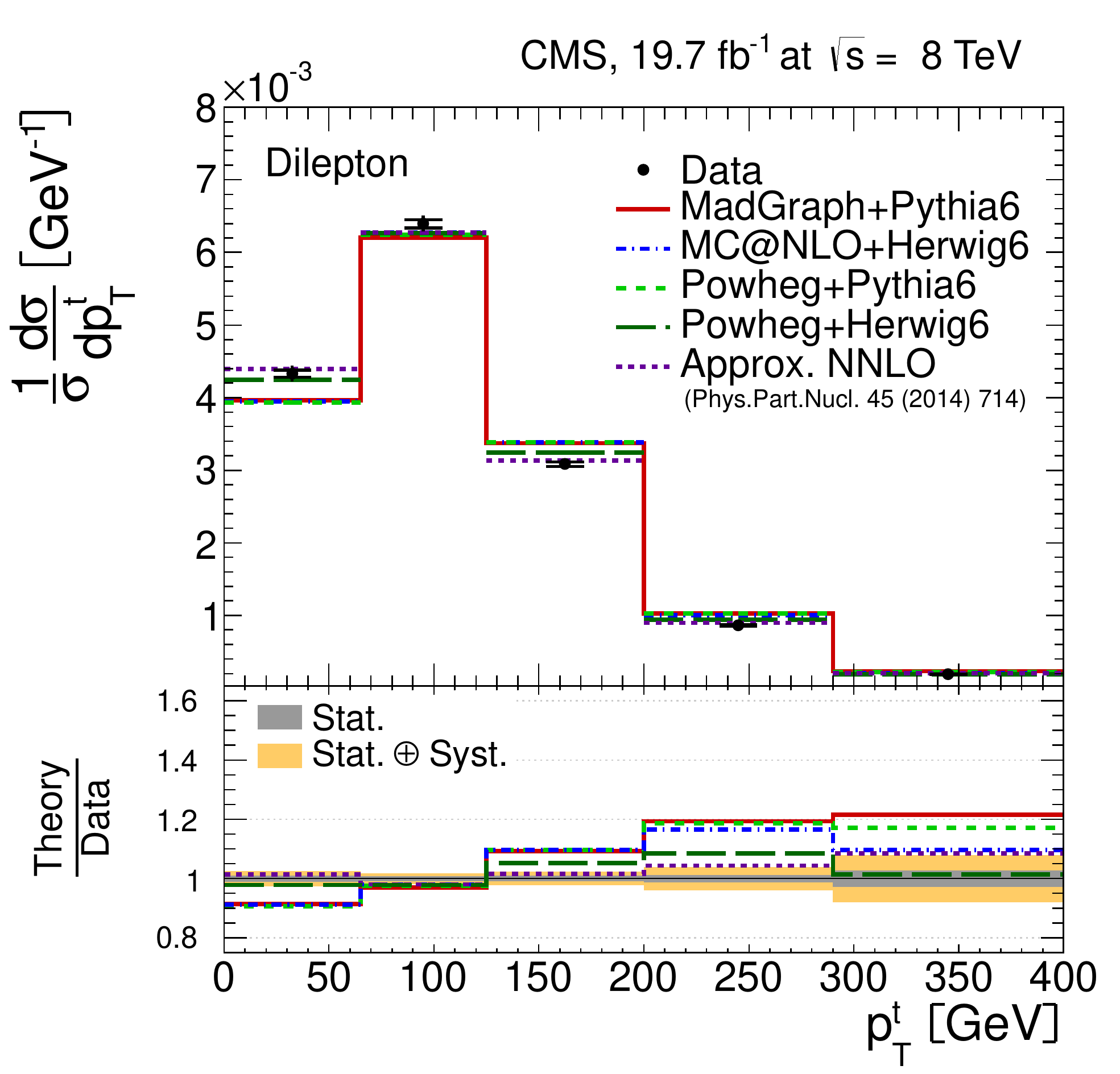}
	\includegraphics[width=0.48\textwidth]{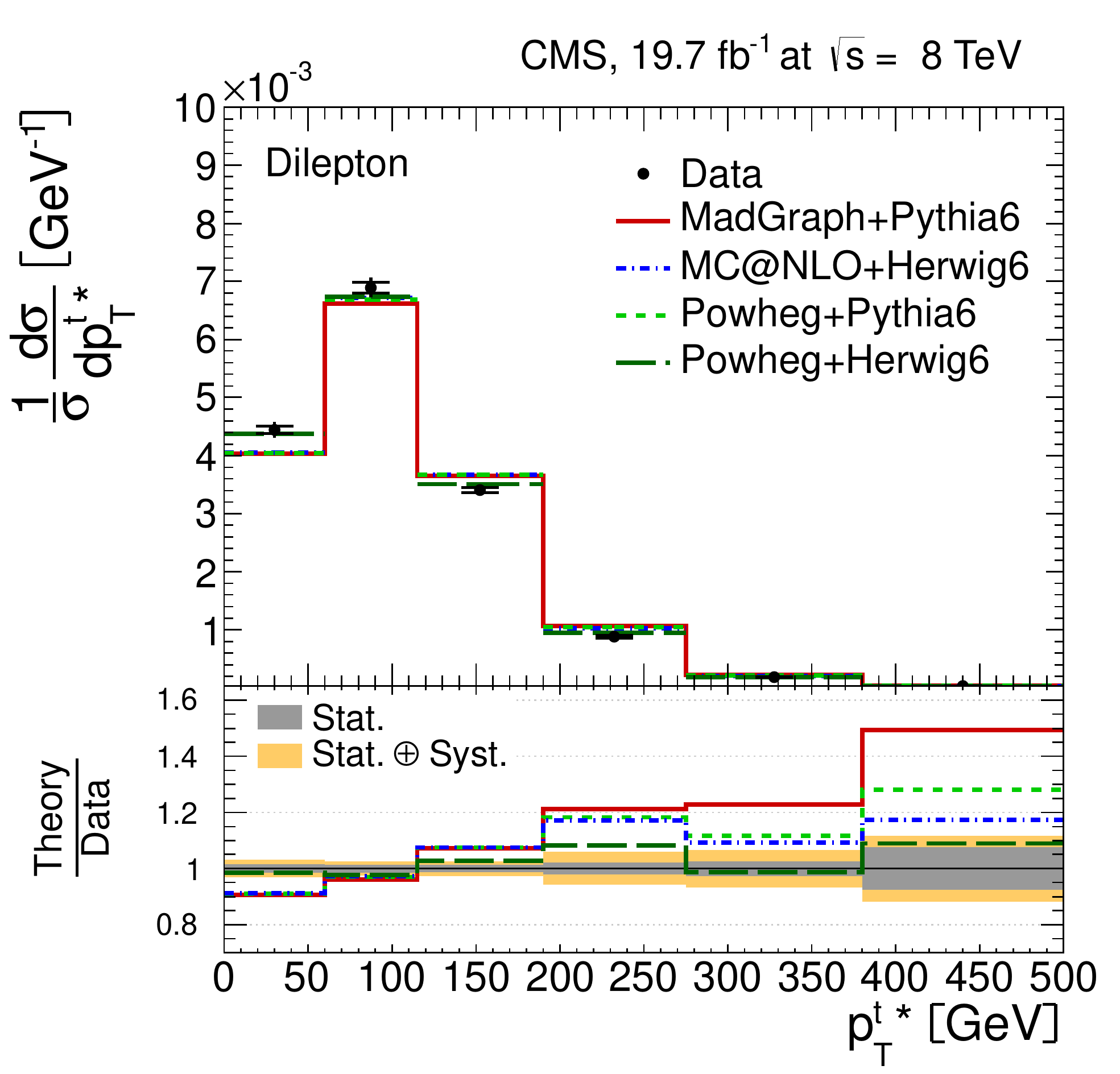}\\
	\includegraphics[width=0.48\textwidth]{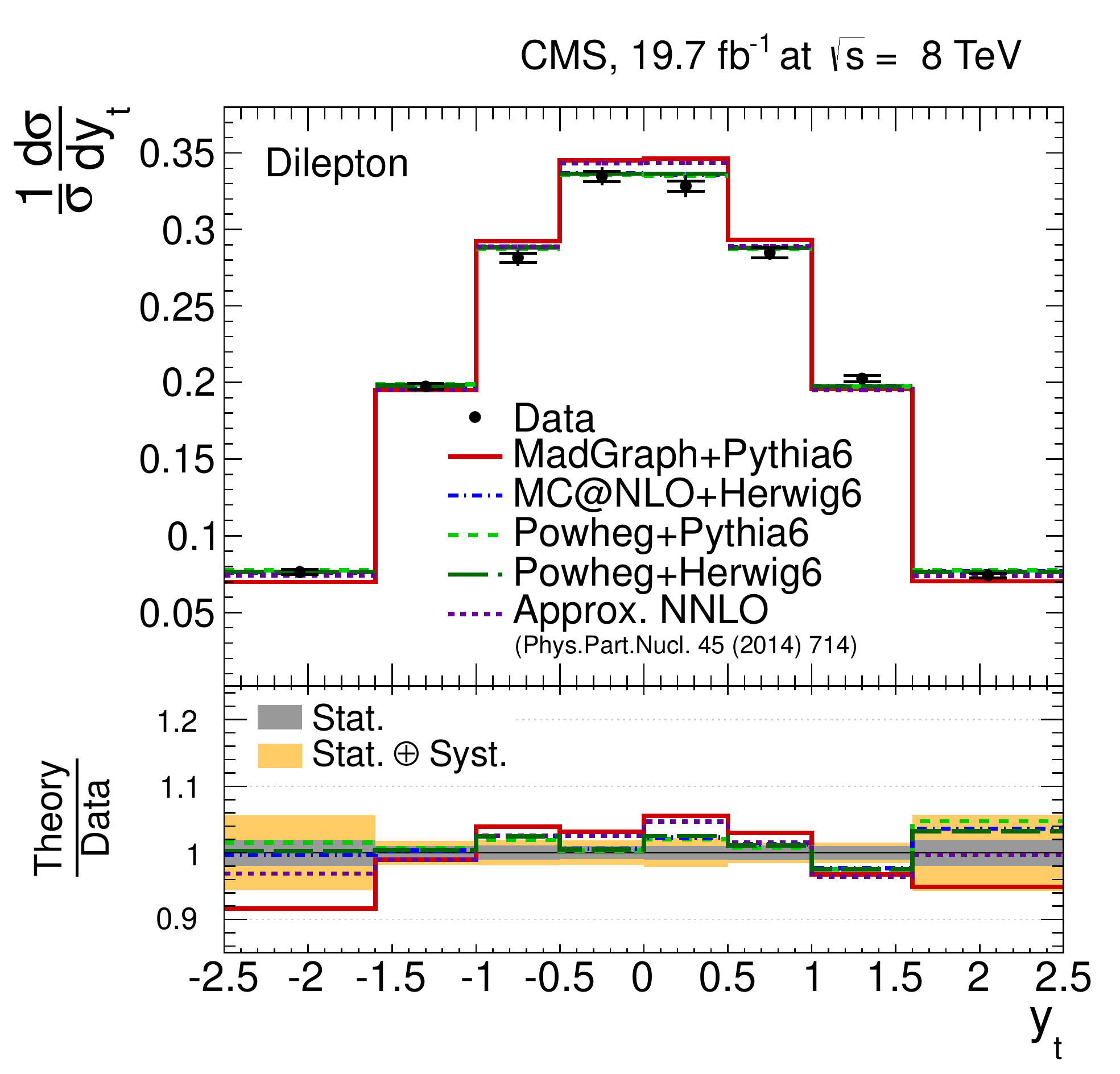}
	\includegraphics[width=0.48\textwidth]{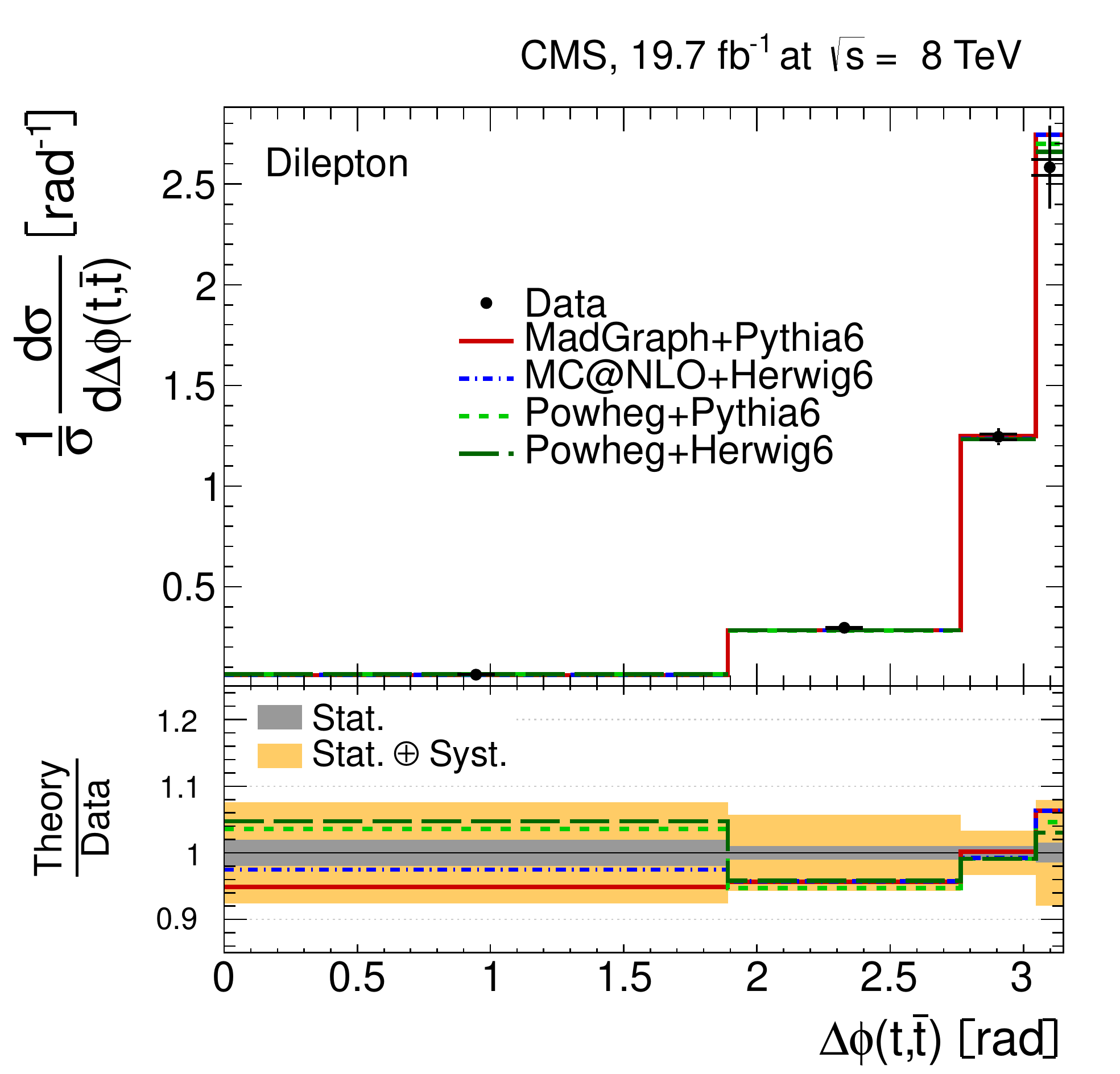}
    \caption{Normalized differential \ttbar production cross section in the dilepton channels as a function of the $\pt^{\PQt}$ (top left), the \ttbar rest frame $\pt^{\PQt\ast}$ (top right), and the rapidity $y_{\PQt}$ (bottom left) of the top quarks or antiquarks, and the difference in the azimuthal angle between the top quark and the antiquark $\Delta \phi(\text{t,}\bar{\PQt})$ (bottom right). The data points are placed at the midpoint of the bins. The inner (outer) error bars indicate the statistical (combined statistical and systematic) uncertainties. The measurements are compared to predictions from \MADGRAPH{}+\PYTHIA{6}, \POWHEG{}+\PYTHIA{6}, \POWHEG{}+\HERWIG{6}, \MCATNLO{}+\HERWIG{6}, and to approximate NNLO~\cite{bib:kidonakis_8TeV} calculations, when available. The lower part of each plot shows the ratio of the predictions to data.}
    \label{fig:diffXSec:top:dilepton}

\end{figure*}

\begin{figure*}[htb]
  \centering
       \includegraphics[width=0.48\textwidth]{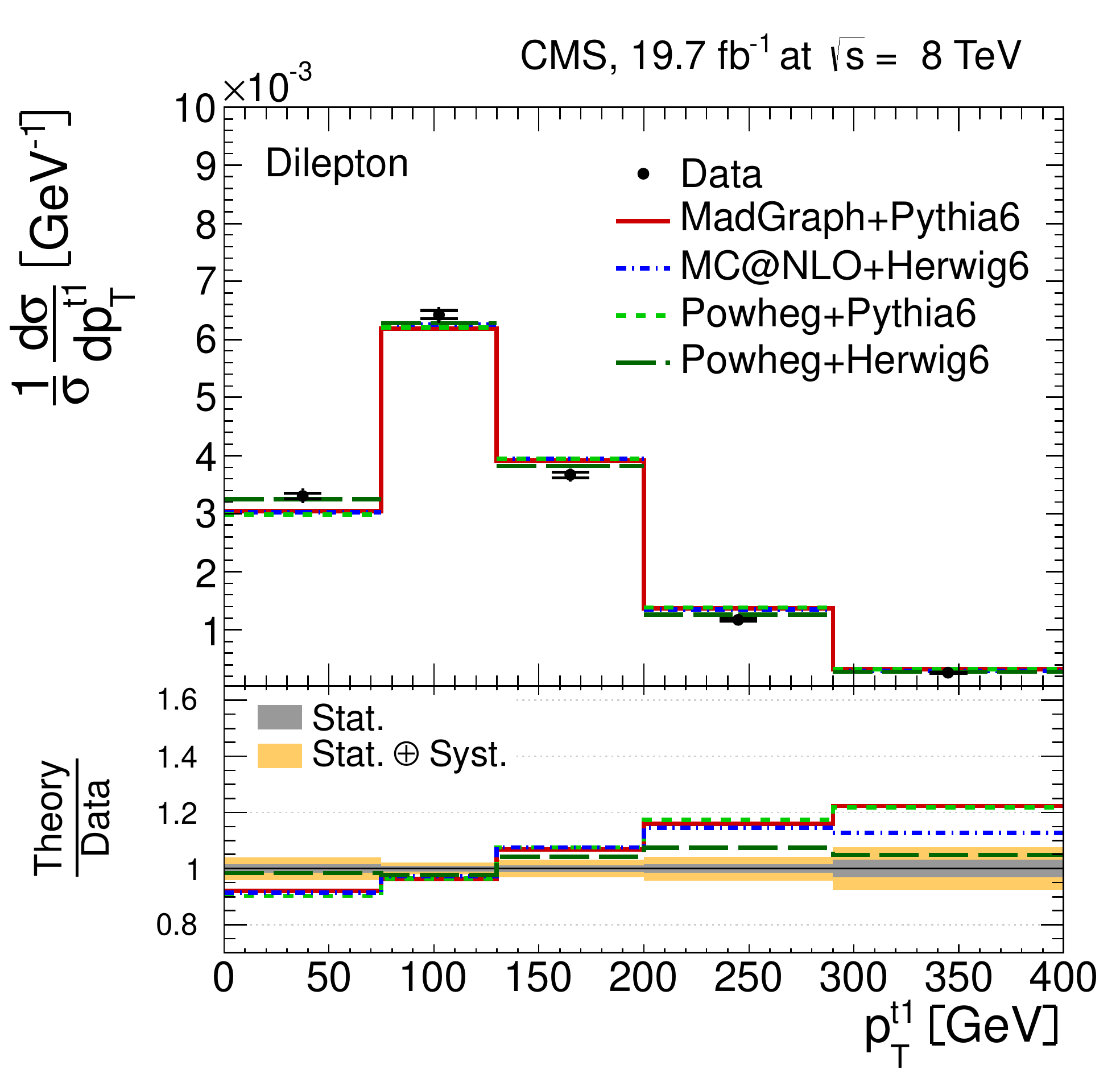}
       \includegraphics[width=0.48\textwidth]{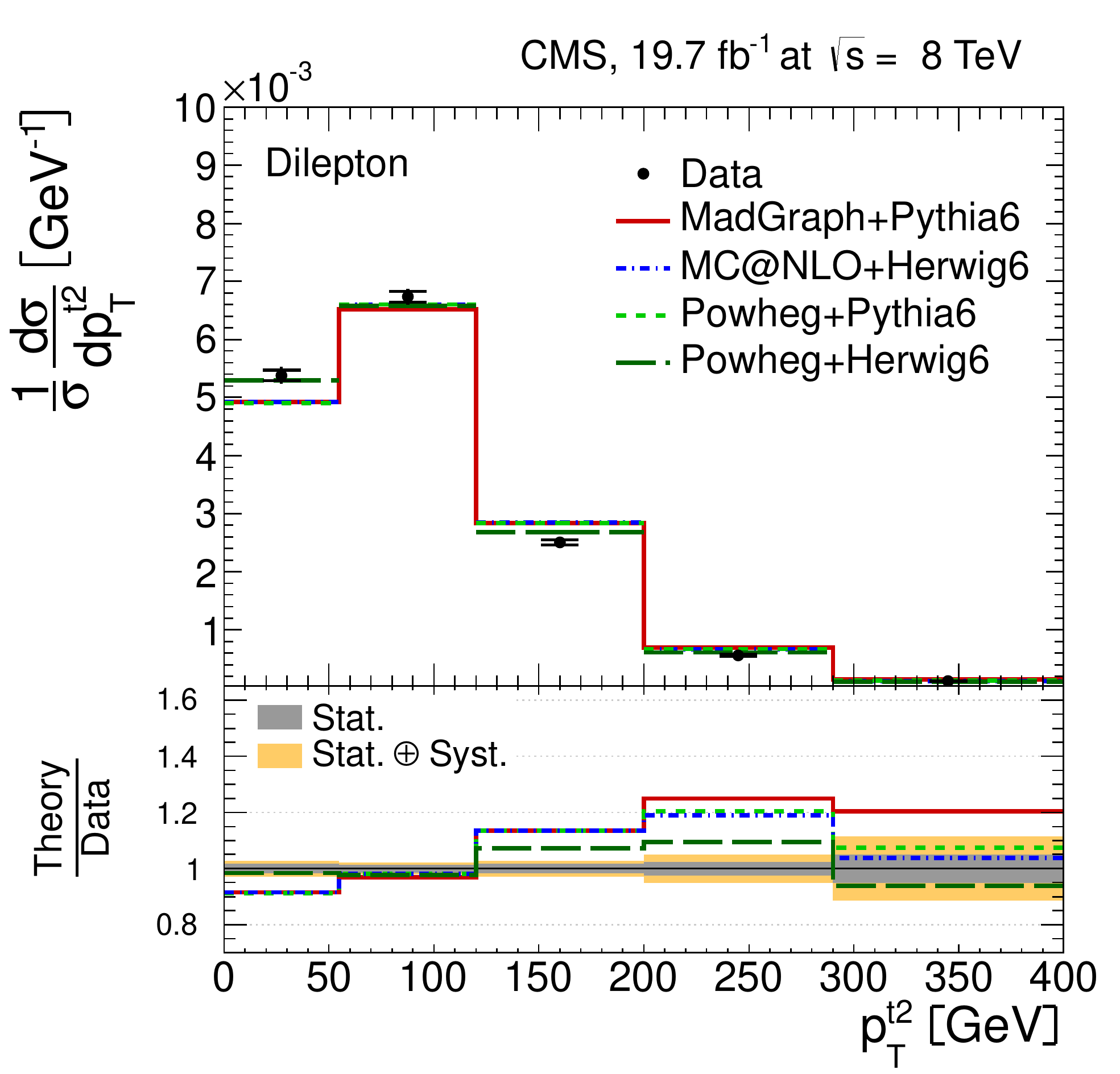}
    \caption{Normalized differential \ttbar production cross section in the dilepton channels as a function of the \pt of the leading (left) and trailing (right) top quarks or antiquarks. The data points are placed at the midpoint of the bins. The inner (outer) error bars indicate the statistical (combined statistical and systematic) uncertainties. The measurements are compared to predictions from \MADGRAPH{}+\PYTHIA{6}, \POWHEG{}+\PYTHIA{6}, \POWHEG{}+\HERWIG{6}, and \MCATNLO{}+\HERWIG{6}. The lower part of each plot shows the ratio of the predictions to data.}
\label{fig:diffXSec:top2:dilepton}

\end{figure*}

\begin{figure*}[!htbp]
  \centering
	\includegraphics[width=0.48\textwidth]{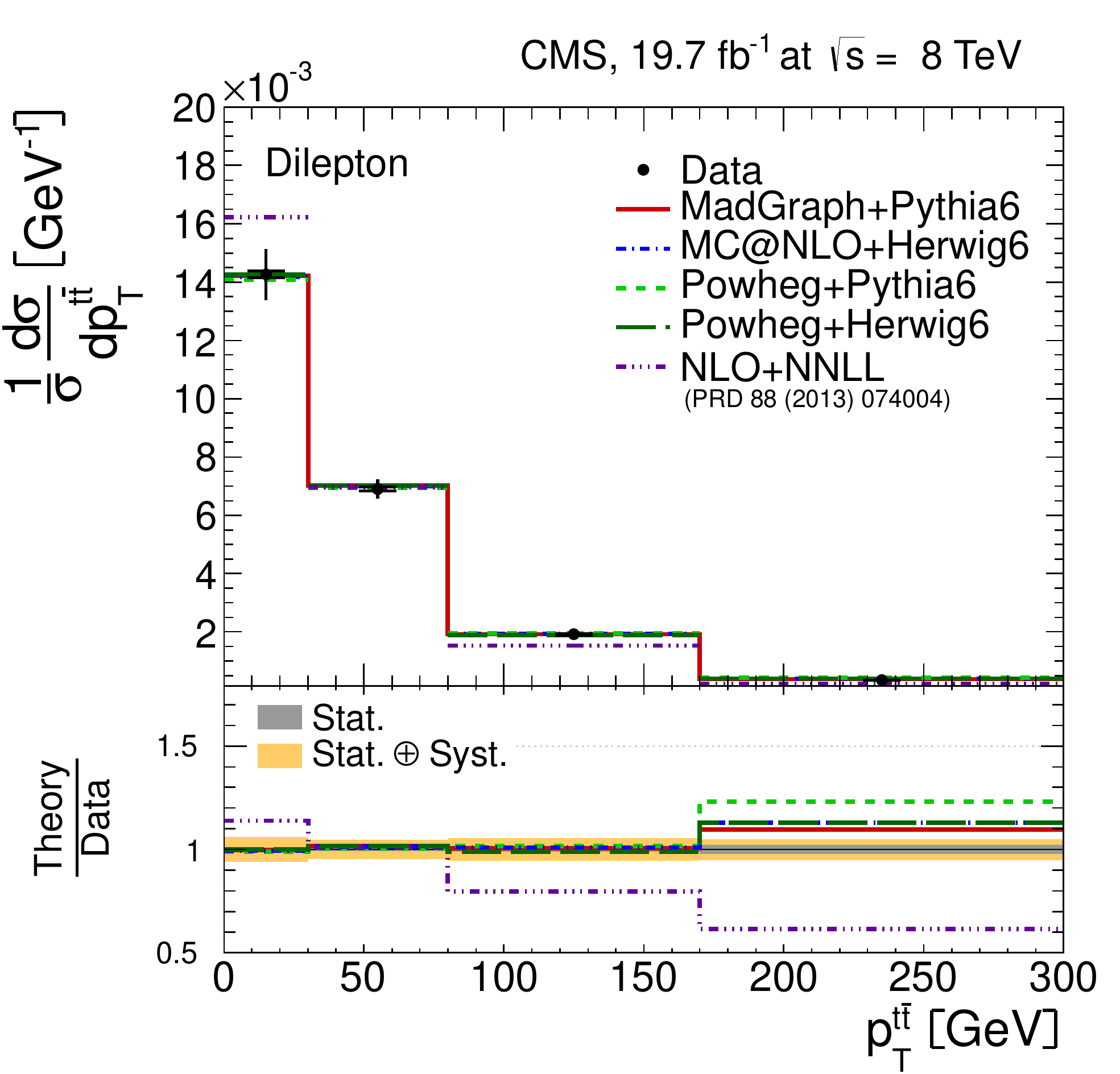}
	\includegraphics[width=0.48\textwidth]{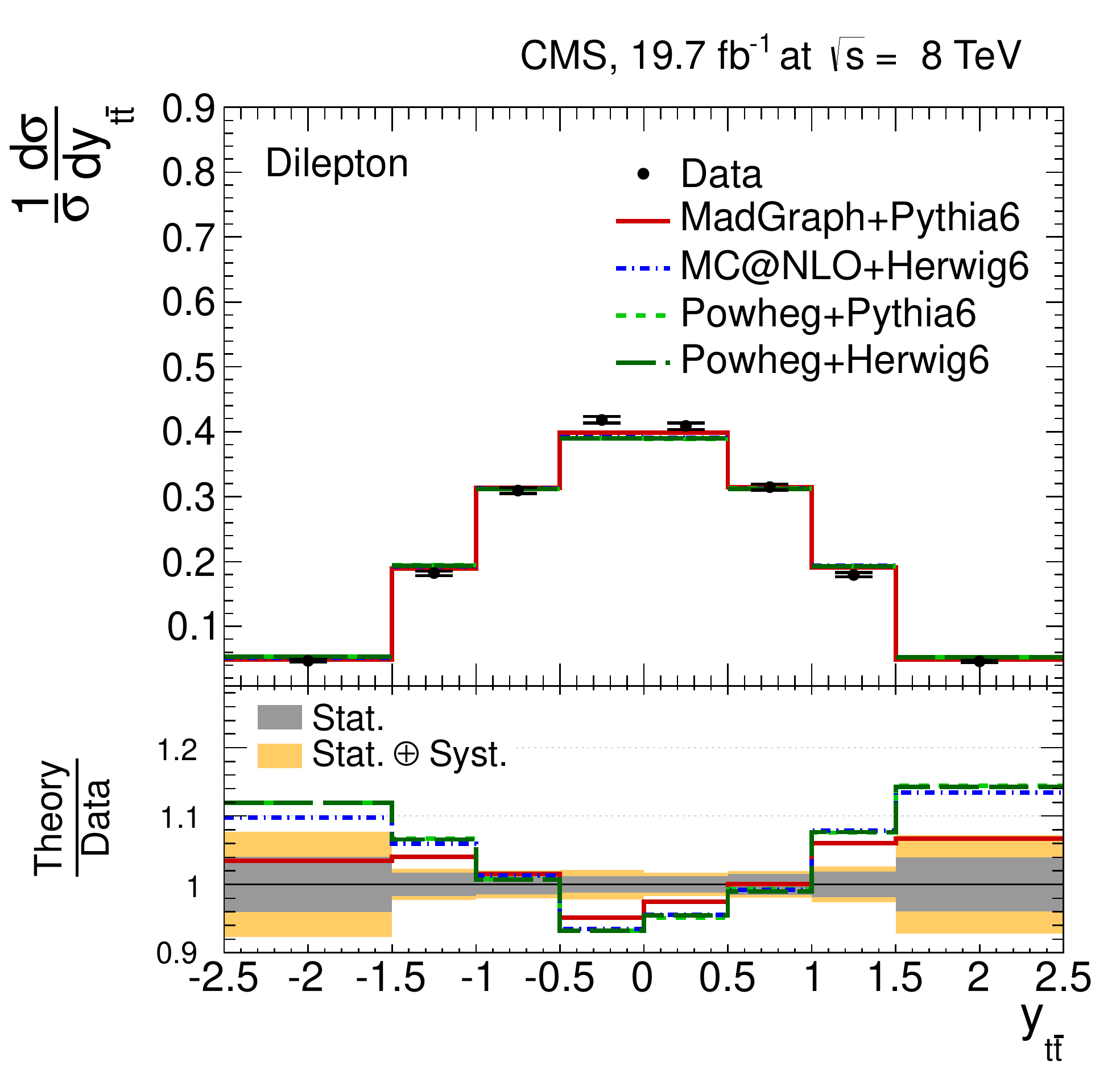}
	\includegraphics[width=0.48\textwidth]{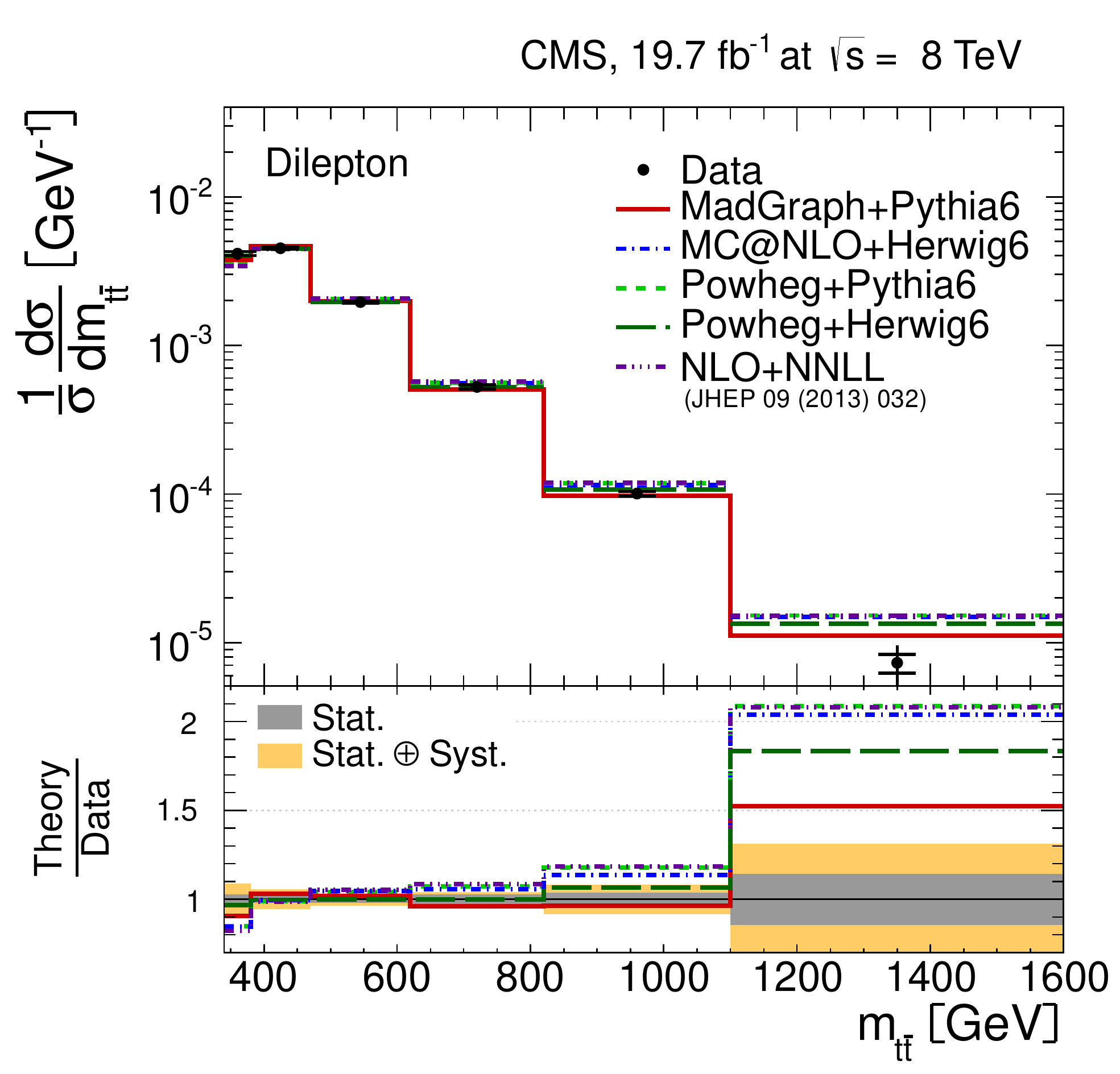}
    \caption{Normalized differential \ttbar production cross section in the dilepton channels as a function of the $\pt^{\ttbar}$ (top left), $y_{\ttbar}$ (top right), and $m_{\ttbar}$ (bottom) of the \ttbar system. The data points are placed at the midpoint of the bins. The inner (outer) error bars indicate the statistical (combined statistical and systematic) uncertainties. The measurements are compared to predictions from \MADGRAPH{}+\PYTHIA{6}, \POWHEG{}+\PYTHIA{6}, \POWHEG{}+\HERWIG{6}, \MCATNLO{}+\HERWIG{6}, and to NLO+NNLL~\cite{bib:ahrens_mttbar,bib:ahrens_ptttbar} calculations, when available. The lower part of each plot shows the ratio of the predictions to data.}
    \label{fig:diffXSec:tt:dilepton}

\end{figure*}

{\tolerance=400
In Figs.~\ref{fig:diffXSec:top:ljets} to~\ref{fig:diffXSec:tt:ljets}, the following distributions are presented for the \ljets channels: the transverse momentum $\pt^{\PQt}$ and the rapidity $y_{\PQt}$ of the top quarks or antiquarks, the transverse momentum $\pt^{\PQt\ast}$ of the top quarks or antiquarks in the \ttbar rest frame, the difference in the azimuthal angle between the top quark and antiquark $\Delta \phi(\text{t,}\bar{\PQt})$, the transverse momentum of the leading ($\pt^{\text{t1}}$) and trailing ($\pt^{\text{t2}}$) top quark or antiquark, and the transverse momentum $\pt^{\ttbar}$, the rapidity $y_{\ttbar}$, and the invariant mass $m_{\ttbar}$ of the \ttbar system. The data are compared to predictions from \MADGRAPH{}+\PYTHIA{6}, \POWHEG{}+\PYTHIA{6}, \POWHEG{}+\HERWIG{6}, and \MCATNLO{}+\HERWIG{6}. In addition, the approximate NNLO calculation~\cite{bib:kidonakis_8TeV} is also shown for the top quark \pt and rapidity results, while the $m_{\ttbar}$ and the $\pt^{\ttbar}$ distributions are compared to the NLO+NNLL predictions from Refs.~\cite{bib:ahrens_mttbar} and~\cite{bib:ahrens_ptttbar}, respectively. Figures~\ref{fig:diffXSec:top:dilepton}--\ref{fig:diffXSec:tt:dilepton} show the corresponding distributions in the dilepton channels. The lower panel in each plot also shows the ratio of each prediction relative to data.

In general, the \POWHEG{}+\HERWIG{6} prediction provides a good description of data for all measured distributions. The shape of the top quark \pt spectrum is softer in data than in the predictions from \MADGRAPH{}+\PYTHIA{6}, \POWHEG{}+\PYTHIA{6}, and \MCATNLO{}+\HERWIG{6} in both channels. The data lie above the predictions for $\pt^{\PQt} < 60$ (65)\GeV in the \ljets (dilepton) channels, while they lie below for $\pt^{\PQt} > 200\GeV$. This effect was also observed at 7\TeV~\cite{bib:TOP-11-013_paper}. The disagreement between data and predictions in the tail of the distributions is also observed in a measurement by the ATLAS Collaboration~\cite{bib:ATLASnew}. In contrast, the prediction from \POWHEG{}+\HERWIG{6} and the approximate NNLO calculation provide a better description of the data, as they predict a slightly softer top quark \pt distribution than the three other simulations. The difference between the \POWHEG{}+\PYTHIA{6} and \POWHEG{}+\HERWIG{6} distributions is attributed to different treatment of the hardest initial state radiation in \PYTHIA{6} and \HERWIG{6}. The same pattern is observed for $\pt^{\PQt\ast}$, indicating that the softer spectrum in data is not caused by the boost of the \ttbar system. It is also present in the $\pt^{\text{t1}}$, and particularly, in the $\pt^{\text{t2}}$ distributions. For all these distributions, the \POWHEG{}+\HERWIG{6} prediction provides a better description of the data. The difference in the shape of the top quark \pt spectrum between data and simulation is observed consistently in the analyses using different event selection requirements or different pileup conditions. The $y_{t}$ distribution is found to be slightly less central in data than in the predictions, particularly in the case of \MADGRAPH{}+\PYTHIA{6} and the approximate NNLO calculation, which are more central than the other predictions. On the contrary, $y_{\ttbar}$ is more central in data, and it is slightly better described by \MADGRAPH{}+\PYTHIA{6}. The $m_{\ttbar}$ distribution in data tends to be lower than the predictions for large $m_{\ttbar}$ values, and is better described by \MADGRAPH{}+\PYTHIA{6} and \POWHEG{}+\HERWIG{6}. The $\pt^{\ttbar}$ spectrum is well described by all the considered predictions, except for the NLO+NNLL calculation, which fails to describe the data for all $\pt^{\ttbar}$ values.
\par}


\begin{figure*}[htbp]
  \centering
  \includegraphics[width=0.48\textwidth]{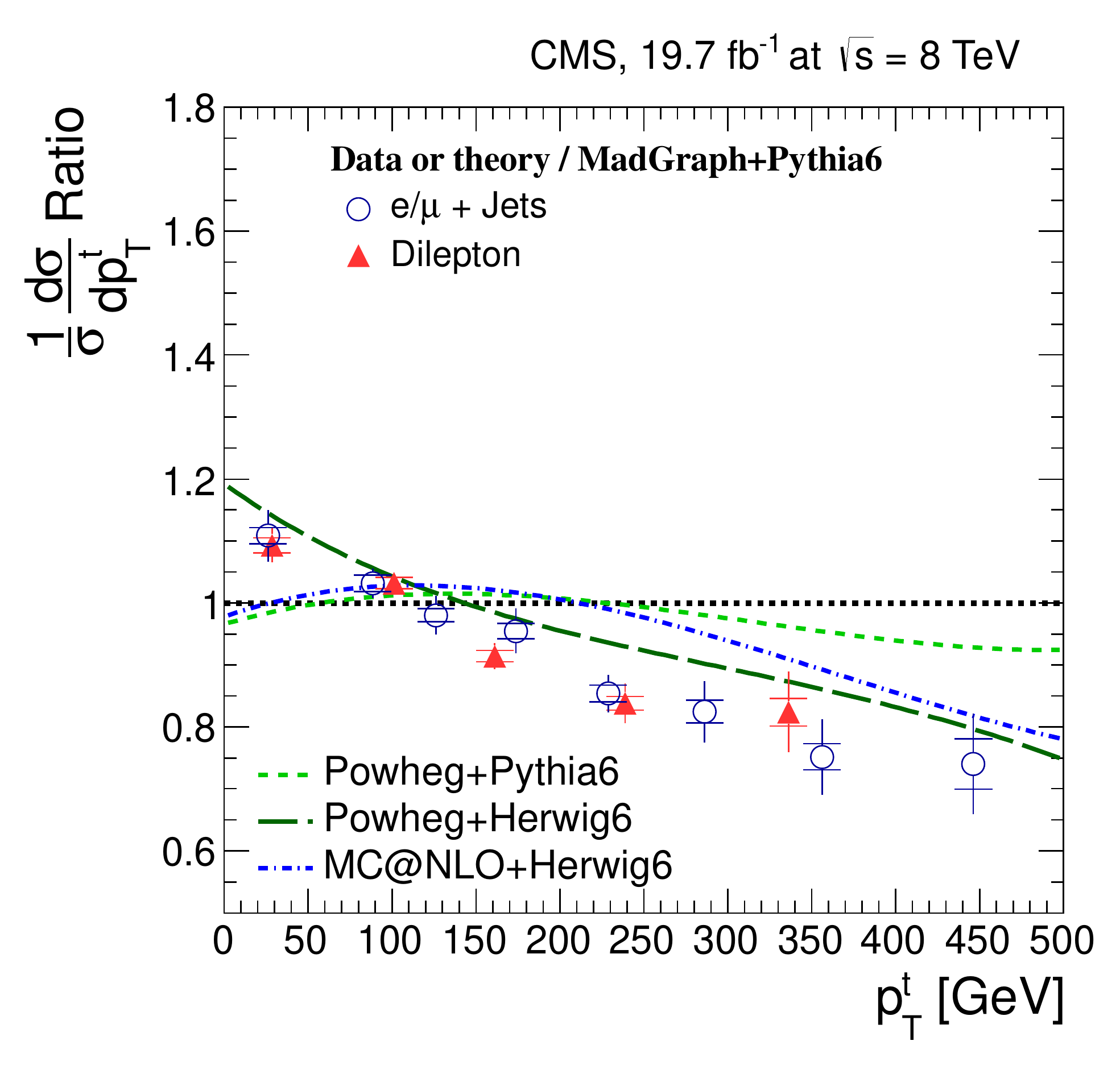} \includegraphics[width=0.48\textwidth]{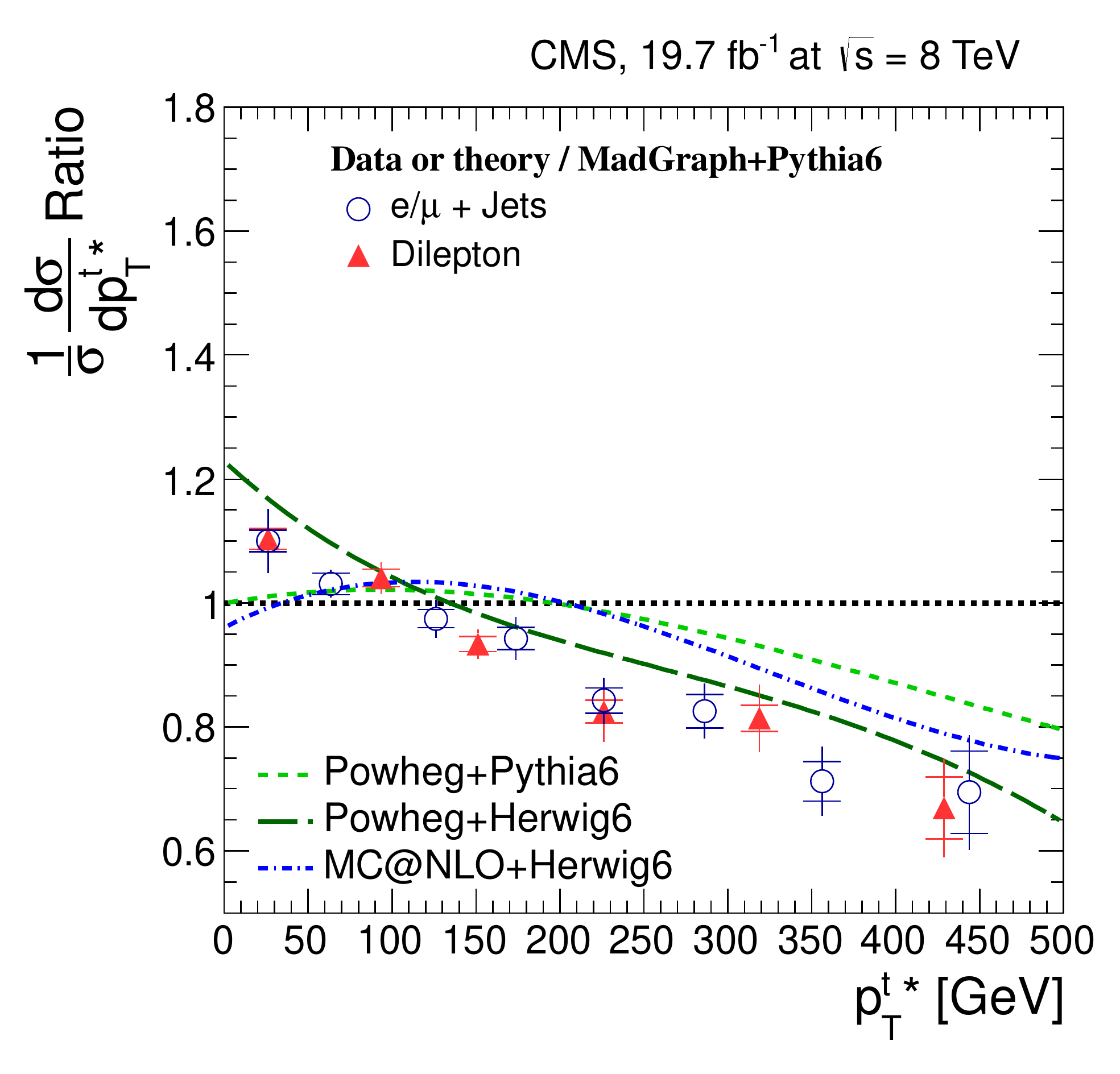}\\
  \includegraphics[width=0.48\textwidth]{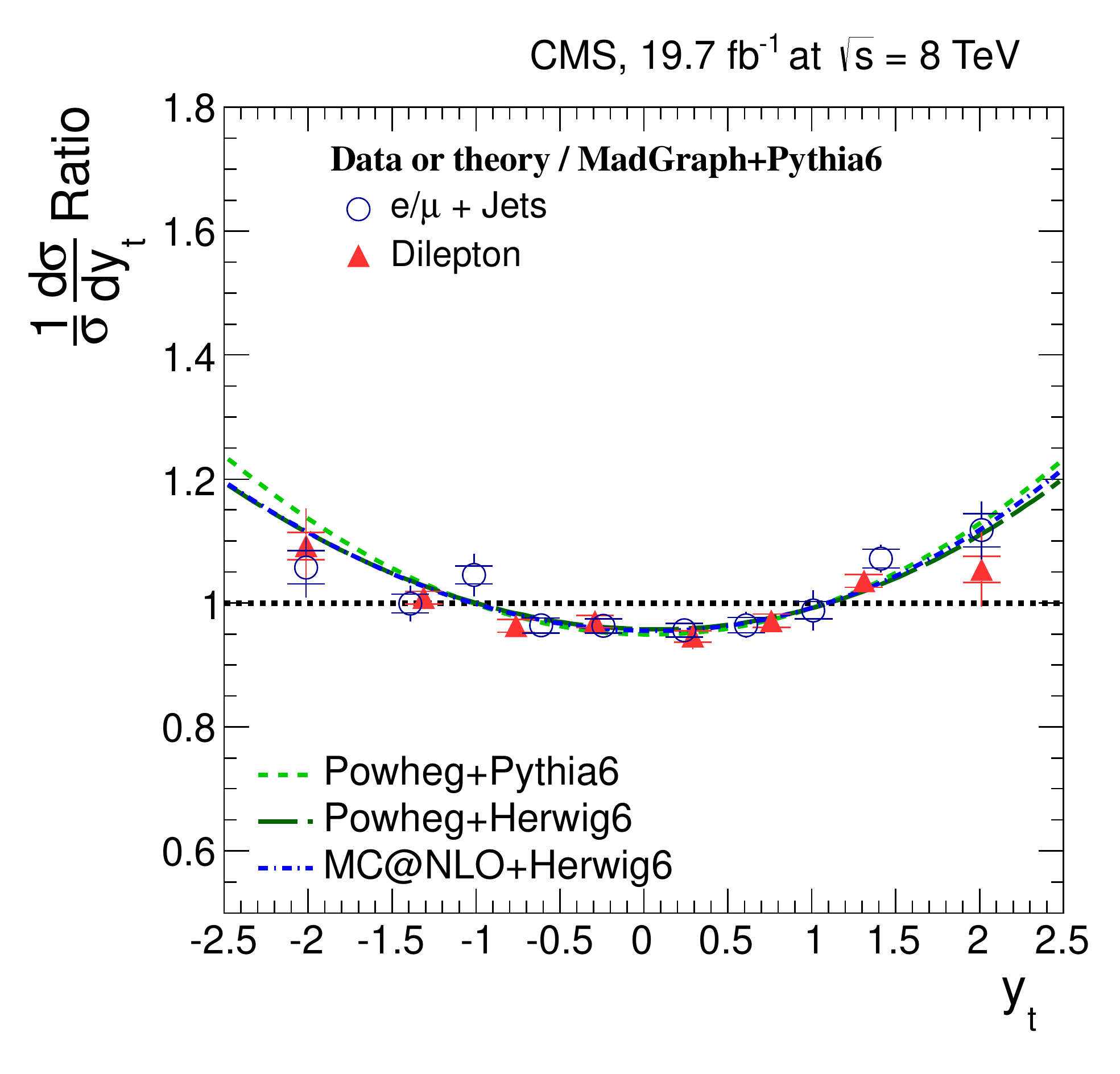}	\includegraphics[width=0.48\textwidth]{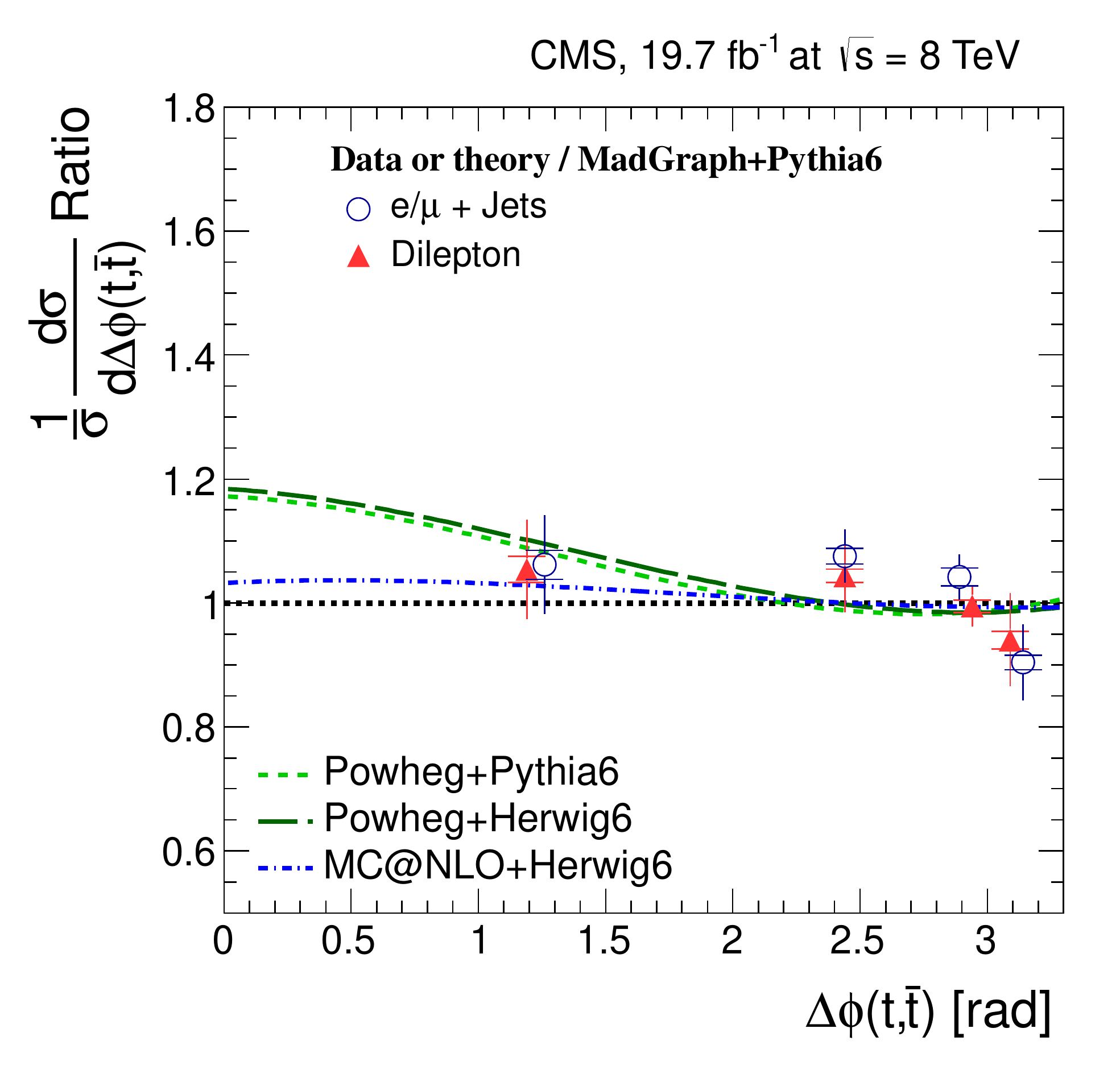}
    \caption{Comparison of normalized differential \ttbar production cross section in the dilepton and \ljets channels as a function of the $\pt^{\PQt}$ (top left), the \ttbar rest frame $\pt^{\PQt\ast}$ (top right), and the rapidity $y_{\PQt}$ (bottom left) of the top quarks or antiquarks, and the difference in the azimuthal angle between the top quark and the antiquark $\Delta \phi(\text{t,}\bar{\PQt})$ (bottom right). The measurements are presented relative to the \MADGRAPH{}+\PYTHIA{6} prediction. A horizontal bin-centre correction is applied to all data points (cf. Section~\ref{subsec:fullPS}). The inner (outer) error bars indicate the statistical (combined statistical and systematic) uncertainties. The predictions from \POWHEG{}+\PYTHIA{6}, \POWHEG{}+\HERWIG{6}, and \MCATNLO{}+\HERWIG{6}, also presented relative to \MADGRAPH{}+\PYTHIA{6}, are shown for comparison.}
    \label{fig:diffXSec:top:8TeV}

\end{figure*}

\begin{figure*}[htbp]
  \centering
       \includegraphics[width=0.48\textwidth]{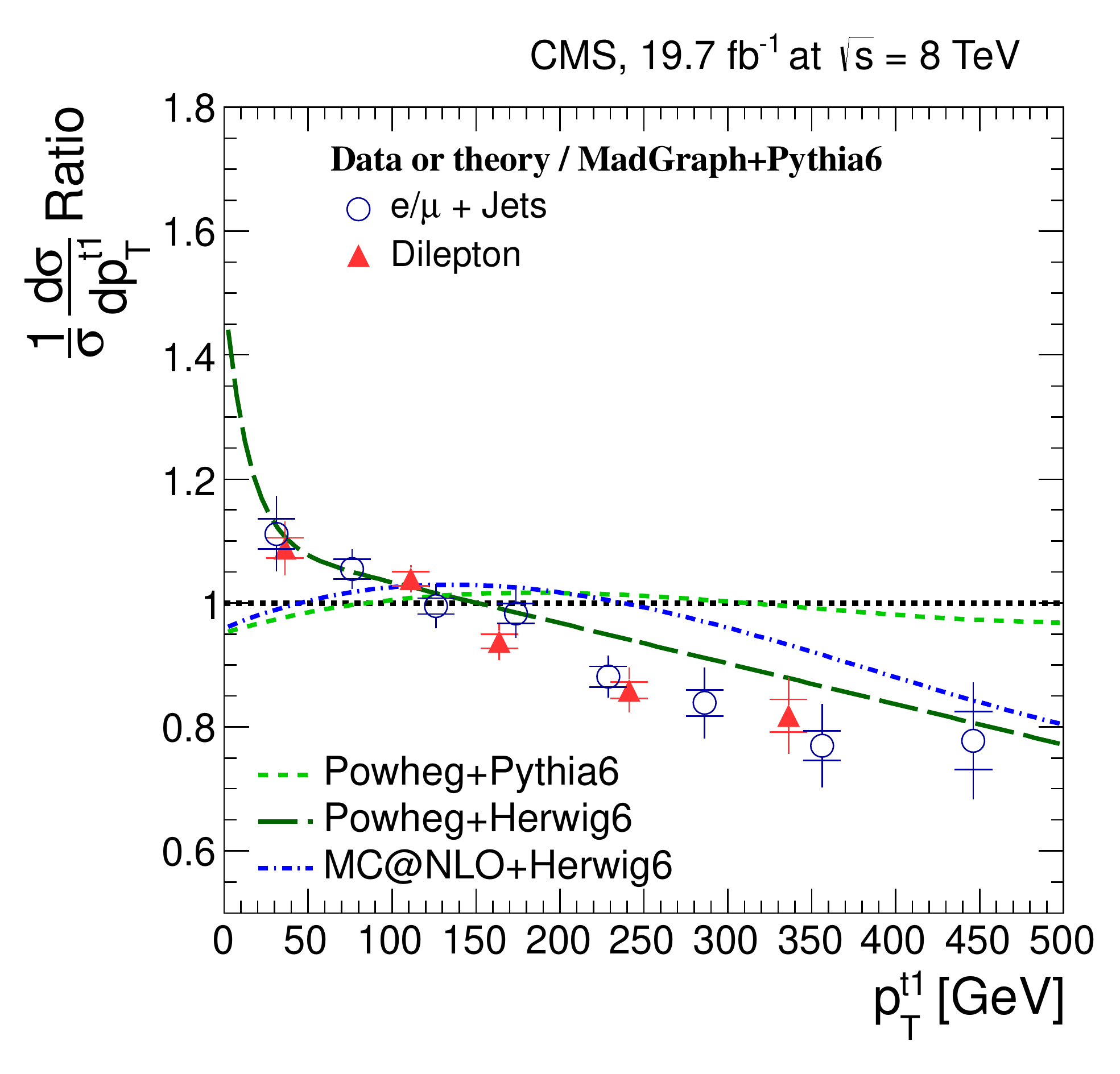}
       \includegraphics[width=0.48\textwidth]{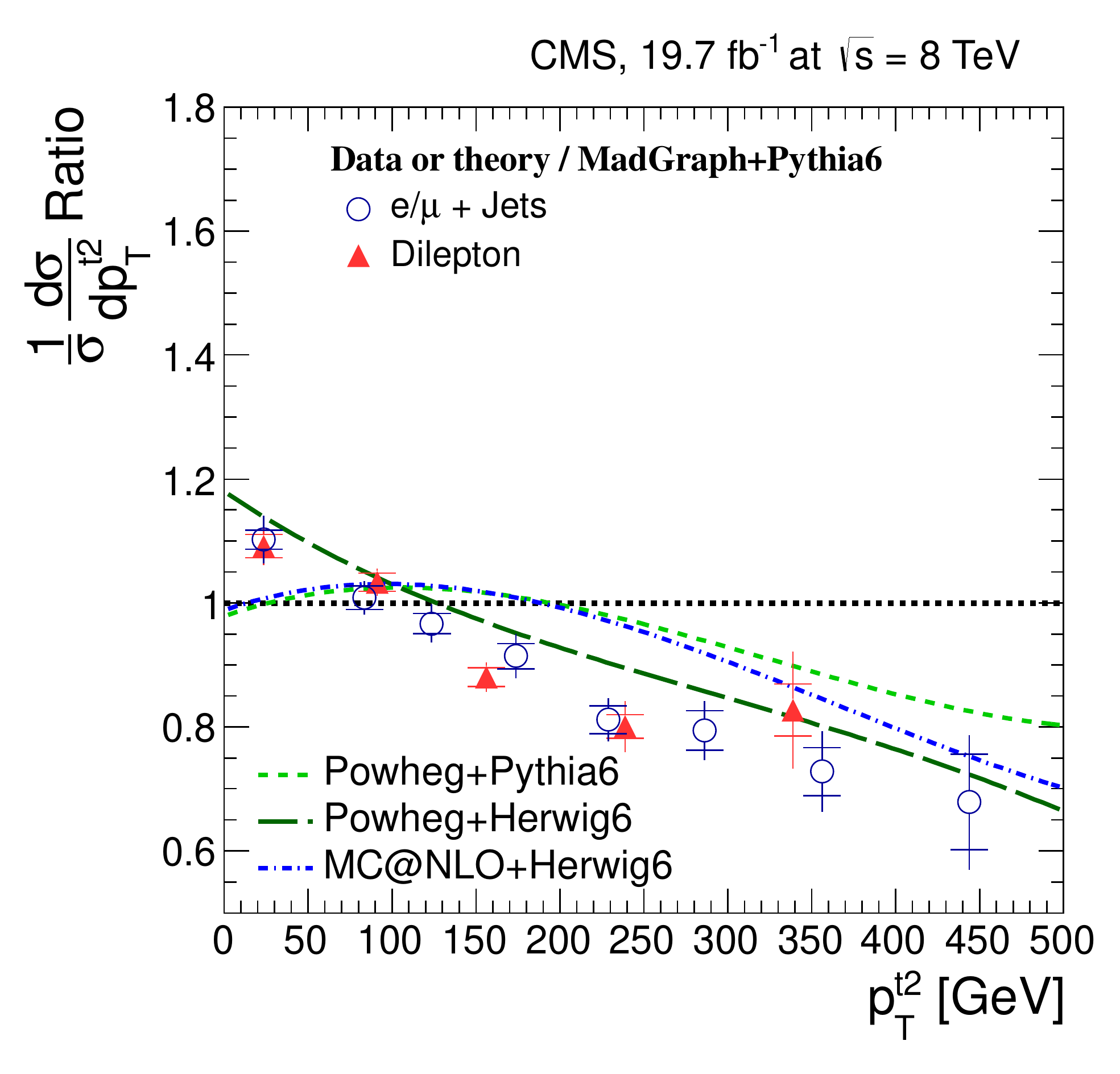}
    \caption{Comparison of normalized differential \ttbar production cross section in the dilepton and \ljets channels as a function of the \pt of the leading (left) and trailing (right) top quarks or antiquarks. The measurements are presented relative to the \MADGRAPH{}+\PYTHIA{6} prediction. A horizontal bin-centre correction is applied to all data points (cf. Section~\ref{subsec:fullPS}). The inner (outer) error bars indicate the statistical (combined statistical and systematic) uncertainties. The predictions from \POWHEG{}+\PYTHIA{6}, \POWHEG{}+\HERWIG{6}, and \MCATNLO{}+\HERWIG{6}, also presented relative to \MADGRAPH{}+\PYTHIA{6}, are shown for comparison.}
\label{fig:diffXSec:top2:8TeV}

\end{figure*}

\begin{figure*}[htbp]
  \centering
	\includegraphics[width=0.48\textwidth]{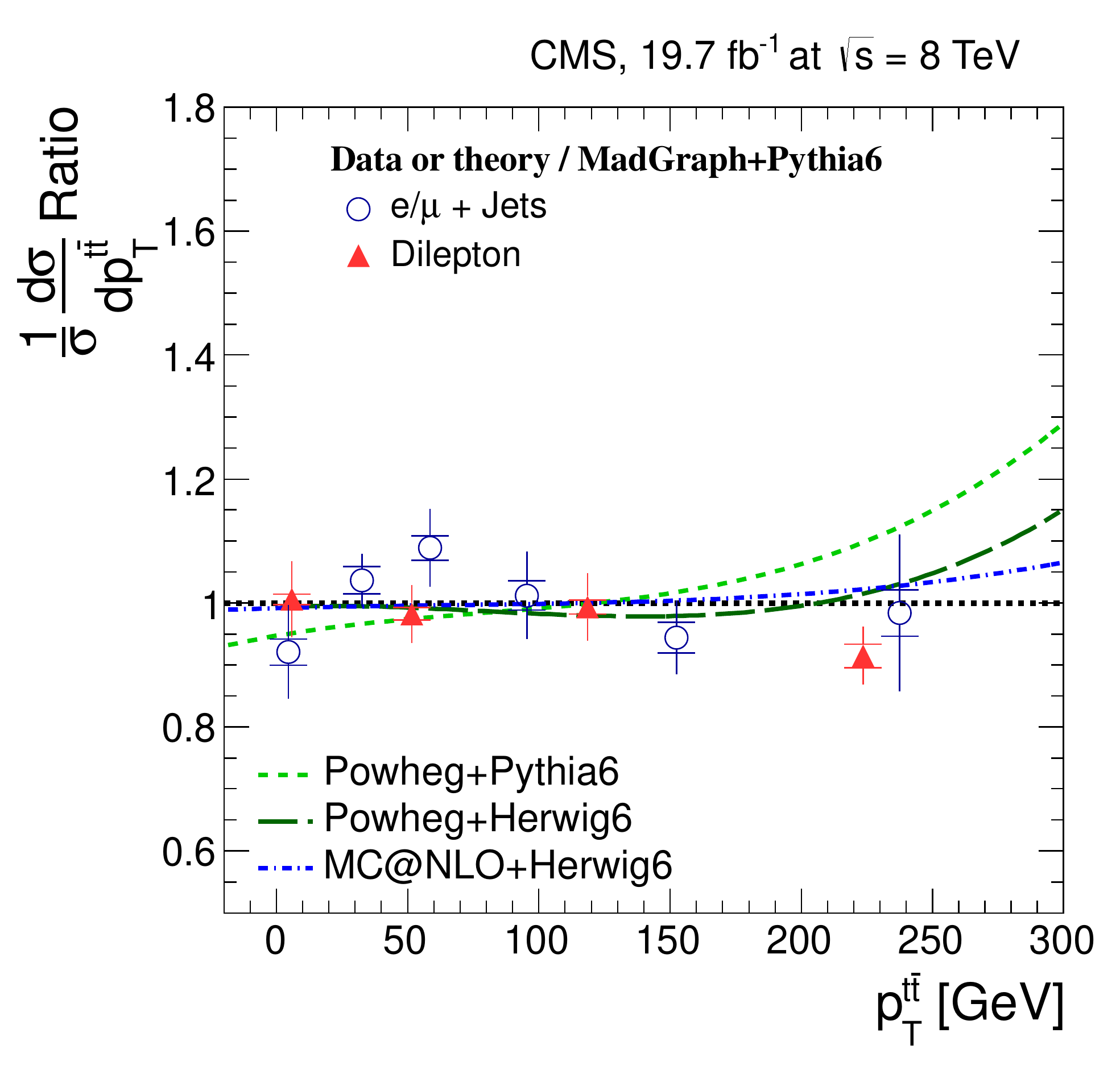}
	\includegraphics[width=0.48\textwidth]{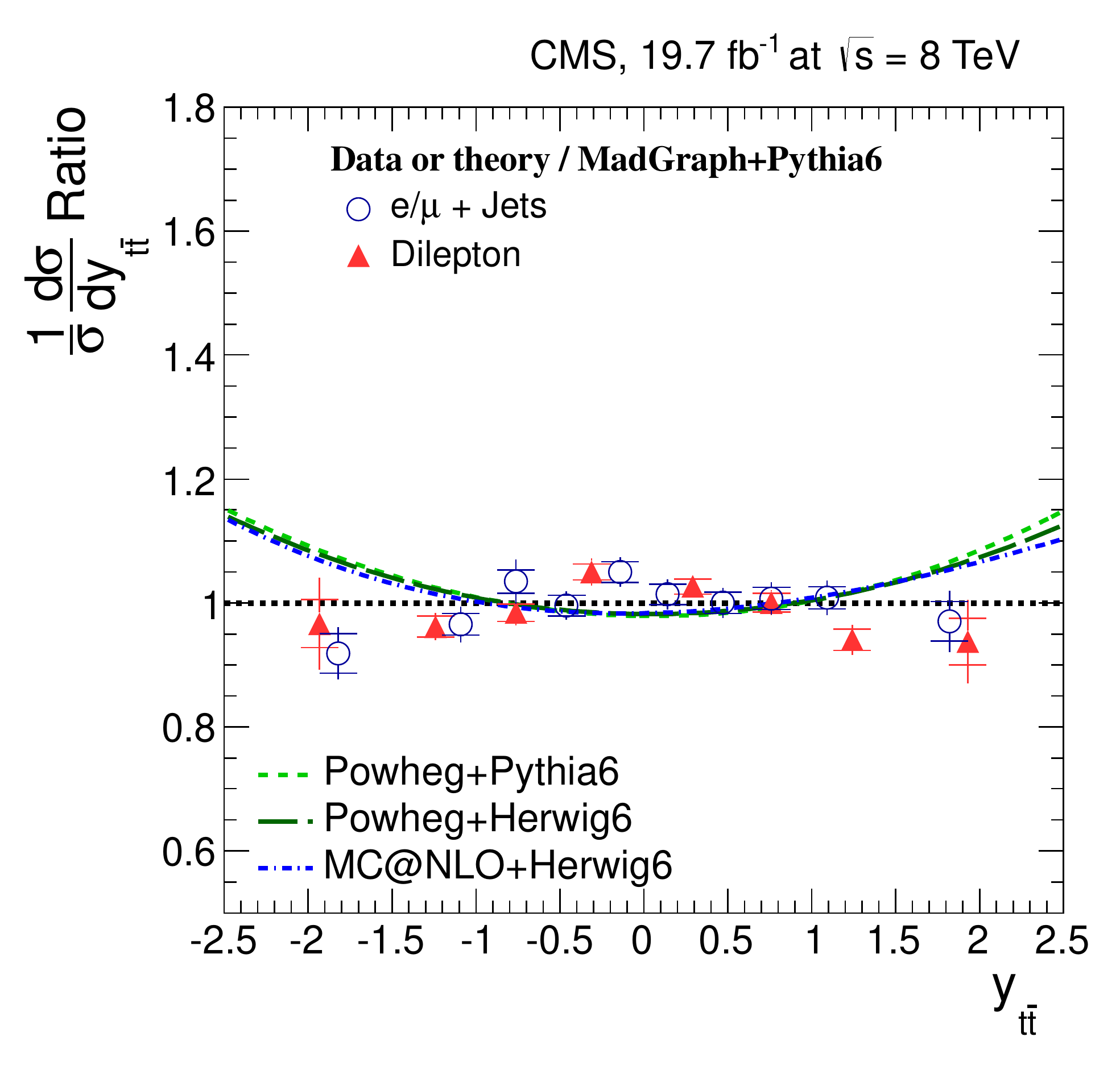}
	\includegraphics[width=0.48\textwidth]{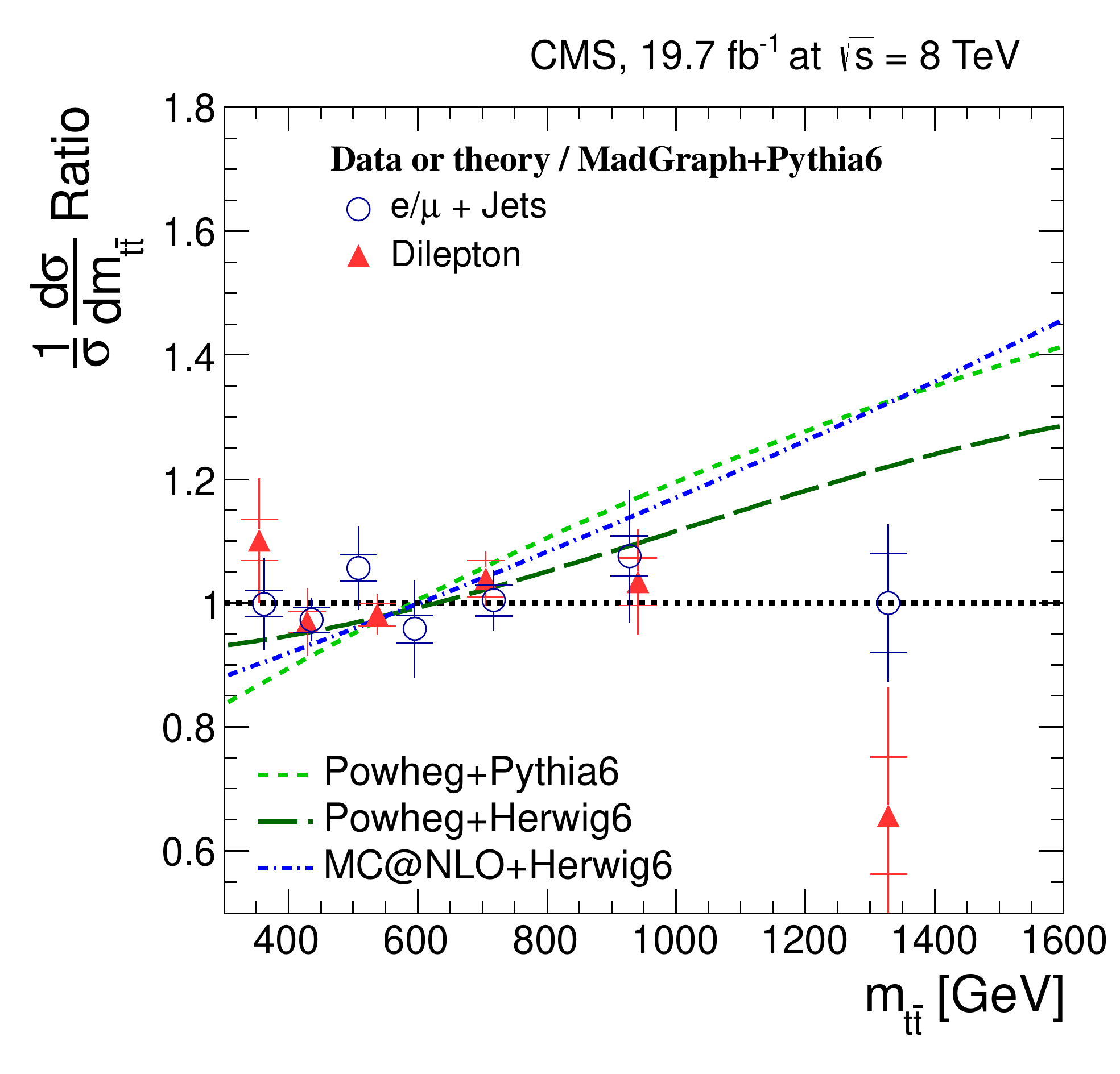}
    \caption{Comparison of normalized differential \ttbar production cross section in the dilepton and \ljets channels as a function of the $\pt^{\ttbar}$ (top left), $y_{\ttbar}$ (top right), and $m_{\ttbar}$ (bottom) of the \ttbar system. The measurements are presented relative to the \MADGRAPH{}+\PYTHIA{6} prediction. A horizontal bin-centre correction is applied to all data points (cf. Section~\ref{subsec:fullPS}). The inner (outer) error bars indicate the statistical (combined statistical and systematic) uncertainties. The predictions from \POWHEG{}+\PYTHIA{6}, \POWHEG{}+\HERWIG{6}, and \MCATNLO{}+\HERWIG{6}, also presented relative to \MADGRAPH{}+\PYTHIA{6}, are shown for comparison. For better visibility, data points with identical bin centres (cf. \supplemental Tables~\suppRef{\ref{tab:ljets:SummaryResultsDiffXSecSemileptonTopFullPS_1} and~\ref{tab:dilepton:SummaryResultsDiffXSecDileptonTopFullPS_1}}{6 and 10}) are shifted horizontally by a negligible amount.}
    \label{fig:diffXSec:tt:8TeV}

\end{figure*}

The results from the \ljets and dilepton channels are compared to each other in Figs.~\ref{fig:diffXSec:top:8TeV} to~\ref{fig:diffXSec:tt:8TeV}. This is only feasible for the top quark and \ttbar quantities, since they are measured in the same phase space (\ie the full parton level phase space) for both channels. The results are presented relative to the \MADGRAPH{}+\PYTHIA{6} prediction to highlight the level of agreement between data and the default \ttbar simulation. To facilitate the comparison of measurements that are performed using different size and number of bins, a horizontal bin-centre correction is applied to all data points from both channels. In each bin, the measured data points are presented at the horizontal position in the bin where the predicted bin-averaged cross section equals the cross section of the unbinned \MADGRAPH{}+\PYTHIA{6} calculation (cf.~\cite{bib:BCCs}), which is common for both channels. The data are also compared to the predictions from \POWHEG{}+\PYTHIA{6}, \POWHEG{}+\HERWIG{6}, \MCATNLO{}+\HERWIG{6} relative to \MADGRAPH{}+\PYTHIA{6}. The results are consistent between the channels for all quantities, in particular, for all measurements related to the top quark \pt distribution. The softer spectrum in data relative to \MADGRAPH{}+\PYTHIA{6} is clearly visible.

\begin{figure*}[htbp]
  \centering
        \includegraphics[width=0.48\textwidth]{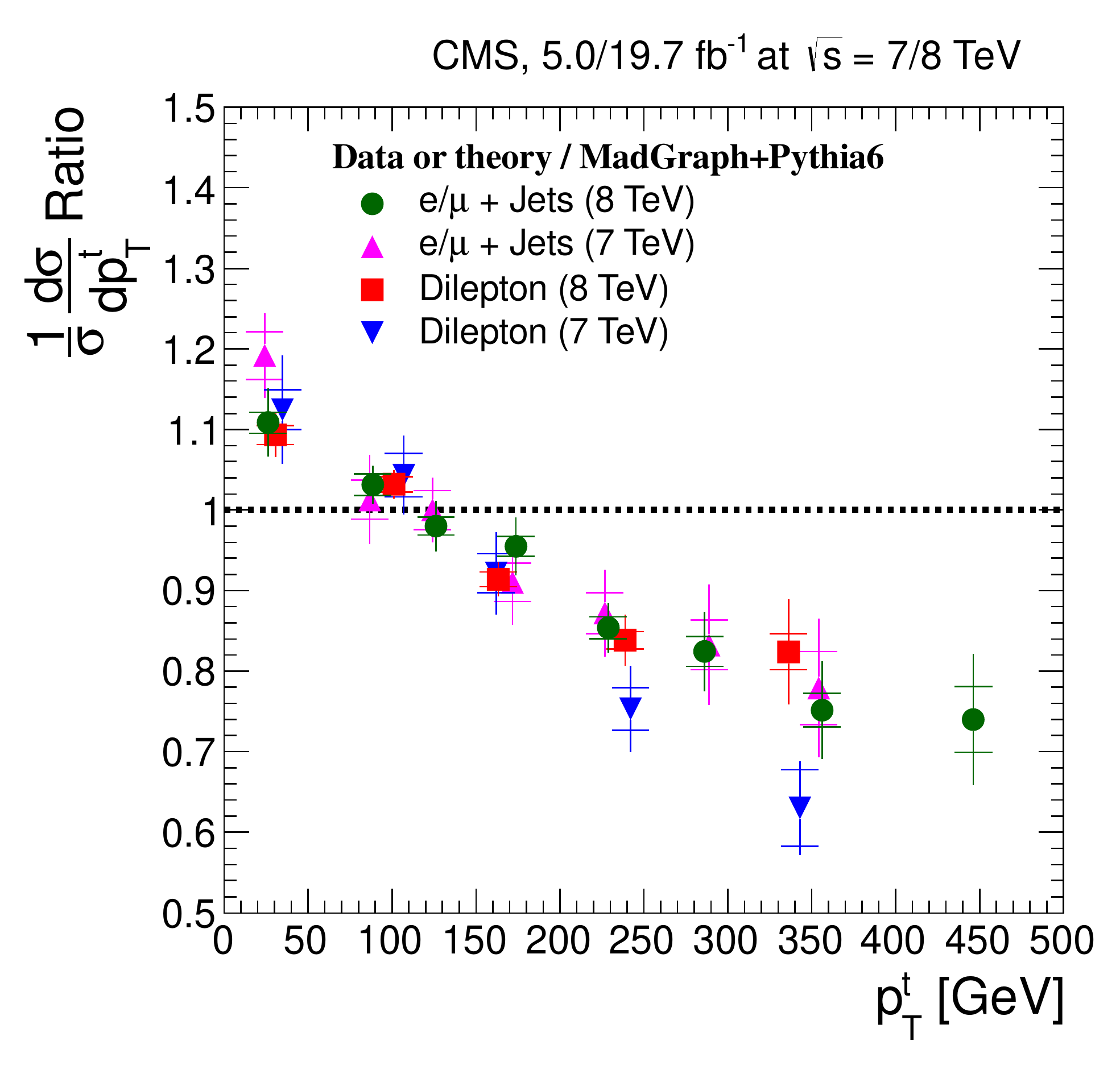}
	\includegraphics[width=0.48\textwidth]{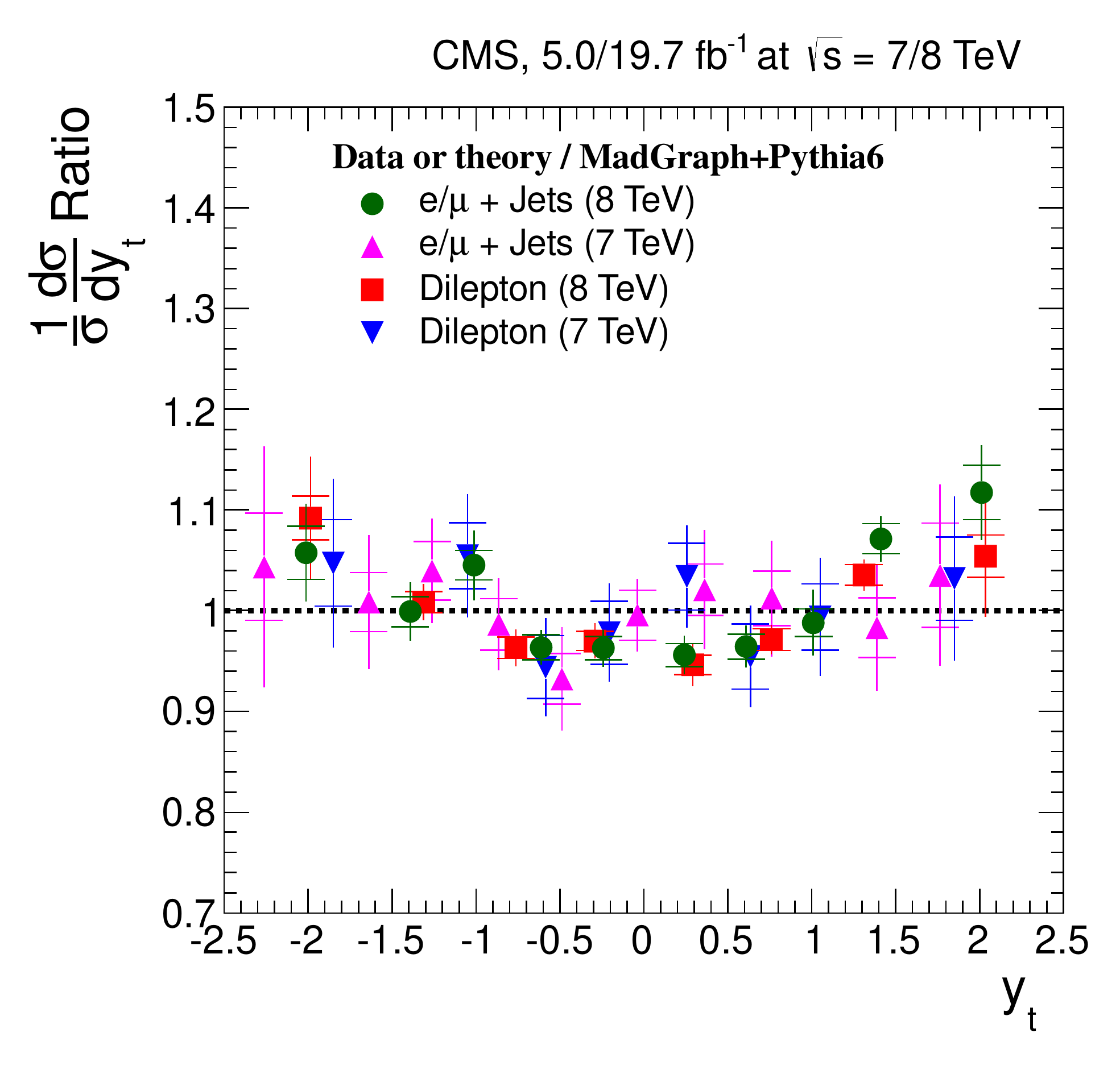}
    \caption{Comparison of normalized differential \ttbar production cross section in the dilepton and \ljets channels at 7\TeV~\cite{bib:TOP-11-013_paper} and 8\TeV, as a function of the $\pt^{\PQt}$ (left) and rapidity $y_{\PQt}$ (right) of the top quarks or antiquarks. The measurements are presented relative to the corresponding \MADGRAPH{}+\PYTHIA{6} predictions. A horizontal bin-centre correction is applied to all data points (cf. Section~\ref{subsec:fullPS}). The inner (outer) error bars indicate the statistical (combined statistical and systematic) uncertainties. For better visibility, data points with identical bin centres (cf. \supplemental Tables~\suppRef{\ref{tab:ljets:SummaryResultsDiffXSecSemileptonTopFullPS_1} and~\ref{tab:dilepton:SummaryResultsDiffXSecDileptonTopFullPS_1}}{6 and 10}) are shifted horizontally by a negligible amount.}
    \label{fig:diffXSec:top:78TeV}

\end{figure*}

\begin{figure*}[htbp]
  \centering
	\includegraphics[width=0.48\textwidth]{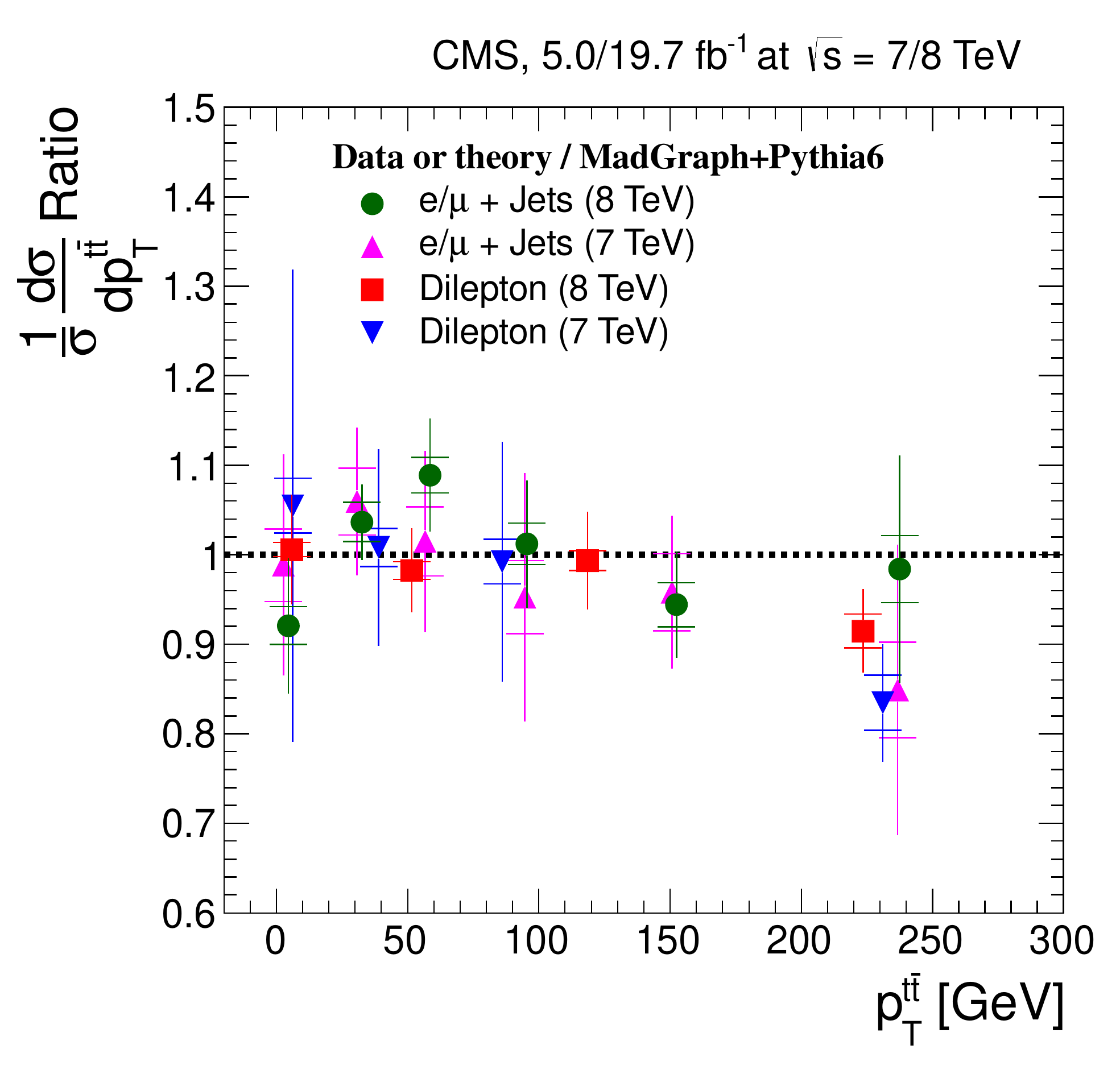}
	\includegraphics[width=0.48\textwidth]{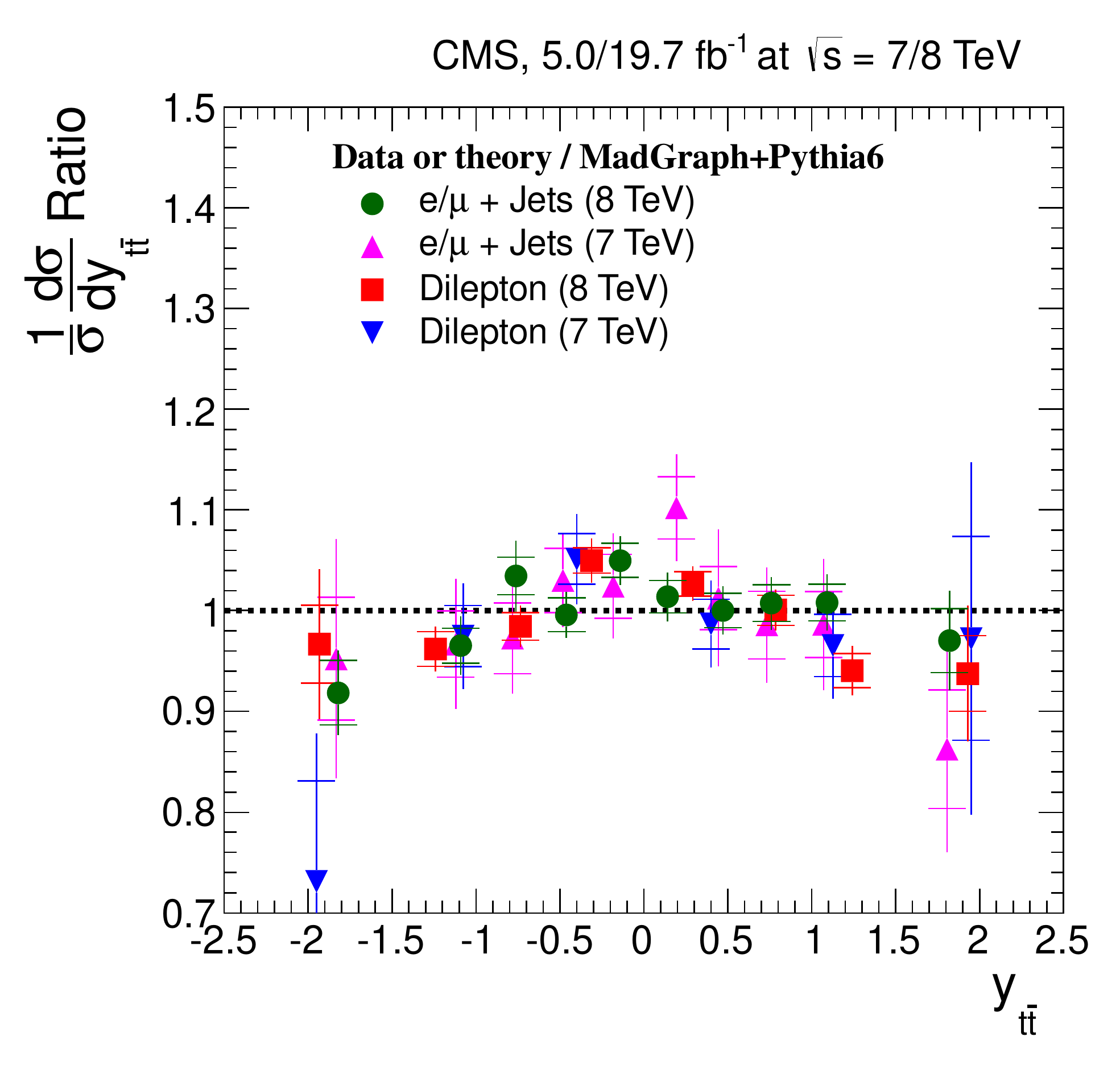}
	\includegraphics[width=0.48\textwidth]{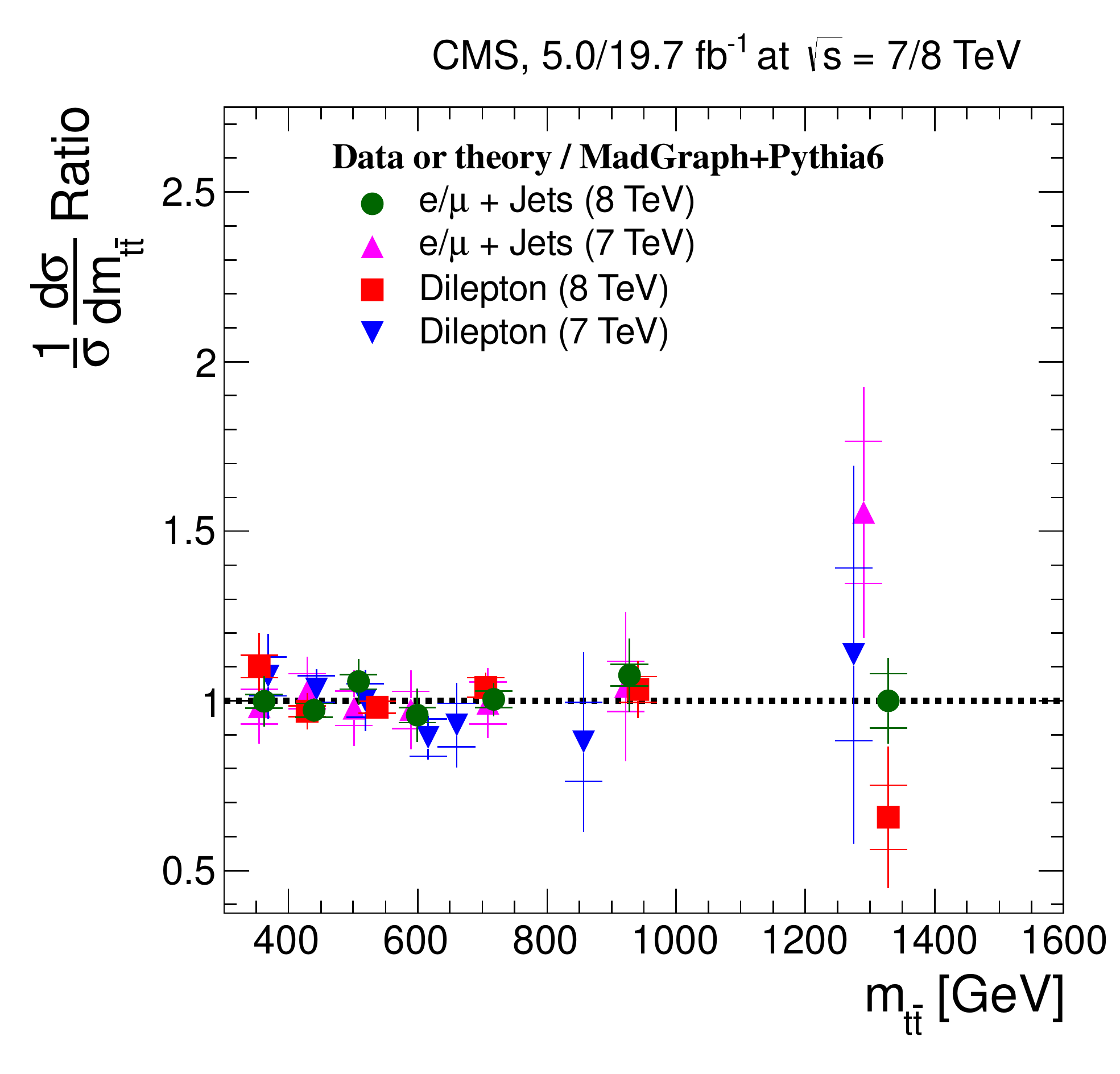}
    \caption{Comparison of normalized differential \ttbar production cross section in the dilepton and \ljets channels at 7\TeV~\cite{bib:TOP-11-013_paper} and 8\TeV, as a function of the $\pt^{\ttbar}$ (top left), $y_{\ttbar}$ (top right), and $m_{\ttbar}$ (bottom) of the \ttbar system. The measurements are presented relative to the corresponding \MADGRAPH{}+\PYTHIA{6} predictions. A horizontal bin-centre correction is applied to all data points (cf. Section~\ref{subsec:fullPS}). The inner (outer) error bars indicate the statistical (combined statistical and systematic) uncertainties. For better visibility, data points with identical bin centres (cf. \supplemental Tables~\suppRef{\ref{tab:ljets:SummaryResultsDiffXSecSemileptonTTbarFullPS} and~\ref{tab:dilepton:SummaryResultsDiffXSecDileptonTTbarFullPS}}{9 and 12}) are shifted horizontally by a negligible amount.}
    \label{fig:diffXSec:tt:78TeV}

\end{figure*}

In addition, a comparison between results obtained at $\sqrt{s}=7$~\cite{bib:TOP-11-013_paper} and 8\TeV is also performed for both the \ljets and dilepton channels, and presented in Figs.~\ref{fig:diffXSec:top:78TeV} and~\ref{fig:diffXSec:tt:78TeV} for $\pt^{\PQt}$, $y_{\PQt}$, $\pt^{\ttbar}$, $y_{\ttbar}$, and $m_{\ttbar}$. Since the fiducial phase space definition for the normalized differential cross sections is also different for each value of $\sqrt{s}$, the comparison is again only possible for top quark and \ttbar quantities. The measurements are presented relative to the corresponding default \MADGRAPH{}+\PYTHIA{6} predictions at 7 and 8\TeV. A horizontal bin-centre correction with respect to the \MADGRAPH{}+\PYTHIA{6} predictions is applied to all data points from both channels and $\sqrt{s}$ values. The results are consistent between the channels for all quantities, both at 7 and 8\TeV. The uncertainties in almost all bins of the distributions are reduced for the 8\TeV results relative to 7\TeV, mainly due to the improvements discussed in Section~\ref{sec:kinfit}. The softer $\pt^{\PQt}$ in data relative to \MADGRAPH{}+\PYTHIA{6} is also visible at 7\TeV.



\section{Summary}
\label{sec:concl}

{\tolerance=400
First measurements are presented of normalized differential \ttbar production cross sections in pp~collisions at $\sqrt{s}=8\TeV$. The measurements are performed with the CMS detector in the \ljets ($\ell = \Pe \text{ or }\mu$) and dilepton (\ee, \mumu, and \emu) \ttbar decay channels. The normalized \ttbar cross section is measured as a function of the transverse momentum, rapidity, and invariant mass of the final-state leptons and b jets in the fiducial phase space, and the top quarks and \ttbar system in the full phase space. The measurements in the different decay channels are in agreement with each other. In general, the data are in agreement with standard model predictions up to approximate NNLO precision. Among the examined predictions, \POWHEG{}+\HERWIG{6} provides the best overall description of the data. However, the \pt spectrum in data for leptons, jets, and top quarks is softer than expected, particularly for \MADGRAPH{}+\PYTHIA{6}, \POWHEG{}+\PYTHIA{6}, and \MCATNLO{}+\HERWIG{6}. The calculation at approximate NNLO precision also provides a good description of the top quark \pt spectrum. The $m_{\ttbar}$ distribution in data tends to be lower than the predictions for large $m_{\ttbar}$ values. The $\pt^{\ttbar}$ spectrum is well described by all the considered predictions, except for the NLO+NNLL calculation, which fails to describe the data for all $\pt^{\ttbar}$ values. The results show the same behaviour as the corresponding CMS measurements at $\sqrt{s}=7\TeV$.
\par}

\begin{acknowledgments}
\hyphenation{Bundes-ministerium Forschungs-gemeinschaft Forschungs-zentren} We congratulate our colleagues in the CERN accelerator departments for the excellent performance of the LHC and thank the technical and administrative staffs at CERN and at other CMS institutes for their contributions to the success of the CMS effort. In addition, we gratefully acknowledge the computing centres and personnel of the Worldwide LHC Computing Grid for delivering so effectively the computing infrastructure essential to our analyses. Finally, we acknowledge the enduring support for the construction and operation of the LHC and the CMS detector provided by the following funding agencies: the Austrian Federal Ministry of Science, Research and Economy and the Austrian Science Fund; the Belgian Fonds de la Recherche Scientifique, and Fonds voor Wetenschappelijk Onderzoek; the Brazilian Funding Agencies (CNPq, CAPES, FAPERJ, and FAPESP); the Bulgarian Ministry of Education and Science; CERN; the Chinese Academy of Sciences, Ministry of Science and Technology, and National Natural Science Foundation of China; the Colombian Funding Agency (COLCIENCIAS); the Croatian Ministry of Science, Education and Sport, and the Croatian Science Foundation; the Research Promotion Foundation, Cyprus; the Ministry of Education and Research, Estonian Research Council via IUT23-4 and IUT23-6 and European Regional Development Fund, Estonia; the Academy of Finland, Finnish Ministry of Education and Culture, and Helsinki Institute of Physics; the Institut National de Physique Nucl\'eaire et de Physique des Particules~/~CNRS, and Commissariat \`a l'\'Energie Atomique et aux \'Energies Alternatives~/~CEA, France; the Bundesministerium f\"ur Bildung und Forschung, Deutsche Forschungsgemeinschaft, and Helmholtz-Gemeinschaft Deutscher Forschungszentren, Germany; the General Secretariat for Research and Technology, Greece; the National Scientific Research Foundation, and National Innovation Office, Hungary; the Department of Atomic Energy and the Department of Science and Technology, India; the Institute for Studies in Theoretical Physics and Mathematics, Iran; the Science Foundation, Ireland; the Istituto Nazionale di Fisica Nucleare, Italy; the Ministry of Science, ICT and Future Planning, and National Research Foundation (NRF), Republic of Korea; the Lithuanian Academy of Sciences; the Ministry of Education, and University of Malaya (Malaysia); the Mexican Funding Agencies (CINVESTAV, CONACYT, SEP, and UASLP-FAI); the Ministry of Business, Innovation and Employment, New Zealand; the Pakistan Atomic Energy Commission; the Ministry of Science and Higher Education and the National Science Centre, Poland; the Funda\c{c}\~ao para a Ci\^encia e a Tecnologia, Portugal; JINR, Dubna; the Ministry of Education and Science of the Russian Federation, the Federal Agency of Atomic Energy of the Russian Federation, Russian Academy of Sciences, and the Russian Foundation for Basic Research; the Ministry of Education, Science and Technological Development of Serbia; the Secretar\'{\i}a de Estado de Investigaci\'on, Desarrollo e Innovaci\'on and Programa Consolider-Ingenio 2010, Spain; the Swiss Funding Agencies (ETH Board, ETH Zurich, PSI, SNF, UniZH, Canton Zurich, and SER); the Ministry of Science and Technology, Taipei; the Thailand Center of Excellence in Physics, the Institute for the Promotion of Teaching Science and Technology of Thailand, Special Task Force for Activating Research and the National Science and Technology Development Agency of Thailand; the Scientific and Technical Research Council of Turkey, and Turkish Atomic Energy Authority; the National Academy of Sciences of Ukraine, and State Fund for Fundamental Researches, Ukraine; the Science and Technology Facilities Council, UK; the US Department of Energy, and the US National Science Foundation.

Individuals have received support from the Marie-Curie programme and the European Research Council and EPLANET (European Union); the Leventis Foundation; the A. P. Sloan Foundation; the Alexander von Humboldt Foundation; the Belgian Federal Science Policy Office; the Fonds pour la Formation \`a la Recherche dans l'Industrie et dans l'Agriculture (FRIA-Belgium); the Agentschap voor Innovatie door Wetenschap en Technologie (IWT-Belgium); the Ministry of Education, Youth and Sports (MEYS) of the Czech Republic; the Council of Science and Industrial Research, India; the HOMING PLUS programme of the Foundation for Polish Science, cofinanced from European Union, Regional Development Fund; the Compagnia di San Paolo (Torino); the Consorzio per la Fisica (Trieste); MIUR project 20108T4XTM (Italy); the Thalis and Aristeia programmes cofinanced by EU-ESF and the Greek NSRF; and the National Priorities Research Program by Qatar National Research Fund.
\end{acknowledgments}
\clearpage

\bibliography{auto_generated}
\ifthenelse{\boolean{cms@external}}{}{
\clearpage
\appendix
\numberwithin{table}{section}
\section{Values of the normalized differential cross sections}
\label{app:suppMat}
\newcommand{\invGeVns}{\ensuremath{\text{Ge\hspace{-.08em}V}^\text{$-$1}}\xspace}
In Tables~\ref{tab:ljets:SummaryResultsDiffXSecSemileptonLepB} to~\ref{tab:dilepton:SummaryResultsDiffXSecDileptonTTbarFullPS}, the results for the normalized differential distributions are summarized. For each distribution, the result of the measurement, together with the bin range, as well as the statistical, systematic, and total uncertainties, are provided. For the top quark or antiquark and \ttbar quantities, the bin centres corrected according to the \MADGRAPH{}+\PYTHIA{6} prediction (cf. Section~\suppRef{\ref{subsec:fullPS}}{6.2 (main document)}) are also presented.

\begin{table*}[htb]
  \centering
    \topcaption{Normalized differential \ttbar cross section in the \ljets channels as a function of the charged lepton transverse momentum ($\pt^{\ell}$) and pseudorapidity ($\eta_{\ell}$). The superscript `$\ell$' refers to both $\ell^{+}$ and $\ell^{-}$. The results are presented at particle level in the fiducial phase space. The statistical and systematic uncertainties are added in quadrature to yield the total uncertainty.}
    \label{tab:ljets:SummaryResultsDiffXSecSemileptonLepB}
    \renewcommand{\arraystretch}{1.3}
    \tabcolsep=0.09cm
    \begin{tabular}{c|c||c|c|c}
       \hline
      $\pt^{\ell}$ bin range [\GeVns{}]
                       & $1/\!\sigma\,\rd\sigma\!/\!\rd \pt^{\ell}$ [\invGeVns{}]
		       & Stat. [\%] & Syst. [\%] & Total [\%] \\
      \hline
     $[30,37)$   & $1.43 \times 10^{-2}$ & 2.2 & 2.4 &   3.2 \\
     $[37,45)$   & $2.21 \times 10^{-2}$ & 1.4 & 2.8 &   3.1 \\
     $[45,55)$   & $1.83 \times 10^{-2}$ & 1.4 & 1.7 &   2.3 \\
     $[55,68)$   & $1.36 \times 10^{-2}$ & 1.5 & 1.4 &   2.1 \\
     $[68,80)$   & $9.42 \times 10^{-3}$ & 2.0 & 1.5 &   2.5 \\
     $[80,100)$  & $5.89 \times 10^{-3}$ & 1.8 & 1.2 &   2.2 \\
     $[100,135)$ & $2.51 \times 10^{-3}$ & 2.1 & 6.3 &   6.7 \\
     $[135,200)$ & $5.73 \times 10^{-4}$& 3.3  & 9.5 &  10.1 \\
      \hline
      $\eta_{\ell}$ bin range
                & $1/\!\sigma\, \rd\sigma\!/\!\rd\eta_{\ell}$
		& Stat. [\%] & Syst. [\%] & Total [\%] \\
      \hline
     $[-2.1,-1.8)$ & $9.76 \times 10^{-2}$ & 4.3 &   4.3 &   6.0 \\
     $[-1.8,-1.5)$ & $1.37 \times 10^{-1}$ & 3.3 &   3.1 &   4.5 \\
     $[-1.5,-1.2)$ & $1.94 \times 10^{-1}$ & 2.7 &   2.4 &   3.6 \\
     $[-1.2,-0.9)$ & $2.39 \times 10^{-1}$ & 2.5 &   2.8 &   3.7 \\
     $[-0.9,-0.6)$ & $3.03 \times 10^{-1}$ & 2.0 &   2.3 &   3.0 \\
     $[-0.6,-0.3)$ & $3.41 \times 10^{-1}$ & 1.9 &   1.4 &   2.3 \\
     $[-0.3,0.0)$  & $3.60 \times 10^{-1}$ & 1.9 &   1.4 &   2.4 \\
     $[0.0,0.3)$   & $3.39 \times 10^{-1}$ & 2.0 &   2.0 &   2.8 \\
     $[0.3,0.6)$   & $3.35 \times 10^{-1}$ & 1.9 &   3.0 &   3.6 \\
     $[0.6,0.9)$   & $3.06 \times 10^{-1}$ & 1.9 &   4.2 &   4.6 \\
     $[0.9,1.2)$   & $2.25 \times 10^{-1}$ & 2.2 &   2.2 &   3.2 \\
     $[1.2,1.5)$   & $2.01 \times 10^{-1}$ & 2.6 &   2.2 &   3.4 \\
     $[1.5,1.8)$   & $1.36 \times 10^{-1}$ & 3.3 &   2.5 &   4.2 \\
     $[1.8,2.1)$   & $9.31 \times 10^{-2}$ & 4.2 &   5.6 &   7.0 \\
      \hline
    \end{tabular}

\end{table*}

\begin{table*}[htb]
  \centering
    \topcaption{Normalized differential \ttbar cross section in the \ljets channels as a function of the b jet transverse momentum ($\pt^{\PQb}$) and pseudorapidity ($\eta_{\PQb}$), and the transverse momentum ($\pt^{\bbbar}$) and the invariant mass ($m_{\bbbar}$) of the \bbbar system. The superscript `b' refers to both b and $\PAQb$ jets. The results are presented at particle level in the fiducial phase space. The statistical and systematic uncertainties are added in quadrature to yield the total uncertainty.}
    \label{tab:ljets:SummaryResultsDiffXSecSemileptonBbbarl}
    \renewcommand{\arraystretch}{1.3}
    \tabcolsep=0.09cm
    \begin{tabular}{c|c||c|c|c}
    \hline
     $\pt^{\PQb}$ bin range [\GeVns{}]
                        & $1/\!\sigma\, \rd\sigma\!/\!\rd\pt^{\PQb}$ [\invGeVns{}]
			& Stat. [\%] & Syst. [\%] & Total [\%] \\
      \hline
     $[30,48)$   & $1.40 \times 10^{-2}$ & 1.2 & 7.0 & 7.1 \\
     $[48,75)$   & $1.24 \times 10^{-2}$ & 0.9 & 1.8 & 2.0 \\
     $[75,180)$  & $3.67 \times 10^{-3}$ & 0.7 & 4.9 & 5.0 \\
     $[180,400)$ & $1.34 \times 10^{-4}$ & 3.3 & 12.4 & 12.9 \\
      \hline
      $\eta_{\PQb}$ bin range
                    & $1/\!\sigma\, \rd\sigma\!/\!\rd\eta_{\PQb}$
		    & Stat. [\%] & Syst. [\%] & Total [\%] \\
      \hline
     $[-2.4,-1.5)$ & $1.01 \times 10^{-1}$ & 1.7 & 2.9 &   3.3 \\
     $[-1.5,-1.0)$ & $2.08 \times 10^{-1}$ & 1.3 & 1.5 &   2.0 \\
     $[-1.0,-0.5)$ & $2.77 \times 10^{-1}$ & 1.1 & 1.2 &   1.7 \\
     $[-0.5,0.0)$  & $3.23 \times 10^{-1}$ & 1.1 & 2.0 &   2.2 \\
     $[0.0,0.5)$   & $3.28 \times 10^{-1}$ & 1.1 & 1.9 &   2.1 \\
     $[0.5,1.0)$   & $2.89 \times 10^{-1}$ & 1.1 & 1.0 &   1.5 \\
     $[1.0,1.5)$   & $2.09 \times 10^{-1}$ & 1.4 & 2.2 &   2.6 \\
     $[1.5,2.4)$   & $1.03 \times 10^{-1}$ & 1.6 & 2.8 &   3.2 \\
      \hline
    $\pt^{\bbbar}$ bin range [\GeVns{}]
                     & $1/\!\sigma\, \rd\sigma\!/\!\rd\pt^{\bbbar}$ [\invGeVns{}]
                     & Stat. [\%] & Syst. [\%] & Total [\%] \\
    \hline
     $[0,35)$    & $3.41 \times 10^{-3}$ & 2.4 &   4.1 &   4.7 \\
     $[35,75)$   & $6.55 \times 10^{-3}$ & 1.7 &   3.0 &   3.5 \\
     $[75,115)$  & $8.51 \times 10^{-3}$ & 1.5 &   3.0 &   3.4 \\
     $[115,155)$ & $5.04 \times 10^{-3}$ & 2.0 &   6.7 &   7.0 \\
     $[155,280)$ & $5.79 \times 10^{-4}$ & 3.8 &  11.3 &  11.9 \\
     $[280,500)$ & $0.17 \times 10^{-4}$ & 14.1 & 23.3 &  27.2 \\
    \hline
    $m_{\bbbar}$ bin range [\GeVns{}]
                     & $1/\!\sigma\, \rd\sigma\!/\!\rd m_{\bbbar}$ [\invGeVns{}]
		     & Stat. [\%] & Syst. [\%] & Total [\%] \\
    \hline
   $[0 , 85)$    & $2.00 \times 10^{-3}$ & 1.9 & 7.6 & 7.9 \\
   $[85 , 135)$  & $5.79 \times 10^{-3}$ & 1.7 & 3.9 & 4.3 \\
   $[135, 190)$ & $4.44 \times 10^{-3}$ & 2.0 & 3.0 & 3.6 \\
   $[190 , 255)$ & $2.37 \times 10^{-3}$ & 2.5 & 4.6 & 5.3 \\
   $[255 , 325)$ & $1.13 \times 10^{-3}$ & 3.7 & 7.6 & 8.5 \\
   $[325 , 415)$ & $4.50 \times 10^{-4}$ & 4.9 & 8.6 & 10.0 \\
   $[415 , 505)$ & $1.57 \times 10^{-4}$ & 8.8 & 10.9 & 14.0 \\
   $[505 , 630)$ & $0.47 \times 10^{-4}$ & 12.6 & 28.4 & 31.1 \\
   $[630 , 800)$ & $0.12 \times 10^{-4}$ & 21.9 & 15.4 & 26.8 \\

    \hline
    \hline
    \end{tabular}

\end{table*}

\begin{table*}[htb]
  \centering
    \topcaption{Normalized differential \ttbar cross section in the dilepton channels as a function of the charged lepton transverse momentum ($\pt^{\ell}$) and pseudorapidity ($\eta_{\ell}$). The superscript `$\ell$' refers to both $\ell^{+}$ and $\ell^{-}$. The results are presented at particle level in the fiducial phase space. The statistical and systematic uncertainties are added in quadrature to yield the total uncertainty.}
    \label{tab:dileptons:SummaryResultsDiffXSecdileptonLepB}
    \renewcommand{\arraystretch}{1.3}
    \tabcolsep=0.09cm
    \begin{tabular}{c|c||c|c|c}
      \hline
      $\pt^{\ell}$ bin range [\GeVns{}]
                       & $1/\!\sigma\, \rd\sigma\!/\!\rd\pt^{\ell}$ [\invGeVns{}]
		       & Stat. [\%] & Syst. [\%] & Total [\%] \\
      \hline
     $[20,  40)$    &  $1.93 \times 10^{-2}$  &   0.5 & 1.6 & 1.7   \\
     $[40,  70)$    &  $1.25 \times 10^{-2}$  &   0.5 & 0.7 & 0.9   \\
     $[70,  120)$   &  $3.79 \times 10^{-3}$  &   0.8 & 1.8 & 2.0   \\
     $[120, 180)$   &  $6.51 \times 10^{-4}$  &    2.0 & 7.2 & 7.4   \\
     $[180, 400)$   &  $3.89 \times 10^{-5}$  &    4.6 & 12.0 & 12.9 \\
      \hline
      $\eta_{\ell}$ bin range
                & $1/\!\sigma\, \rd\sigma\!/\!\rd\eta_{\ell}$
		& Stat. [\%] & Syst. [\%] & Total [\%] \\
      \hline
     $[-2.4 , -2.1)$   &  $6.04 \times 10^{-2}$  &   3.0 & 2.7 & 4.1 \\
     $[-2.1 , -1.8)$   &  $9.49 \times 10^{-2}$  &   2.2 & 2.9 & 3.7 \\
     $[-1.8 , -1.5)$   &  $1.40 \times 10^{-1}$  &   1.9 & 1.3 & 2.3 \\
     $[-1.5 , -1.2)$   &  $1.98 \times 10^{-1}$  &   1.5 & 1.5 & 2.2 \\
     $[-1.2 , -0.9)$   &  $2.43 \times 10^{-1}$  &   1.4 & 0.6 & 1.5 \\
     $[-0.9 , -0.6)$   &  $2.86 \times 10^{-1}$  &   1.2 & 2.3 & 2.6 \\
     $[-0.6 , -0.3)$   &  $3.22 \times 10^{-1}$  &   1.2 & 1.0 & 1.5 \\
     $[-0.3 , 0.0)$    &  $3.36 \times 10^{-1}$  &   1.2 & 0.7 & 1.4 \\
     $[0.0 , 0.3)$     &  $3.17 \times 10^{-1}$  &   1.2 & 1.2 & 1.7 \\
     $[0.3 , 0.6)$     &  $3.23 \times 10^{-1}$  &   1.2 & 1.2 & 1.7 \\
     $[0.6 , 0.9)$     &  $2.90 \times 10^{-1}$  &   1.3 & 1.1 & 1.7 \\
     $[0.9 , 1.2)$     &  $2.43 \times 10^{-1}$  &   1.4 & 1.0 & 1.7 \\
     $[1.2 , 1.5)$     &  $1.97 \times 10^{-1}$  &   1.6 & 1.2 & 2.0 \\
     $[1.5 , 1.8)$     &  $1.30 \times 10^{-1}$  &   2.1 & 2.6 & 3.4 \\
     $[1.8 , 2.1)$     &  $9.70 \times 10^{-2}$  &   2.3 & 3.1 & 3.8 \\
     $[2.1 , 2.4)$     &  $5.38 \times 10^{-2}$  &   3.5 & 5.2 & 6.3 \\
      \hline
    \end{tabular}

\end{table*}

\begin{table*}[htb]
  \centering
    \topcaption{Normalized differential \ttbar cross section in the dilepton channels as a function of the transverse momentum ($\pt^{\ell\ell}$) and the invariant mass ($m_{\ell\ell}$) of the dilepton pair. The superscript `$\ell$' refers to both $\ell^{+}$ and $\ell^{-}$. The results are presented at particle level in the fiducial phase space. The statistical and systematic uncertainties are added in quadrature to yield the total uncertainty.}
    \label{tab:dileptons:SummaryResultsDiffXSecdileptonLepB_2}
    \renewcommand{\arraystretch}{1.3}
    \tabcolsep=0.09cm
    \begin{tabular}{c|c||c|c|c}
      \hline
      $\pt^{\ell\ell}$ bin range [\GeVns{}]
                       & $1/\!\sigma\, \rd\sigma\!/\!\rd\pt^{\ell\ell}$ [\invGeVns{}]
		       & stat. [\%] & Syst. [\%] & Total [\%] \\
      \hline
      $[0 , 10)$      &  $1.84 \times 10^{-3}$  &   3.7 & 3.2 & 4.9 \\
      $[10 , 20)$     &  $4.82 \times 10^{-3}$  &   2.0 & 2.8 & 3.5 \\
      $[20 , 40)$     &  $7.74 \times 10^{-3}$  &   1.2 & 2.8 & 3.0 \\
      $[40 , 60)$     &  $1.10 \times 10^{-2}$  &   1.0 & 2.2 & 2.5 \\
      $[60 , 100)$    &  $9.70 \times 10^{-3}$  &   0.7 & 0.9 & 1.2 \\
      $[100 , 150)$   &  $2.87 \times 10^{-3}$  &   1.3 & 5.0 & 5.2 \\
      $[150 , 400)$   &  $1.03 \times 10^{-4}$  &   3.4 & 7.0 & 7.8 \\
      \hline
      $m_{\ell\ell}$ bin range [\GeVns{}]
                & $1/\!\sigma\, \rd\sigma\!/\!\rd m_{\ell\ell}$ [\invGeVns{}]
		& Stat. [\%] & Syst. [\%] & Total [\%] \\
      \hline
      $[20 , 30)$      &  $3.74 \times 10^{-3}$  &   2.6 & 2.3 & 3.5 \\
      $[30 , 50)$      &  $5.28 \times 10^{-3}$  &   1.5 & 2.1 & 2.6 \\
      $[50 , 76)$      &  $7.61 \times 10^{-3}$  &   1.0 & 1.6 & 1.9 \\
      $[76 , 106)$     &  $7.19 \times 10^{-3}$  &   1.0 & 2.1 & 2.4 \\
      $[106 , 130)$   &  $5.41 \times 10^{-3}$  &    1.3 & 1.4 & 1.9 \\
      $[130 , 170)$   &  $3.30 \times 10^{-3}$  &    1.4 & 1.7 & 2.2 \\
      $[170 , 260)$   &  $1.23 \times 10^{-3}$  &    1.5 & 3.4 & 3.7 \\
      $[260 , 400)$   &  $2.29 \times 10^{-4}$  &    2.7 & 6.9 & 7.4 \\
      \hline
    \end{tabular}

\end{table*}

\begin{table*}[htb]
  \centering
    \topcaption{Normalized differential \ttbar cross section in the dilepton channels as a function of the b jet transverse momentum ($\pt^{\PQb}$) and pseudorapidity ($\eta_{\PQb}$), and the transverse momentum ($\pt^{\bbbar}$) and the invariant mass ($m_{\bbbar}$) of the \bbbar system. The superscript `b' refers to both b and $\PAQb$ jets. The results are presented at particle level in the fiducial phase space. The statistical and systematic uncertainties are added in quadrature to yield the total uncertainty.}
    \label{tab:dilepton:SummaryResultsDiffXSecDileptonBbbarl}
    \renewcommand{\arraystretch}{1.3}
    \tabcolsep=0.09cm
    \begin{tabular}{c|c||c|c|c}
    \hline
     $\pt^{\PQb}$ bin range [\GeVns{}]
                        & $1/\!\sigma\, \rd\sigma\!/\!\rd\pt^{\PQb}$ [\invGeVns{}]
			& Stat. [\%] & Syst. [\%] & Total [\%] \\
      \hline
     $[30 , 50)$     &  $1.16 \times 10^{-2}$  &   1.2 & 7.7  & 7.8  \\
     $[50 , 80)$     &  $1.23 \times 10^{-2}$  &   1.2 & 3.8  & 4.0  \\
     $[80 , 130)$    &  $5.94 \times 10^{-3}$  &   1.4 &  3.4  & 3.7  \\
     $[130 , 210)$   &  $1.11 \times 10^{-3}$  &   2.2 &  5.1  & 5.5  \\
     $[210 , 400)$   &  $6.88 \times 10^{-5}$  &   6.8 &  13.5 & 15.1 \\
      \hline
      $\eta_{\PQb}$ bin range
                    & $1/\!\sigma\, \rd\sigma\!/\!\rd\eta_{\PQb}$
		    & Stat. [\%] & Syst. [\%] & Total [\%] \\
      \hline
     $[-2.4 , -1.5)$   &  $1.08 \times 10^{-1}$  &   2.3 & 6.9 & 7.3 \\
     $[-1.5 , -1.0)$   &  $2.16 \times 10^{-1}$  &   2.0 & 1.8 & 2.7 \\
     $[-1.0 , -0.5)$   &  $2.74 \times 10^{-1}$  &   1.9 & 3.2 & 3.7 \\
     $[-0.5 , 0.0)$    &  $3.01 \times 10^{-1}$  &   1.9 & 3.2 & 3.7 \\
     $[0.0 , 0.5)$     &  $3.20 \times 10^{-1}$  &  1.8 & 2.6 & 3.2 \\
     $[0.5 , 1.0)$     &  $2.78 \times 10^{-1}$  &  1.9 & 2.6 & 3.2 \\
     $[1.0 , 1.5)$     &  $2.19 \times 10^{-1}$  &  2.0 & 1.9 & 2.7 \\
     $[1.5 , 2.4)$     &  $1.09 \times 10^{-1}$  &  2.4 & 5.8 & 6.2 \\
      \hline
    $\pt^{\bbbar}$ bin range [\GeVns{}]
                    & $1/\!\sigma\, \rd\sigma\!/\!\rd\pt^{\bbbar}$ [\invGeVns{}]
                     & Stat. [\%] & Syst. [\%] & Total [\%] \\
    \hline
    $[0 , 30)$      &  $3.49 \times 10^{-3}$  &   2.3 &  3.6 & 4.3 \\
    $[30 , 60)$     &  $6.50 \times 10^{-3}$  &   1.5 &  2.9 & 3.3 \\
    $[60 , 100)$    &  $8.07 \times 10^{-3}$  &   1.2 &  2.1 & 2.4 \\
    $[100 , 180)$   &  $4.27 \times 10^{-3}$   &   1.0 &  3.2 & 3.4 \\
    $[180 , 400)$   &  $1.54 \times 10^{-4}$   &   4.4 &  9.8 & 10.7 \\
    \hline
    $m_{\bbbar}$ bin range [\GeVns{}]
                     & $1/\!\sigma\, \rd\sigma\!/\!\rd m_{\bbbar}$ [\invGeVns{}]
		     & Stat. [\%] & Syst. [\%] & Total [\%] \\
    \hline
    $[0 , 60)$      &  $8.48 \times 10^{-4}$  &   1.9 & 3.9 & 4.3 \\
    $[60 , 120)$    &  $4.57 \times 10^{-3}$  &   0.9 & 2.7 & 2.8 \\
    $[120 , 240)$   &  $3.97 \times 10^{-3}$   &   0.6 & 1.0 & 1.1 \\
    $[240 , 600)$   &  $5.35 \times 10^{-4}$   &   1.3 & 4.6 & 4.8 \\
    \hline
    \end{tabular}

\end{table*}

\begin{table*}[htb]
  \centering
    \topcaption{Normalized differential \ttbar cross section in the \ljets channels as a function of top quark or antiquark observables: the transverse momentum ($\pt^{\PQt}$) and the transverse momentum in the \ttbar rest frame ($\pt^{\PQt\ast}$) of the top quarks or antiquarks. The horizontally-corrected bin centres according to the \MADGRAPH{}+\PYTHIA{6} prediction (cf. Section~\suppRef{\ref{subsec:fullPS}}{6.2 (main document)}) are also provided. The results are presented at parton level in the full phase space. The statistical and systematic uncertainties are added in quadrature to yield the total uncertainty.}
    \label{tab:ljets:SummaryResultsDiffXSecSemileptonTopFullPS_1}
    \renewcommand{\arraystretch}{1.3}
    \tabcolsep=0.09cm
    \begin{tabular}{c|c|c||c|c|c}
      \hline
      $\pt^{\PQt}$ bin range [\GeVns{}]  & Bin centre [\GeVns{}]
                       & $1/\!\sigma\, \rd\sigma\!/\!\rd\pt^{\PQt}$ [\invGeVns{}]
		       & Stat. [\%] & Syst. [\%] & Total [\%] \\
      \hline
     $[0,60)$    &  26.25 & $4.14 \times 10^{-3}$ & 1.2 & 3.6 &   3.8 \\
     $[60,100)$  &  88.75 & $6.69 \times 10^{-3}$ & 1.3 & 1.7 &   2.1 \\
     $[100,150)$ & 126.25 & $4.96 \times 10^{-3}$ & 1.1 & 3.0 &   3.2 \\
     $[150,200)$ & 173.75 & $2.66 \times 10^{-3}$ & 1.3 & 3.5 &   3.7 \\
     $[200,260)$ & 228.75 & $1.06 \times 10^{-3}$ & 1.6 & 3.2 &   3.6 \\
     $[260,320)$ & 286.25 & $3.99 \times 10^{-4}$ & 2.2 & 5.6 &   6.0 \\
     $[320,400)$ & 356.25 & $1.30 \times 10^{-4}$ & 2.8 & 7.6 &   8.1 \\
     $[400,500)$ & 446.25 & $0.37 \times 10^{-4}$ & 5.5 & 9.5 &  10.9 \\
      \hline
      $\pt^{\PQt\ast}$ bin range [\GeVns{}]  & Bin centre [\GeVns{}]
                       & $1/\!\sigma\, \rd\sigma\!/\!\rd\pt^{\PQt\ast}$ [\invGeVns{}]
		       & Stat. [\%] & Syst. [\%] & Total [\%] \\
      \hline
     $[0,60)$    &  26.25 & $4.44 \times 10^{-3}$ & 1.6 & 4.4 & 4.7 \\
     $[60,100)$  &  63.75 & $7.03 \times 10^{-3}$ & 1.7 & 1.4 & 2.2 \\
     $[100,150)$ & 126.25 & $4.93 \times 10^{-3}$ & 1.5 & 2.9 & 3.2 \\
     $[150,200)$ & 173.75 & $2.44 \times 10^{-3}$ & 1.9 & 3.2 & 3.7 \\
     $[200,260)$ & 226.25 & $9.00 \times 10^{-4}$ & 2.4 & 3.8 & 4.4 \\
     $[260,320)$ & 286.25 & $3.21 \times 10^{-4}$ & 3.3 & 4.2 & 5.4 \\
     $[320,400)$ & 356.25 & $0.94 \times 10^{-4}$ & 4.5 & 6.3 & 7.8 \\
     $[400,500)$ & 443.75 & $0.25 \times 10^{-4}$ & 9.6 & 9.3 & 13.4 \\
      \hline
    \end{tabular}

\end{table*}

\begin{table*}[htb]
  \centering
    \topcaption{Normalized differential \ttbar cross section in the \ljets channels as a function of top quark or antiquark observables: the rapidity ($y_{\PQt}$) of the top quarks or antiquarks, and the difference in the azimuthal angle between the top quark and antiquark ($\Delta \phi(\text{t,}\bar{\PQt})$). The horizontally-corrected bin centres according to the \MADGRAPH{}+\PYTHIA{6} prediction (cf. Section~\suppRef{\ref{subsec:fullPS}}{6.2 (main document)}) are also provided. The results are presented at parton level in the full phase space. The statistical and systematic uncertainties are added in quadrature to yield the total uncertainty.}
    \label{tab:ljets:SummaryResultsDiffXSecSemileptonTopFullPS_2}
    \renewcommand{\arraystretch}{1.3}
    \tabcolsep=0.09cm
    \begin{tabular}{c|c|c||c|c|c}
      \hline
      $y_{\PQt}$ bin range & Bin centre
                & $1/\!\sigma\, \rd\sigma\!/\!\rd y_{\PQt}$
		& Stat. [\%] & Syst. [\%] & Total [\%] \\
      \hline
     $[-2.5,-1.6)$ & $-2.01$ & $7.36 \times 10^{-2}$ & 2.5 & 3.8 &   4.5 \\
     $[-1.6,-1.2)$ & $-1.39$ & $1.75 \times 10^{-1}$ & 1.5 & 2.4 &   2.9 \\
     $[-1.2,-0.8)$ & $-1.01$ & $2.61 \times 10^{-1}$ & 1.4 & 3.0 &   3.3 \\
     $[-0.8,-0.4)$ & $-0.61$ & $3.00 \times 10^{-1}$ & 1.3 & 1.3 &   1.8 \\
     $[-0.4,0.0)$  & $-0.24$ & $3.33 \times 10^{-1}$ & 1.2 & 1.5 &   1.9 \\
     $[0.0,0.4)$   &  0.24   & $3.31 \times 10^{-1}$ & 1.2 & 1.6 &   2.0 \\
     $[0.4,0.8)$   &  0.61   & $3.00 \times 10^{-1}$ & 1.3 & 1.8 &   2.2 \\
     $[0.8,1.2)$   &  1.01   & $2.47 \times 10^{-1}$ & 1.4 & 3.0 &   3.3 \\
     $[1.2,1.6)$   &  1.41   & $1.88 \times 10^{-1}$ & 1.4 & 1.6 &   2.1 \\
     $[1.6,2.5)$   &  2.01   & $7.77 \times 10^{-2}$ & 2.4 & 3.4 &   4.1 \\
      \hline
      $\Delta \phi(\text{t,}\bar{\PQt})$ bin range [rad] & Bin centre [rad]
                & $1/\!\sigma\, \rd\sigma\!/\!\rd\Delta \phi(\text{t,}\bar{\PQt})$ [rad$^{-1}$]
		& Stat. [\%] & Syst. [\%] & Total [\%] \\
      \hline
     $[0.00,2.00)$ & 1.26 & $6.83 \times 10^{-2}$ & 2.2 & 7.0 &   7.3 \\
     $[2.00,2.75)$ & 2.44 & $3.22 \times 10^{-1}$ & 1.2 & 3.8 &   4.0 \\
     $[2.75,3.00)$ & 2.89 & 1.13                 & 1.4 & 3.2 &   3.5 \\
     $[3.00,3.15)$ & 3.14 & 2.27                 & 1.3 & 6.6 &   6.8 \\
      \hline
    \end{tabular}

\end{table*}

\begin{table*}[htb]
  \centering
    \topcaption{Normalized differential \ttbar cross section in the \ljets channels as a function of the \pt of the leading ($\pt^{\text{t1}}$) and trailing ($\pt^{\text{t2}}$) top quarks or antiquarks. The horizontally-corrected bin centres according to the \MADGRAPH{}+\PYTHIA{6} prediction (cf. Section~\suppRef{\ref{subsec:fullPS}}{6.2 (main document)}) are also provided. The results are presented at parton level in the full phase space. The statistical and systematic uncertainties are added in quadrature to yield the total uncertainty.}
    \label{tab:ljets:SummaryResultsDiffXSecSemileptonTopFullPS_3}
    \renewcommand{\arraystretch}{1.3}
    \tabcolsep=0.09cm
    \begin{tabular}{c|c|c||c|c|c}
      \hline
      $\pt^{\text{t1}}$ bin range [\GeVns{}]  & Bin centre [\GeVns{}]
                       & $1/\!\sigma\, \rd\sigma\!/\!\rd\pt^{\text{t1}}$ [\invGeVns{}]
		       & Stat. [\%] & Syst. [\%] & Total [\%] \\
      \hline
      $[0, 60)$    &  31.25 & $2.61 \times 10^{-3}$ & 2.2 &  5.0 &   5.4 \\
      $[60, 100)$  &  76.25 & $6.39 \times 10^{-3}$ & 1.5 &  2.3 &   2.8 \\
      $[100, 150)$ & 126.25 & $5.56 \times 10^{-3}$ & 1.3 &  3.3 &   3.6 \\
      $[150, 200)$ & 173.75 & $3.36 \times 10^{-3}$ & 1.6 &  3.6 &   3.9 \\
      $[200, 260)$ & 228.75 & $1.43 \times 10^{-3}$ & 1.9 &  3.4 &   3.9 \\
      $[260, 320)$ & 286.25 & $5.56 \times 10^{-4}$ & 2.5 &  6.3 &   6.8 \\
      $[320, 400)$ & 356.25 & $1.87 \times 10^{-4}$ & 3.1 &  8.2 &   8.8 \\
      $[400, 500)$ & 446.25 & $0.56 \times 10^{-4}$ & 6.0 & 10.5 &  12.1 \\
      \hline
      $\pt^{\text{t2}}$ bin range [\GeVns{}]  & Bin centre [\GeVns{}]
                       & $1/\!\sigma\, \rd\sigma\!/\!\rd\pt^{\text{t2}}$ [\invGeVns{}]
		       & Stat. [\%] & Syst. [\%] & Total [\%] \\
      \hline
      $[0,60)$    &  23.75 & $5.64 \times 10^{-3}$ & 1.4 &   3.2 &   3.5 \\
      $[60,100)$  &  83.75 & $6.98 \times 10^{-3}$ & 1.9 &   1.8 &   2.6 \\
      $[100,150)$ & 123.75 & $4.39 \times 10^{-3}$ & 1.7 &   2.6 &   3.1 \\
      $[150,200)$ & 173.75 & $1.97 \times 10^{-3}$ & 2.2 &   3.2 &   3.9 \\
      $[200,260)$ & 228.75 & $6.93 \times 10^{-4}$ & 2.8 &   3.3 &   4.3 \\
      $[260,320)$ & 286.25 & $2.43 \times 10^{-4}$ & 4.0 &   4.5 &   6.0 \\
      $[320,400)$ & 356.25 & $0.75 \times 10^{-4}$ & 5.3 &   7.3 &   9.0 \\
      $[400,500)$ & 443.75 & $0.19 \times 10^{-4}$ & 11.4 & 11.1 &  16.0 \\
      \hline
    \end{tabular}

\end{table*}

 \begin{table*}[htb]
  \centering
    \topcaption{Normalized differential \ttbar cross section in the \ljets channels as a function of top quark pair observables: the transverse momentum ($\pt^{\ttbar}$), the rapidity ($y_{\ttbar}$) and the invariant mass ($m_{\ttbar}$) of the \ttbar system. The horizontally-corrected bin centres according to the \MADGRAPH{}+\PYTHIA{6} prediction (cf. Section~\suppRef{\ref{subsec:fullPS}}{6.2 (main document)}) are also provided. The results are presented at parton level in the full phase space. The statistical and systematic uncertainties are added in quadrature to yield the total uncertainty. }
    \label{tab:ljets:SummaryResultsDiffXSecSemileptonTTbarFullPS}
    \renewcommand{\arraystretch}{1.3}
    \tabcolsep=0.09cm
    \begin{tabular}{c|c|c||c|c|c}
      \hline
      $\pt^{\ttbar}$ bin range [\GeVns{}]  & Bin centre [\GeVns{}]
                       & $1/\!\sigma\, \rd\sigma\!/\!\rd\pt^{\ttbar}$ [\invGeVns{}]
		       & Stat. [\%] & Syst. [\%] & Total [\%] \\
      \hline
      $[0,20)$    &   4.50 & $1.32 \times 10^{-2}$ & 2.3 &  7.9 &   8.2 \\
      $[20,45)$   &  32.50 & $1.18 \times 10^{-2}$ & 2.1 &  3.4 &   3.9 \\
      $[45,75)$   &  58.50 & $6.40 \times 10^{-3}$ & 1.8 &  5.5 &   5.8 \\
      $[75,120)$  &  95.50 & $2.84 \times 10^{-3}$ & 2.3 &  6.6 &   7.0 \\
      $[120,190)$ & 152.50 & $1.07 \times 10^{-3}$ & 2.6 &  5.7 &   6.2 \\
      $[190,300)$ & 237.50 & $3.06 \times 10^{-4}$ & 3.8 & 12.2 &  12.8 \\
      \hline
      $y_{\ttbar}$ bin range & Bin centre
                & $1/\!\sigma\, \rd\sigma\!/\!\rd y_{\ttbar}$
		& Stat. [\%] & Syst. [\%] & Total [\%] \\
      \hline
      $[-2.5,-1.3)$  & $-1.82$ & $6.07 \times 10^{-2}$  & 3.5 & 2.9 &   4.5 \\
      $[-1.3,-0.9)$  & $-1.09$ & $2.20 \times 10^{-1}$  & 1.8 & 2.3 &   3.0 \\
      $[-0.9,-0.6)$  & $-0.76$ & $3.27 \times 10^{-1}$   & 1.8 & 2.9 &   3.4 \\
      $[-0.6,-0.3)$  & $-0.46$ & $3.73 \times 10^{-1}$   & 1.7 & 1.6 &   2.3 \\
      $[-0.3,0.0)$   & $-0.14$ & $4.27 \times 10^{-1}$   & 1.6 & 1.6 &   2.2 \\
      $[0.0,0.3)$    &  0.14   & $4.13 \times 10^{-1}$  & 1.6 & 1.9 &   2.4 \\
      $[0.3,0.6)$    &  0.47   & $3.74 \times 10^{-1}$  & 1.7 & 1.7 &   2.4 \\
      $[0.6,0.9)$    &  0.76   & $3.17 \times 10^{-1}$  & 1.8 & 1.7 &   2.5 \\
      $[0.9,1.3)$    &  1.09   & $2.30 \times 10^{-1}$ & 1.8 & 2.2 &   2.8 \\
      $[1.3,2.5)$    &  1.82   & $6.41 \times 10^{-2}$ & 3.3 & 3.8 &   5.1 \\
      \hline
      $m_{\ttbar}$ bin range [\GeVns{}]  & Bin centre [\GeVns{}]
                       & $1/\!\sigma\, \rd\sigma\!/\!\rd m_{\ttbar}$ [\invGeVns{}]
		       & Stat. [\%] & Syst. [\%] & Total [\%] \\
      \hline
      $[345,400)$   &  362.5 0& $4.69 \times 10^{-3}$ & 2.1 & 7.1 &   7.5 \\
      $[400,470)$   &  435.50 & $4.30 \times 10^{-3}$ & 2.1 & 2.9 &   3.6 \\
      $[470,550)$   &  508.50 & $2.67 \times 10^{-3}$ & 2.0 & 6.1 &   6.4 \\
      $[550,650)$   &  595.50 & $1.17 \times 10^{-3}$ & 2.3 & 7.3 &   7.7 \\
      $[650,800)$   &  717.50 & $4.66 \times 10^{-4}$ & 2.5 & 4.2 &   4.9 \\
      $[800,1100)$  &  927.50 & $1.14 \times 10^{-4}$ & 3.0 & 9.5 &   10.0 \\
      $[1100,1600)$ & 1328.50 & $0.11 \times 10^{-4}$ & 8.0 & 9.8 &   12.7 \\
      \hline
     \end{tabular}

   \end{table*}

\begin{table*}[htb]
  \centering
    \topcaption{Normalized differential \ttbar cross section in the dilepton channels as a function of top quark or antiquark observables: the transverse momentum ($\pt^{\PQt}$), the transverse momentum in the \ttbar rest frame ($\pt^{\PQt\ast}$), and the rapidity ($y_{\PQt}$) of the top quarks or antiquarks, and the difference in the azimuthal angle between the top quark and antiquark ($\Delta \phi(\text{t,}\bar{\PQt})$). The horizontally-corrected bin centres according to the \MADGRAPH{}+\PYTHIA{6} prediction (cf. Section~\suppRef{\ref{subsec:fullPS}}{6.2 (main document)}) are also provided. The results are presented at parton level in the full phase space. The statistical and systematic uncertainties are added in quadrature to yield the total uncertainty.}
    \label{tab:dilepton:SummaryResultsDiffXSecDileptonTopFullPS_1}
    \renewcommand{\arraystretch}{1.3}
    \tabcolsep=0.09cm
    \begin{tabular}{c|c|c||c|c|c}
      \hline
      $\pt^{\PQt}$ bin range [\GeVns{}]  & Bin centre [\GeVns{}]
                       & $1/\!\sigma\, \rd\sigma\!/\!\rd\pt^{\PQt}$ [\invGeVns{}]
		       & Stat. [\%] & Syst. [\%] & Total [\%] \\
      \hline
      $[0 , 65)$      &  28.75 &  $4.33 \times 10^{-3}$  &   1.1 & 2.3 & 2.5 \\
      $[65 , 125)$    & 101.25 &  $6.40 \times 10^{-3}$  &   0.9 & 1.4 & 1.7 \\
      $[125 , 200)$   & 161.25 &  $3.08 \times 10^{-3}$  &   1.0 & 2.1 & 2.3 \\
      $[200 , 290)$   & 238.75 &  $8.62 \times 10^{-4}$  &   1.3 & 3.6 & 3.8 \\
      $[290 , 400)$   & 336.25 &  $1.88 \times 10^{-4}$  &   2.7 & 7.4 & 7.9 \\
      \hline
      $\pt^{\PQt\ast}$ bin range [\GeVns{}]  & Bin centre [\GeVns{}]
                       & $1/\!\sigma\, \rd\sigma\!/\!\rd\pt^{\PQt\ast}$ [\invGeVns{}]
		       & Stat. [\%] & Syst. [\%] & Total [\%] \\
      \hline
      $[0 , 60)$      &  26.25 & $4.45\times 10^{-3}$  &   1.5 & 2.7 & 3.1  \\
      $[60 , 115)$    &  93.75 & $6.89\times 10^{-3}$  &   1.4 & 2.1 & 2.5  \\
      $[115 , 190)$   & 151.25 & $3.41\times 10^{-3}$  &   1.3 & 2.2 & 2.6  \\
      $[190 , 275)$   & 226.25 & $8.78\times 10^{-4}$  &   2.2 & 5.5 & 5.9  \\
      $[275 , 380)$   & 318.75 & $1.87\times 10^{-4}$  &   2.6 & 6.2 & 6.7  \\
      $[380 , 500)$   & 428.75 & $2.91\times 10^{-5}$  &   7.5 & 9.1 & 11.8 \\
      \hline
      $y_{\PQt}$ bin range & Bin centre
                & $1/\!\sigma\, \rd\sigma\!/\!\rd y_{\PQt}$
		& Stat. [\%] & Syst. [\%] & Total [\%] \\
      \hline
      $[-2.5 , -1.6)$   &$-2.01$ & $7.63\times 10^{-2}$  &   2.0 & 5.3 & 5.6 \\
      $[-1.6 , -1.0)$   &$-1.31$ & $1.97\times 10^{-1}$  &   1.0 & 1.5 & 1.8 \\
      $[-1.0 , -0.5)$   &$-0.76$ & $2.82\times 10^{-1}$  &   1.1 & 1.5 & 1.9 \\
      $[-0.5 , 0.0)$    &$-0.29$ & $3.35\times 10^{-1}$  &   1.0 & 1.5 & 1.8 \\
      $[0.0 , 0.5)$     & 0.29   & $3.28\times 10^{-1}$  &   1.0 & 1.9 & 2.2 \\
      $[0.5 , 1.0)$     & 0.76   & $2.85\times 10^{-1}$  &   1.1 & 1.1 & 1.5 \\
      $[1.0 , 1.6)$     & 1.31   & $2.03\times 10^{-1}$  &   1.1 & 1.1 & 1.5 \\
      $[1.6 , 2.5)$     & 2.01   & $7.40\times 10^{-2}$  &   2.0 & 5.3 & 5.7 \\
      \hline
      $\Delta \phi(\text{t,}\bar{\PQt})$ bin range [rad]  & Bin centre [rad]
                & $1/\!\sigma\, \rd\sigma\!/\!\rd\Delta \phi(\text{t,}\bar{\PQt})$ [rad$^{-1}$]
		& Stat. [\%] & Syst. [\%] & Total [\%] \\
      \hline
      $[0 , 1.89)$    & 1.19 & $6.40\times 10^{-2}$  &   2.0 & 7.3 & 7.6 \\
      $[1.89 , 2.77)$ & 2.44 & $2.96\times 10^{-1}$  &   1.0 & 5.6 & 5.7 \\
      $[2.77 , 3.04)$ & 2.94 & $1.24$               &   1.1 & 3.2 & 3.3 \\
      $[3.04 , 3.15)$ & 3.09 & $2.58$               &   1.5 & 7.8 & 8.0 \\
      \hline
    \end{tabular}

\end{table*}

\begin{table*}[htb]
  \centering
    \topcaption{Normalized differential \ttbar cross section in the dilepton channels as a function of the \pt of the leading ($\pt^{\text{t1}}$) and trailing ($\pt^{\text{t2}}$) top quarks or antiquarks. The results are presented at parton level in the full phase space. The statistical and systematic uncertainties are added in quadrature to yield the total uncertainty.}
    \label{tab:dilepton:SummaryResultsDiffXSecDileptonTopFullPS_2}
    \renewcommand{\arraystretch}{1.3}
    \tabcolsep=0.09cm
    \begin{tabular}{c|c|c||c|c|c}
      \hline
      $\pt^{\text{t1}}$ bin range [\GeVns{}]  & Bin centre [\GeVns{}]
                       & $1/\!\sigma\, \rd\sigma\!/\!\rd\pt^{\text{t1}}$ [\invGeVns{}]
		       & Stat. [\%] & Syst. [\%] & Total [\%] \\
      \hline
      $[0 , 75)$      &  36.25 & $3.31 \times 10^{-3}$  &   1.5 & 3.7 & 4.0 \\
      $[75 , 130)$    & 111.25 & $6.43 \times 10^{-3}$  &   1.1 & 1.8 & 2.1 \\
      $[130 , 200)$   & 163.75 & $3.67 \times 10^{-3}$  &   1.2 & 2.9 & 3.2 \\
      $[200 , 290)$   & 241.25 & $1.17 \times 10^{-3}$  &   1.5 & 3.9 & 4.2 \\
      $[290 , 400)$   & 336.25 & $2.61 \times 10^{-4}$  &   3.2 & 6.9 & 7.6 \\
      \hline
      $\pt^{\text{t2}}$ bin range [\GeVns{}]  & Bin centre [\GeVns{}]
                       & $1/\!\sigma\, \rd\sigma\!/\!\rd\pt^{\text{t2}}$ [\invGeVns{}]
		       & Stat. [\%] & Syst. [\%] & Total [\%] \\
      \hline
      $[0 , 55)$      &  23.7 5& $5.38 \times 10^{-3}$  &   1.7 & 2.2  & 2.8 \\
      $[55 , 120)$    &  91.25 & $6.74 \times 10^{-3}$  &   1.4 & 1.6  & 2.1 \\
      $[120 , 200)$   & 156.25 & $2.50 \times 10^{-3}$  &   1.7 & 2.1  & 2.7 \\
      $[200 , 290)$   & 238.75 & $5.58 \times 10^{-4}$  &   2.4 & 4.6  & 5.2 \\
      $[290 , 400)$   & 338.75 & $1.14 \times 10^{-4}$  &   5.1 & 10.2 & 11.4 \\
      \hline
    \end{tabular}

\end{table*}

 \begin{table*}[htb]
  \centering
    \topcaption{Normalized differential \ttbar cross section in the dilepton channels as a function of top quark pair observables: the transverse momentum ($\pt^{\ttbar}$), the rapidity ($y_{\ttbar}$) and the invariant mass ($m_{\ttbar}$) of the \ttbar system. The results are presented at parton level in the full phase space. The statistical and systematic uncertainties are added in quadrature to yield the total uncertainty.}
    \label{tab:dilepton:SummaryResultsDiffXSecDileptonTTbarFullPS}
    \renewcommand{\arraystretch}{1.3}
    \tabcolsep=0.09cm
    \begin{tabular}{c|c|c||c|c|c}
      \hline
      $\pt^{\ttbar}$ bin range [\GeVns{}]  & Bin centre [\GeVns{}]
                       & $1/\!\sigma\, \rd\sigma\!/\!\rd\pt^{\ttbar}$ [\invGeVns{}]
		       & Stat. [\%] & Syst. [\%] & Total [\%] \\
      \hline
      $[0 , 30)$    &   4.50 & $1.43\times 10^{-2}$  &   0.8 & 6.1 & 6.1 \\
      $[30 , 80)$   &  51.50 & $6.90\times 10^{-3}$  &   1.0 & 4.7 & 4.8 \\
      $[80 , 170)$  & 118.50 & $1.91\times 10^{-3}$  &   1.1 & 5.4 & 5.5 \\
      $[170 , 300)$ & 223.50 & $3.47\times 10^{-4}$  &   2.1 & 4.7 & 5.1 \\
      \hline
      $y_{\ttbar}$ bin range & Bin centre
                & $1/\!\sigma\, \rd\sigma\!/\!\rd y_{\ttbar}$
		& Stat. [\%] & Syst. [\%] & Total [\%] \\
      \hline
      $[-2.5 , -1.5)$  & $-1.93$ & $4.71\times 10^{-2}$  &   4.0 & 6.6 & 7.7 \\
      $[-1.5 , -1.0)$  & $-1.24$ & $1.82\times 10^{-1}$  &   1.8 & 1.5 & 2.3 \\
      $[-1.0 , -0.5)$  & $-0.76$ & $3.09\times 10^{-1}$  &   1.4 & 1.5 & 2.1 \\
      $[-0.5 , 0.0)$   & $-0.31$ & $4.18\times 10^{-1}$  &   1.2 & 1.7 & 2.1 \\
      $[0.0 , 0.5)$    &  0.29   & $4.09\times 10^{-1}$  &   1.2 & 1.2 & 1.7 \\
      $[0.5 , 1.0)$    &  0.76   & $3.15\times 10^{-1}$  &   1.5 & 1.3 & 2.0 \\
      $[1.0 , 1.5)$    &  1.24   & $1.79\times 10^{-1}$  &   1.8 & 1.8 & 2.6 \\
      $[1.5 , 2.5)$    &  1.93   & $4.59\times 10^{-2}$  &   4.0 & 6.0 & 7.2 \\
      \hline
      $m_{\ttbar}$ bin range [\GeVns{}]  & Bin centre [\GeVns{}]
                       & $1/\!\sigma\, \rd\sigma\!/\!\rd m_{\ttbar}$ [\invGeVns{}]
		       & Stat. [\%] & Syst. [\%] & Total [\%] \\
      \hline
      $[340 , 380)$    &  354.50 & $4.14\times 10^{-3}$  &   3.0 &  8.6  & 9.1 \\
      $[380 , 470)$    &  428.50 & $4.50\times 10^{-3}$  &   1.7 &  5.3  & 5.6 \\
      $[470 , 620)$    &  537.50 & $1.95\times 10^{-3}$  &   1.8 &  2.9  & 3.4 \\
      $[620 , 820)$    &  705.50 & $5.25\times 10^{-4}$  &   2.8 &  3.2  & 4.2 \\
      $[820 , 1100)$   &  940.50 & $1.00\times 10^{-4}$  &   3.7 &  7.3  & 8.2 \\
      $[1100 , 1600)$  & 1328.50 & $7.28\times 10^{-6}$  &   14.4 & 28.2 & 31.6 \\
      \hline
     \end{tabular}

   \end{table*} 
}

\cleardoublepage \section{The CMS Collaboration \label{app:collab}}\begin{sloppypar}\hyphenpenalty=5000\widowpenalty=500\clubpenalty=5000\textbf{Yerevan Physics Institute,  Yerevan,  Armenia}\\*[0pt]
V.~Khachatryan, A.M.~Sirunyan, A.~Tumasyan
\vskip\cmsinstskip
\textbf{Institut f\"{u}r Hochenergiephysik der OeAW,  Wien,  Austria}\\*[0pt]
W.~Adam, T.~Bergauer, M.~Dragicevic, J.~Er\"{o}, M.~Friedl, R.~Fr\"{u}hwirth\cmsAuthorMark{1}, V.M.~Ghete, C.~Hartl, N.~H\"{o}rmann, J.~Hrubec, M.~Jeitler\cmsAuthorMark{1}, W.~Kiesenhofer, V.~Kn\"{u}nz, M.~Krammer\cmsAuthorMark{1}, I.~Kr\"{a}tschmer, D.~Liko, I.~Mikulec, D.~Rabady\cmsAuthorMark{2}, B.~Rahbaran, H.~Rohringer, R.~Sch\"{o}fbeck, J.~Strauss, W.~Treberer-Treberspurg, W.~Waltenberger, C.-E.~Wulz\cmsAuthorMark{1}
% ...
\vskip\cmsinstskip
% ...
1:~~Also at Vienna University of Technology, Vienna, Austria\\
2:~~Also at CERN, European Organization for Nuclear Research, Geneva, Switzerland\\
% ...

\end{sloppypar}
\end{document}